%% file: sn-article.tex
\documentclass[journal=jctcce,manuscript=article,layout=traditional,hyperref=true]{achemso}

\usepackage{graphicx}%
\usepackage{multirow}%
\usepackage{amsmath,amssymb,amsfonts}%
\usepackage{amsthm}%
\usepackage{mathrsfs}%
\usepackage[title]{appendix}%
\usepackage{xcolor}%
\usepackage{textcomp}%
\usepackage{manyfoot}%
\usepackage{booktabs}%
\usepackage{algorithm}%
\usepackage{algorithmicx}%
\usepackage{algpseudocode}%
\usepackage{listings}%
\usepackage{makecell} 
\usepackage{placeins}
\usepackage{tabularx}
\usepackage[version=4]{mhchem}
\usepackage{longtable}
\usepackage{soul}
\usepackage{microtype}
\hypersetup{hidelinks}

\newcommand{\tabnote}[2]{%
  \par\noindent\hangindent=0.7em\hangafter=1
  \makebox[0.7em][l]{#1}#2%
}

\ifdefined\affilsize
\setkeys{acs}{chaptertitle=true}

  \renewcommand{\affilsize}{\footnotesize\renewcommand{\baselinestretch}{1.05}\selectfont}
  
\else
  \providecommand{\affiliation}[2][]{}
  \providecommand{\alsoaffiliation}[2][]{}
  \providecommand{\altaffiliation}[2][]{}
  \renewcommand{\affiliation}[2][]{\affil{#2}}
  \renewcommand{\alsoaffiliation}[2][]{\affil{#2}}
  \renewcommand{\altaffiliation}[2][]{\affil{#2}}
  \renewcommand{\email}[1]{\emailaddr{#1}}
\fi

\title{A System-Independent Metadynamics Strategy for Reactive Training Data: Application to Gas-Phase Organic Reactions}

\author{Wanrun Jiang}
\affiliation{AI for Science Institute, Beijing 100080, P. R. China}
\author{Jinzhe Zeng}
\affiliation{School of Artificial Intelligence and Data Science, State Key Laboratory of Precision and Intelligent Chemistry, University of Science and Technology of China, Hefei 230026, P. R. China}
\alsoaffiliation{Suzhou Institute for Advanced Research, University of Science and Technology of China, Suzhou 215123, P. R. China}
\alsoaffiliation{Suzhou Big Data {\protect\&} AI Research and Engineering Center, Suzhou 215123, P. R. China}
\author{Manyi Yang}
\affiliation{State Key Laboratory of Coordination Chemistry, Key Laboratory of Mesoscopic Chemistry of Ministry of Education, Nanjing University, Nanjing 210093, P. R. China}
\alsoaffiliation{The Institute of Green Chemistry and Engineering, Nanjing University, Suzhou 215163, P. R. China}
\email{manyi.yang@nju.edu.cn}
\author{Tong Zhu}
\affiliation{Shanghai Frontiers Science Center of Molecule Intelligent Syntheses, Shanghai Engineering Research Center of Molecular Therapeutics and New Drug Development, SII--ECNU Joint Autonomous Chemistry Laboratory, School of Chemistry and Molecular Engineering, East China Normal University, Shanghai 200062, P. R. China}
\alsoaffiliation{AI for Science Institute, Beijing 100080, P. R. China}
\alsoaffiliation{Shanghai Innovation Institute, Shanghai 200003, P. R. China}
\email{tzhu@lps.ecnu.edu.cn}
\author{Han Wang}
\affiliation{National Key Laboratory of Computational Physics, Institute of Applied Physics and Computational Mathematics, Fenghao East Road 2, Beijing 100094, P. R. China}
\alsoaffiliation{HEDPS, CAPT, College of Engineering, Peking University, Beijing 100871, P. R. China}
\email{wang\_han@iapcm.ac.cn}

\begin{document}
\ifdefined\affilsize\else\maketitle\fi

\begin{abstract}
General-purpose machine-learning interatomic potentials (MLIPs) for organic reactions need to accurately describe both minimum-energy paths (MEPs) for evaluating reaction energetics and the broader configurational space for simulating reaction dynamics. 
Existing general datasets for gas-phase reactions largely rely on quasi-static optimization near MEPs, leaving off-MEP regions under-covered and model reliability in reactive molecular dynamics weakly supported; the gap is methodological, not a question of dataset size.
We introduce a spatiotemporally resolved, system-independent collective variable (CV): Cartesian RMSD within randomly partitioned local domains against an expanding list of time-averaged reference geometries.
The CV drives metadynamics as the exploration engine, supplemented by transition state (TS) optimization to augment coverage around TSs.
Within a concurrent-learning workflow, this produces OpenRxn26, a dataset of 1.8~M DFT-labeled configurations covering neutral singlet unimolecular reactions in the H/C/N/O chemical space ($N_\mathrm{heavy} \leq 30$), containing reactive atomic environments underrepresented in community datasets.
Trained on OpenRxn26, a DPA4 model (DPA4\_rxn26) is the only model in the benchmark suite to achieve transferable sub-kcal/mol energy accuracy on barrier heights, reaction energies and off-MEP reactive trajectories, relative to the training DFT reference, demonstrating the efficiency and generality of the sampling strategy.
Domain MLIPs performing well on static benchmarks degrade several-fold on off-MEP reactive trajectories (e.g.\ MACE-OMol25), showing the insufficiency of MEP-anchored benchmarks for guaranteeing dynamics reliability.
The sampling strategy enables efficient development of reliable general-purpose reactive MLIPs, with OpenRxn26 as a demonstration of MD-ready training data for gas-phase neutral singlet organic reactions.

\end{abstract}

\section{Introduction}\label{sec1}

Organic chemical reactions mediate structural and energy transformations across the vast space of organic molecules\cite{Ruddigkeit2012}, underlying biochemical processes, materials synthesis, and energy conversion.
Accurate and efficient prediction of elementary-step reaction behaviors underpins mechanism discovery under operative conditions, rational design of synthetic routes, and high-throughput \emph{in silico} screening.

Reliable prediction of reaction behavior requires both static analysis of the potential energy surface (PES) and direct simulation of reaction dynamics on it.
The PES picture was first introduced by Marcelin to rationalize the Arrhenius equation within a statistical-mechanical framework\cite{Marcelin1915}.
Transition state theory (TST)\cite{Eyring1935} sharpened this picture by relating the rate, under the statistical quasi-equilibrium assumption, to the free-energy barrier between the reactant minimum and the transition state (TS) saddle point.
The minimum energy path (MEP)\cite{Fukui1970} that connects reactant, TS and product on the PES then anchors static analysis of the reaction's basic properties.
These include rate constants (from the barrier height), equilibrium constants (from the reaction free energy), reaction mechanisms (from geometry evolution on MEP), and selectivities (from barriers of competing MEPs).
Refinements built on the same statistical foundation, such as variational transition state theory (VTST)\cite{Wigner1938,Keck1967,Truhlar1980} and RRKM theory\cite{Rice1927,Kassel1928,Marcus1952}, inherit the MEP-anchored analysis.
However, real reactions do not always satisfy this statistical quasi-equilibrium assumption.
Capturing transient non-equilibrated kinetics and trajectories that stray from the MEP is then critical and requires direct simulation of the dynamic process.
Since the 1990s, non-statistical effects have been systematically documented in organic chemistry\cite{Tantillo2021,IUPAC2022}, including dynamic matching\cite{Carpenter1992,Carpenter2005}, post-transition-state bifurcations\cite{Ess2008}, roaming\cite{Suits2020}, and recrossing\cite{GonzalezJames2012}. These effects have reshaped the understanding of reaction mechanisms, selectivities, and timescales in both fundamental and industrial contexts.

For instance, formal [1,3]-sigmatropic rearrangements can proceed through a biradical intermediate whose lifetime is shorter than the intramolecular vibrational energy redistribution (IVR) timescale\cite{Carpenter1995}.
Trajectories therefore retain momentum from the entry channel, biasing products toward configurations with inverted migrating carbon.
The resulting dynamic-matching stereoselectivity overturns the classical concerted picture governed by Woodward-Hoffmann rules\cite{Berson1968,Baldwin2008}.
In atmospheric formaldehyde dissociation, H-atom roaming locks energy and angular momentum into long-range orbiting modes, preventing statistical randomization.
The resulting modulation of the radical-vs-molecular branching ratio and the vibrational activation of products lies beyond statistical rate theories\cite{Townsend2004,Lahankar2008}.
Diels-Alder cycloaddition, prototypical in pharmaceutical synthesis, can exhibit post-TS bifurcations in which a single TS connects to multiple products, invalidating static barrier comparisons for predicting branching\cite{Caramella2002,Wang2009}.
\ce{C60}-modified variants of Diels-Alder cycloaddition make the static-dynamic gap even
wider: although the MEP profile is similar and the barrier is 7.8~kcal/mol lower than the unmodified reaction, the trajectory traverses the TS region 50\% more slowly because long-range \ce{C60} interactions induce complex motions that hinder reactant alignment\cite{Hou2024}.

Addressing these challenges requires PES descriptions that are accurate and efficient, covering not only the MEP but also the configurational space sampled by reaction dynamics.
High-level first-principles calculations\cite{Mardirossian2017} can reach chemical accuracy ($\leq$1~kcal/mol) that guarantees room-temperature rates within a 5-fold margin of experiment values\cite{Truhlar1996}, but their cost rules out the long-time or large-scale molecular dynamics needed to capture reactive processes.
Empirical reactive force fields such as AIREBO\cite{Stuart2000} and ReaxFF\cite{vanDuin2001} trade accuracy for efficiency, becoming unsuitable for static PES analysis and unreliable under mild conditions, where small energy errors are amplified by the exponential temperature dependence of Eyring rates.
Machine-learning interatomic potentials (MLIPs) aim to overcome this trade-off: trained on first-principles data, they promise near-DFT accuracy at force-field-like cost\cite{Behler2007,MLIPReview2025}.
However, MLIP applicability is bounded by training-data coverage, which for organic reactions must span both the MEP skeleton across the chemical space and the configurational space sampled by reaction dynamics.
Existing general datasets focus on the former, leaving the dynamic regime systematically under-covered and reactive trajectories beyond the reach of current MLIPs.

This gap is methodological, not a question of dataset size.
Current general data-sampling strategies for gas-phase reactions share a common engine, quasi-static structural relaxation, which by design converges to the MEP vicinities and cannot reach dynamic configurations.
One family relaxes the global path along enumerated driving coordinates\cite{Grambow2020} or between enumerated reactant-product pairs\cite{YARP2021}, via nudged elastic band (NEB)\cite{NEB1998}, growing string method (GSM)\cite{GSM2013}, and intrinsic reaction coordinate (IRC)\cite{Gonzalez1989} refinement, underpinning datasets such as Transition1x\cite{Grambow2020,Transition1x}, RGD1\cite{RGD1}, RXN-xTB-AL\cite{RXNxTBAL}, and path-allocation data of AIMNet2-rxn\cite{AIMNet2rxn}.
A second family relaxes individual structures under constraints of external force or fixed distance to trace low-energy paths approximating MEPs, as in ANI-1xBB\cite{ANI1xBB} (13.1 million bond-stretching configurations) and the OMol25\cite{OMol25} subset that applies artificial force induced reaction (AFIR)\cite{AFIR} on established mechanism databases.
This dependence on relaxation cannot be patched at the architectural level.
Auxiliary equilibrium sampling on relaxed configurations, such as normal mode sampling (NMS) in AIMNet2-rxn or fixed-bond molecular dynamics in ANI-1xBB and AIMNet2-rxn, only performs local vibrational expansion and cannot reach trajectories that stray from MEPs.
Bias-driven relaxation such as stochastic surface walking (SSW)\cite{SSW2013} maps reaction-network skeletons along low-energy paths rather than dynamic trajectories.
Hybrid workflows such as MDCD\cite{MDCD2018,MDCD20} combine molecular dynamics over non-reactive configurations with coordinate-driven relaxed scans for bond rearrangement, but each resulting pathway remains confined to MEP vicinity.
The clearest evidence is hydrogen combustion\cite{HeadGordon2023}: a system small enough (six atoms per channel) that all 19 reaction channels have been enumerated and each MEP sampled by IRC with NMS augmentation.
Even so, the trained MLIP fails on direct MD below 500~K, because trajectories released from the transition states diverge from the MEPs and fall outside the training configurational space.
Only after switching to metadynamics sampling, a dynamics-working MLIP was obtained.

Among enhanced-sampling MD techniques that simultaneously explore chemical and conformational spaces of reactions\cite{enhancesamplingReview2019}, metadynamics\cite{Huber1994,Laio2002} fits training-data generation well: by repelling the system from already-visited regions, it naturally produces the broad coverage of diverse atomic environments that training sets benefit from, which is mechanistically implemented through depositing history-dependent bias potentials along selected collective variables (CVs)\cite{Bussi2020Metadynamics}.
Metadynamics has been validated for training-set generation in specific gas-phase 
or solution reactions\cite{Yang2022,Vitartas2025,HeadGordon2023}, but its broader application as a general sampling engine remains constrained by the difficulty of designing transferable CVs.
Traditional CVs are system-dependent: they capture the reaction coordinates of particular channels and require prior knowledge of targeted mechanisms or outcomes, whether crafted manually or extracted by data-driven approaches\cite{DiscriminantsCV2018,HLDA2019,GNNCV2024}.
General sampling, by contrast, is open-ended exploration towards unknown reactions\cite{EndExploreNetwork2020}, where such prior knowledge is unavailable.
Attempts to harvest partial knowledge such as reactant-product pairs by graphical enumeration face combinatorial explosion with molecular size, with no kinetic basis for down-sampling.
The bottleneck is therefore a general, system-independent CV that drives diverse reactions from arbitrary molecules, operates under mild conditions with kinetic relevance to practice, and is scalable to large systems.
Grimme's 2019 scheme is the first system-independent attempt on organic reactions, using Cartesian RMSD against a dynamically accumulating reference library\cite{Grimme2019}.
In gas-phase scenarios, however, it targets rapid product screening, not the sampling of dynamic configurations.
The CV functions under extreme conditions (e.g., 13~eV biases per deposition on benzene, average temperature about 1700~K), driving high-energy fragmentation and recombination that fall outside the kinetically relevant elementary-step regime.
To date, no general organic-reaction dataset mainly built by a metadynamics-based strategy has been reported.

This paper presents a sampling methodology that produces reactive training data with sufficient off-MEP coverage to make general MLIPs MD-ready for gas-phase organic reactions.
The core is a spatiotemporally resolved, system-independent CV: Cartesian RMSD within randomly partitioned local domains (spatial resolution) against an expanding list of time-averaged reference geometries (temporal noise filtering).
Scanning the partition radius captures reaction freedoms over a range of spatial scales, from local bond-changing to diffuse-motion-gated events.
The design supports general, efficient metadynamics sampling under mild conditions and scales to large molecules.
The sampling strategy combines metadynamics as the main exploration engine with TS-targeted structural relaxations that augment the coverage around TS.
The strategy is integrated into the DPGEN2\cite{DPGENZHANG2020107206,dpgen2_github}
concurrent-learning workflow, leveraging Druglike2025\cite{Yang2025DPA2Drug} as non-reactive pretraining data.
The outcome is OpenRxn26, a dataset of 1.8~M DFT-labeled configurations covering reactions from
neutral singlet unimolecular reactants in the H/C/N/O chemical space with $N_\mathrm{heavy} \leq 30$, containing reactive atomic environments underrepresented in community datasets.
Trained on OpenRxn26, a DPA4 model (denoted DPA4\_rxn26) serves to evaluate the data-model pipeline and achieves transferable accuracy on static evaluation of reaction properties across representative test reactions.
Crucially, on reactive trajectories away from MEPs, it reaches sub-kcal/mol energy accuracy with respect to the labeling method of training data.
OpenRxn26 thus provides MD-ready reactive training data, establishing the metadynamics-based sampling strategy as an efficient data solution for building general-purpose MLIPs on gas-phase organic reactions.

The remainder of the paper is organized as follows.
Section~\ref{sec2} describes the methodology, including the sampling strategy, concurrent-learning workflow, CV formulation, and training protocol of PES models.
Section~\ref{sec3} presents results benchmarking the atomic-environment coverage of OpenRxn26, the transferable accuracy of DPA4\_rxn26 on basic reaction properties, and DPA4\_rxn26's reliability on reactive dynamics trajectories.
Section~\ref{sec:conclusion} concludes the work.

\section{Methods}\label{sec2}

\begin{figure}[!htbp]
    \centering
    \includegraphics[width=1.0\textwidth]{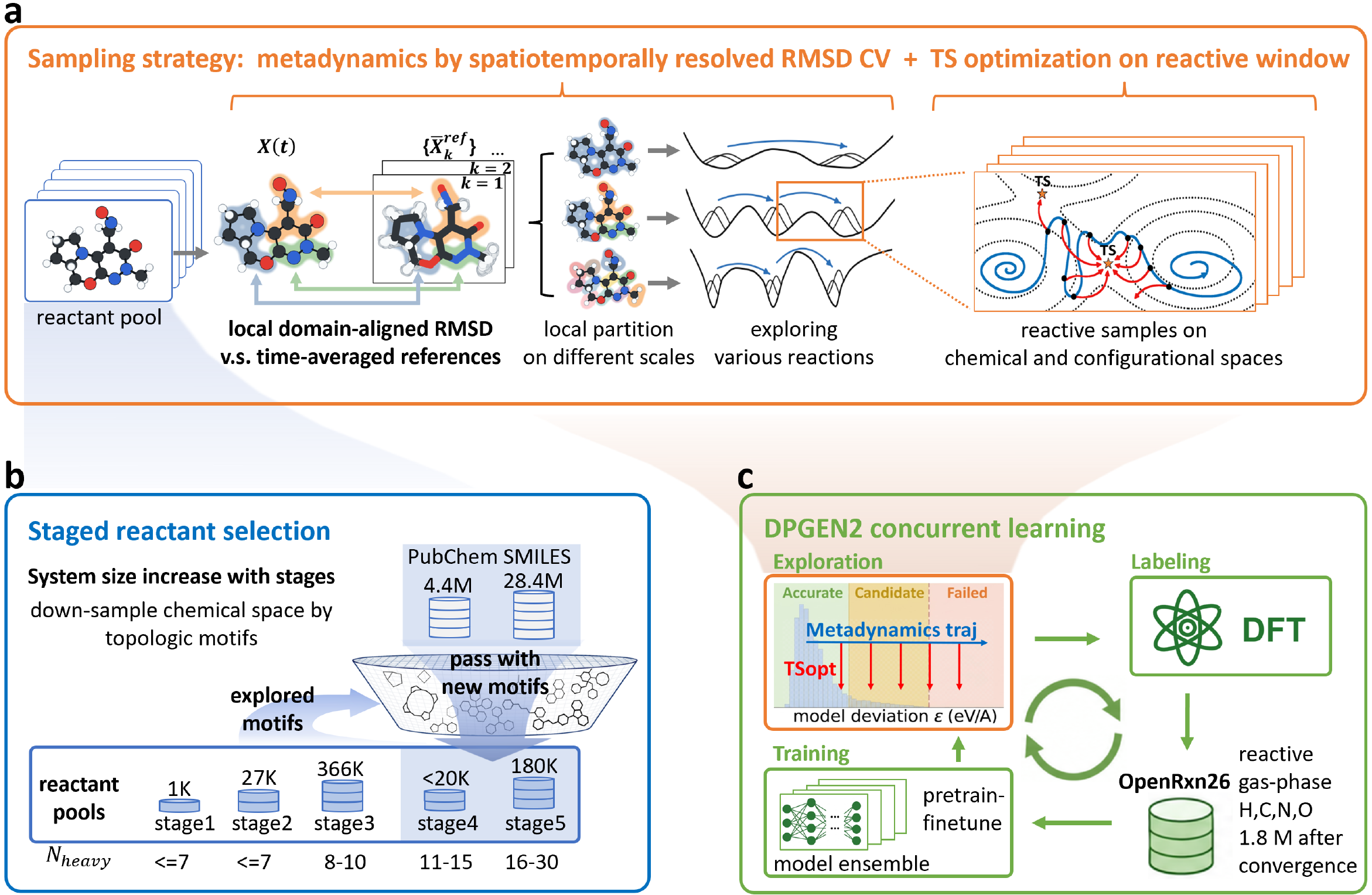}
    \caption{Overview of the OpenRxn26 data-generation workflow.
    a, Sampling strategy. Metadynamics drives reactive exploration using a spatiotemporally resolved, system-independent CV: Cartesian RMSD computed within randomly partitioned local domains against an expanding list of time-averaged reference geometries. Scanning the partition radius captures reactions over a range of spatial scales, from local 
    bond-changing to diffuse-motion-gated events.
    Selected snapshots from reactive windows seed TS optimizations (dimer method);
    metadynamics seeds and sparse configurations along TS-relaxation trajectories are passed to the uncertainty evaluation in the exploration step of DPGEN2.
    b, Staged reactant selection. Five stages cover progressively larger $N_\mathrm{heavy}$ ranges (Stage 1: Transition1x reactants; Stages 2--5: PubChem unimolecular SMILES). For Stages 4--5, a topology-motif-driven down-sampling retains reactants that introduce motifs absent from earlier-stage pools, reducing millions of SMILES to tens of thousands while preserving motif coverage.
    c, DPGEN2 concurrent learning. Each iteration cycles through training (an ensemble of MLIPs is fine-tuned on the current dataset), exploration (sampling strategy in a, runs on PES of one MLIP in the ensemble; sampled candidates are classified by ensemble deviation $\epsilon$ into accurate / candidate / failed), and labeling (selected candidates labeled by DFT and added to the training set). The cycle converges on each stage in order, yielding OpenRxn26.}
    \label{fig1}

\end{figure}

General and efficient sampling of the chemical and conformational spaces of gas-phase organic reactions exceeds the capability of traditional structure-relaxation based schemes.
In particular, these schemes cannot effectively cover the configurations visited by the dynamic reactive processes away from MEPs.
In this work, we develop a general sampling strategy that uses metadynamics as the sampling engine, supplemented by downstream structural relaxations to augment sampling around TSs (Fig.~\ref{fig1}a).
The strategy is then integrated into the concurrent learning platform DPGEN2, which uses disagreement among an ensemble of trained models to automatically identify under-sampled configurations and select them for first-principles labeling (Fig.~\ref{fig1}c).
Because of the local nature of atomic environments and bonding changes in organic reactions, samples from small molecules carry useful representativeness for local motifs in larger molecules and their reactive events.
We therefore start the exploration from the reaction space of small reactants in Transition1x (number of heavy atoms $N_\mathrm{heavy} \leq 7$) and gradually extend it to larger systems drawn from PubChem with a broader chemical variety, up to $N_\mathrm{heavy} \leq 30$ and elements H/C/N/O.
To address the combinatorial explosion of species at larger sizes, a chemically intuitive down-sampling scheme is applied, reducing the number of starting reactants by orders of magnitude while preserving the topological diversity (Fig.~\ref{fig1}b).

\subsection{Sampling strategy}\label{subsec:sampling_strategy}

In metadynamics\cite{Laio2002,Bussi2020Metadynamics}, MD trajectories propagate on a PES augmented by an accumulating sum of Gaussian bias kernels (hills); each hill is centered at a collective variable (CV) value visited earlier in the trajectory, so the bias progressively raises the currently occupied free-energy basin until the system escapes over a barrier in CV space.
This mechanism not only accelerates rare reactive events but also samples chemical and conformational spaces simultaneously, making metadynamics well-suited as the engine of a general-purpose reactive-data curation workflow.

The bias potential is constructed as a sum of Gaussian hills,
\begin{equation}
V_{\mathrm{bias}}(\vec{s},t)
=
\sum_{t_j<t} G\!\left(\vec{s};t_j\right),
\label{eq:metad_sum}
\end{equation}
where the CV $\vec{s}=\vec{s}(X)=(s_1(X),\dots,s_{d_{\mathrm{cv}}}(X))$ is a $d_{\mathrm{cv}}$-dimensional function of the atomic configuration $X$ chosen to capture the slow modes of interest, $t$ is the current simulation time, and $\{t_j\}$ are the discrete times at which hills are deposited.
Each hill takes the Gaussian form
\begin{equation}
G\!\left(\vec{s};t_j\right)
=
W(t_j)\,
\exp\!\left[
-\sum_{i=1}^{d_{\mathrm{cv}}}
\frac{\left(s_i(X)-s_i^{(0)}(t_j)\right)^2}{2\sigma_i^2}
\right],
\label{eq:metad_gauss}
\end{equation}
where $X \equiv X(t)$ is the current configuration, $s_i^{(0)}(t_j)$ is the $i$-th CV component
of the hill center deposited at time $t_j$, $W(t_j)$ is the hill height, and $\sigma_i$ is the width along the $i$-th CV component.

A general-purpose metadynamics engine for arbitrary molecules should be (i) system-independent, so the CV transfers without molecule-specific tuning, and (ii) spatially tunable, so it can resolve reactive motions ranging from local bonding change to collective deformation. 
For efficient sampling, it should additionally be (iii) noise-suppressed, so the deposited bias acts on slow reactive modes rather than on high-frequency vibrations.
In this work, we adopt a CV that achieves all three characteristics.
Conceptually, it consists of a Cartesian RMSD metric (system-independent), locally partitioned molecular domains at a controllable spatial scale (spatially tunable), and time-averaged bias references (noise-suppressed); 
the full definition is given in Sec.~\ref{subsec:cv_formulation}.
This design further supports efficient sampling under moderate hill heights and temperatures, and parameters transferable to larger molecules with little adjustment that is well-suited for high-throughput exploration (Fig.~\ref{fig1}a).

The transition state is a first-order saddle point on the PES, traversed rapidly by reactive trajectories and narrow in the directions orthogonal to the reaction coordinate.
Consequently, even enhanced-sampling dynamics rarely resolves it with precision.
To concentrate sampling on configurations near the TS, we incorporate a geometry-refinement step.
Valuable near-reactive configurations are identified from the metadynamics trajectory by two criteria applied jointly: (i) its bias or potential energy is within a steepest-descent window (reactive window), and (ii) bonding atom pairs differ from the initial reactant by pair changes not observed in preceding reactive windows.
A dimer-method\cite{Dimer1999} saddle-point optimization is launched from each identified configuration, with the initial dimer image generated by Gaussian perturbations of the detected bond-changing atoms.
Finally, sparsified samples are obtained by equal arc-length intervals along the projected arc of each optimization trajectory onto the space spanned by bond-changing distances, containing both the first-frame metadynamics seed and the last frame.
These samples are passed to the concurrent-learning platform for further evaluation.

\subsection{Concurrent learning}\label{subsec:concurrent_learning}

Concurrent learning is an iterative scheme that adaptively constructs training data for MLIPs by focusing first-principles labeling on configurations where the current model has high uncertainty.
This avoids spending expensive labels on configurations the model already handles and progressively expands the model's coverage of the target PES.
We use the DPGEN2 implementation\cite{DPGENZHANG2020107206,dpgen2_github} of this scheme, in which each iteration cycles through three steps (model training, exploration, and first-principles labeling) described in turn below (Fig.~\ref{fig1}c).

\textit{Training.} In each iteration, an ensemble of MLIPs is trained on the same data with the same hyperparameters but different random seeds.
In this work, the ensemble consists of four models; they are first pretrained on non-reactive data and then fine-tuned in subsequent iterations on a mixture of pretrain and newly collected reactive data, using DPA's fine-tuning scheme\cite{DPA2}.

The model ensemble uses the DPA-3\cite{DPA3zhang2025graph} architecture with 6 layers, whose vertex, edge, and angle feature dimensions are set to 128, 64, and 32, respectively.
The vertex-edge and edge-angle graph cutoffs are 6 and 4~\AA, respectively, with the smaller edge-angle cutoff chosen for efficiency.
Smooth switching of graph cutoffs is applied over 5--6~\AA{} and 3.5--4~\AA, respectively.
The fitting network has three layers with 240 neurons each.
The model ensemble is pretrained on the combination of two datasets:
(1) the non-reactive druglike conformation dataset proposed by Yang et al.\cite{Yang2025DPA2Drug}, denoted as Druglike2025, containing 1.4~M conformations of 0.1~M neutral closed-shell molecules, and labeled at a DFT level consistent with this work;
(2) $\sim$1.2~K reactants from Transition1x\cite{Grambow2020,Transition1x}, relabeled at the same DFT level as item (1).
Before concurrent learning starts, the model ensemble is pretrained for 1~M steps, with the number of configurations per batch chosen so that the total atom count is $\sim$1024.
The learning rate decays exponentially from $1 \times 10^{-3}$ to $1 \times 10^{-5}$, and the energy and force loss prefactors change linearly with the learning rate from 0.2 and 100 to 20 and 20, respectively.
At each concurrent learning iteration, each ensemble member is fine-tuned from its corresponding model in the previous iteration (or from the pretrained model at iteration 1) for 400~K steps, with the number of configurations per batch chosen so that the total atom count is $\sim$256.
The learning rate decays exponentially from $5 \times 10^{-4}$ to $1 \times 10^{-5}$, and the energy and force loss prefactors change linearly with the learning rate from 2 and 200 to 20 and 20, respectively.

\textit{Exploration.} In each iteration, the metadynamics-dimer sampling scheme of Sec.~\ref{subsec:sampling_strategy} is run on the PES of one trained model from the previous step, generating sampled configurations.
The ensemble's disagreement on each sample is quantified by the model deviation $\epsilon$, defined as the maximum, over all atoms, of the standard deviation of the predicted force across the ensemble.
Since ensemble members agree on well-covered configurations and disagree on novel ones, $\epsilon$ is used as the metric to select candidates for first-principles labeling.
Configurations are classified by two trust levels $\sigma_{\mathrm{lo}}$ and $\sigma_{\mathrm{hi}}$: those with $\epsilon < \sigma_{\mathrm{lo}}$ are skipped as already well covered, those with $\epsilon > \sigma_{\mathrm{hi}}$ are discarded as failed (likely unphysical), and those with $\sigma_{\mathrm{lo}} \leq \epsilon \leq \sigma_{\mathrm{hi}}$ are retained as candidates for labeling.
For data efficiency, the retained candidates are further down-sampled before labeling: each candidate is kept with probability inversely proportional to its bin population in the model-deviation histogram, biasing selection toward the high-uncertainty tail.
The self-adaptive $\sigma_{\mathrm{lo}}$ scheme of DPGEN2 is adopted: in each iteration, $\sigma_{\mathrm{lo}}$ is set to the lower bound of the top $P\%$ of model deviations among non-failed samples.
The exploration continues until $\sigma_{\mathrm{lo}}$ saturates, defined as the change in $\sigma_{\mathrm{lo}}$ between 2 neighbor iterations remaining below threshold $H$ for $N$ consecutive iterations.
In this work, $P = 10$, $N = 3$, and $H = 0.01$~eV/{\AA}.
$\sigma_{\mathrm{hi}}$ is set to 0.5~eV/{\AA} in early iterations and gradually increased to 2.0~eV/{\AA} in later iterations.

Metadynamics simulations are conducted in the NVT ensemble at 300~$\mathrm{K}$ with a 1~$\mathrm{fs}$ time step, using a Nosé--Hoover thermostat with a 100~$\mathrm{fs}$ damping time.
A static spherical nanoreactor confines the system to prevent dissociated fragments from escaping and to raise their collision probability, with the cavity centered at the reactant's geometric center and radius equal to the distance from that center to its outermost atom.
Confinement is imposed by an outer harmonic wall that is zero at the cavity boundary and grows outward, with wall coefficient 10~eV/\AA$^2$ and thickness 6~\AA.
For each molecule, multiple metadynamics runs are conducted simultaneously across a matrix of sampling parameters to diversify the explored reaction pathways.
All combinations (one value per parameter dimension) are evaluated for Stages 1--3 exploration stages on gradually enlarged reactant pools, defined in Sec.~\ref{subsec:staged_reactant_selection}; for Stages 4--5, the number of combinations is equal to the number of $r_\mathrm{cut}$ candidates (the local-domain partition radius, defined in Sec.~\ref{subsec:cv_formulation}): each $r_\mathrm{cut}$ value is used once, and the other parameter dimensions are sampled randomly.
Hill height $W \in \{1, 2, 3, 4\}$~eV for Stages 1--4 and $\{2, 4, 6\}$~eV for Stage 5.
Hill width $\sigma \in \{0.2, 0.3, 0.35, 0.5\}$~\AA, with a 500~fs deposit interval.
Metadynamics-run length is self adaptive between 3.0 and 11.5~ps, with longer runs for smaller $W$.
Each run is preceded by a 0.5~ps unbiased MD relaxation under the same thermostat and confinement.
The hydrogen mass is set to 8.0~u for numerical stability during long metadynamics runs, without altering the configurational Boltzmann distribution.
Simulations are performed with custom-modified LAMMPS\cite{LAMMPS2022} and PLUMED\cite{PLUMED2} workflows.

\textit{Labeling.} Selected candidates are labeled by first-principles calculations and added to the training set for the next iteration.
Energies and forces are labeled by density functional theory (DFT) using the range-separated hybrid generalized-gradient functional $\omega$B97X-D\cite{wB97XD2008} and the 6-31G(d,p) basis set in Gaussian 16 Revision C.01\cite{g16}.
Self-consistent field (SCF) convergence requires the root-mean-squared (RMS) change in the density matrix to be below $1 \times 10^{-8}$ and the maximum change to be below $1 \times 10^{-6}$.
Numerical integration uses a (99, 590) grid: 99 radial shells per atom with 590 angular points each.

Each calculation is first performed as a closed-shell singlet, followed by stability analysis\cite{Stable_WF_Seeger1977,StabLe_DFT_Bauernschmitt1996} of the restricted Kohn-Sham wavefunction.
If an external instability towards a broken-symmetry (BS) wavefunction is detected, an unrestricted Kohn-Sham optimization is performed to produce the lower-energy BS singlet state with spin-polarized diradical character.
The BS state is essential for properly describing bond cleavage and formation in organic reactions\cite{BS_Isobe2003DielsAlder,BS_good_approxi_MR_but_kink,BS_rxn_mechanism_plannar_2015,BS_MD_nonstatic}.

\subsection{Staged reactant selection}\label{subsec:staged_reactant_selection}

The concurrent-learning iterations are organized into five stages, each sampling reactions from
molecules within a specific $N_\mathrm{heavy}$ range (Fig.~\ref{fig1}b).
Stage 1 uses 90\% of the Transition1x reactants, about 1\,K H/C/N/O molecules with $N_\mathrm{heavy} \leq 7$.
The remaining 10\% are held out as the OOD\_SMILES split of the Transition1x\_1900 test set (Sec.~\ref{subsec3.2}).
Stage 2 takes unimolecular SMILES from the PubChem database restricted to H/C/N/O $N_\mathrm{heavy} \leq 7$, and neutral singlet state.
After de-duplication, 3D-geometry generation, and rough optimization with RDKit, about 27\,K reactants are obtained.
Stage 3 takes PubChem SMILES under the same conditions as Stage 2 but with $8 \leq N_\mathrm{heavy} \leq 10$, yielding about 366\,K reactants after the same treatment.
Stages 4 and 5 take PubChem SMILES with $11 \leq N_\mathrm{heavy} \leq 15$ and $16 \leq N_\mathrm{heavy} \leq 30$, comprising 4.4\,M and 28.4\,M SMILES records, respectively.

For Stages 1--3, the reactants are used directly as configuration pools, where starting points
of exploration are randomly selected at each iteration.
For Stages 4 and 5 ($N_\mathrm{heavy} >10$), however, direct exploration over all reactants is no longer efficient.
As molecules grow larger, the density of novel local atomic environments decreases sharply, because most larger molecules can be viewed as assemblies of structural motifs already explored in smaller systems.
The novelty of the explored reactive trajectories is similarly diluted, given the local nature of bonding changes in organic reactions.

We therefore develop a chemically intuitive down-sampling strategy that maintains a reference library of motifs from earlier stages.
For each SMILES in the new stage's pool, motifs are extracted from both the original SMILES and the SMILES of inferred products obtained by matching 26 typical SMARTS reaction templates (manually-collected in this work, see Table~\ref{tabS:SMARTS}).
Each motif absent from the reference library defines a new-motif group consisting of the SMILES whose original or inferred-product motifs contain it.
An equal quota of SMILES is sampled from each group, and their union forms the final down-sampled pool, maximizing the coverage of new motifs.
For Stage 4, the reference library is built from the Stage 2 and Stage 3 pools, and the filter retains fewer than 20\,K SMILES, reduced to about 13\,K after holding out the OOD\_SMILES test split (Sec.~\ref{subsec3.2}).
For Stage 5, motifs from the Stage 4 pool are added to the reference, and the filter retains about 180\,K SMILES.
The retention ratio is below 1\% in both stages.

The motifs themselves are extracted through a four-step pipeline.
SMILES are first decomposed using the BRICS\cite{BRICS} and RECAP\cite{RECAP} retrosynthetic methods, which break molecules into chemically meaningful fragments.
Each fragment is then reduced to its Murcko scaffold\cite{Murcko}, retaining ring systems and the linkers between them.
Each scaffold is converted to a pure topological motif by replacing all heavy atoms with carbon and setting all bond orders to single.
Finally, the canonical SMILES of each topological motif is stored for comparison.
This pipeline emphasizes cyclic topology, which captures the novel features that emerge at larger sizes; acyclic chains, by contrast, are typically trivial extensions of smaller-scale prototypes.

\subsection{Collective variable formulation}\label{subsec:cv_formulation}

In its global form, the CV is a one-dimensional ($d_{\mathrm{cv}} = 1$ in Eq.~\eqref{eq:metad_gauss}), system-independent RMSD metric in Cartesian coordinates, computed against a dynamically expanding library of reference structures collected from the trajectory:
\begin{equation}
\vec{s}
=
s^{\mathrm{global}}(X,t)
=
\left(
\sum_{k=1}^{N_{\mathrm{ref}}(t)} d\!\left(X,X^{\mathrm{ref}}_k\right)^2
\right)^{1/2},
\label{eq:lambda}
\end{equation}
where $\{X^{\mathrm{ref}}_k\}_{k=1}^{N_{\mathrm{ref}}(t)}$ is the reference library, $N_{\mathrm{ref}}(t)$ is its current size, and $d(X, X^{\mathrm{ref}})$ is the pairwise distance between two structures, defined as
\begin{equation}
d(X,X^{\mathrm{ref}})
=
\left(
\frac{1}{N}\sum_{i=1}^{N}
\left\|\tilde{\mathbf r}_i-\mathbf r^{\mathrm{ref}}_i\right\|^2
\right)^{1/2}.
\label{eq:d_rmsd}
\end{equation}
Here $X = \{\mathbf r_i\}_{i=1}^{N}$ is the current structure, $X^{\mathrm{ref}} = \{\mathbf r^{\mathrm{ref}}_i\}_{i=1}^{N}$ is a reference structure (with consistent atom ordering), and $\tilde{\mathbf r}_i$ denotes the coordinate of atom $i$ after optimal rigid-body alignment of $X$ onto $X^{\mathrm{ref}}$.

To make the CV adaptable to both localized and collective reactive events, we partition the structure into $M(r_{\mathrm{cut}})$ domains controlled by a cutoff radius $r_{\mathrm{cut}}$:
\begin{equation}
X \mapsto \left\{X_{\mathcal D_m}\right\}_{m=1}^{M(r_{\mathrm{cut}})},
\quad
X_{\mathcal D_m} = \{\mathbf r_i\}_{i \in \mathcal D_m},
\quad
\bigcup_{m=1}^{M(r_{\mathrm{cut}})}\mathcal D_m(r_{\mathrm{cut}})=\{1,\dots,N\},
\label{eq:domain_decomp_index}
\end{equation}
where each domain $\mathcal D_m$ is constructed iteratively: (i) the first seed is drawn at random from the unassigned atoms, and subsequent seeds are chosen as the atom farthest from the geometric center of the unassigned set; (ii) $\mathcal D_m$ comprises the seed together with all unassigned atoms within $r_{\mathrm{cut}}$ of it; (iii) the procedure repeats until every atom is assigned; (iv) any single-atom domain is merged into its nearest neighbor
$\mathcal D_m$.
The domain-aggregated distance to a reference $X^{\mathrm{ref}}_k$ is then defined as
\begin{equation}
\tilde{s}\!\left(X, X^{\mathrm{ref}}_k;r_{\mathrm{cut}}\right)
=
\left(
\sum_{m=1}^{M(r_{\mathrm{cut}})}
d\!\left(X_{\mathcal D_m}, X^{\mathrm{ref}}_{k,\mathcal D_m}\right)^2
\right)^{1/2}.
\label{eq:lambda_domain}
\end{equation}

Since reactive progress is usually slow compared to trivial thermal vibrations, each reference is constructed as a time-averaged structure over a window of $n_w$ MD frames, suppressing high-frequency vibrational noise in the bias centers:
\begin{equation}
\bar{\mathbf r}^{(k)}_i
=
\frac{1}{n_w}\sum_{n=1}^{n_w}\mathbf r^{(k,n)}_i,
\qquad
\bar X^{\mathrm{ref}}_k=\{\bar{\mathbf r}^{(k)}_i\}_{i=1}^{N},
\label{eq:time_avg_ref_discrete}
\end{equation}
where $\mathbf r^{(k,n)}_i$ is the coordinate of atom $i$ in the $n$-th frame contributing to reference $k$, and $n_w$ is the averaging window length.

Combining Eqs.~\eqref{eq:domain_decomp_index} and \eqref{eq:time_avg_ref_discrete}, the final CV is
\begin{equation}
s(X,t)
=
\left(
\sum_{k=1}^{N_{\mathrm{ref}}(t)}
\tilde{s}\!\left(X,\bar X^{\mathrm{ref}}_k;r_{\mathrm{cut}}\right)^2
\right)^{1/2}
=
\left(
\sum_{k=1}^{N_{\mathrm{ref}}(t)}
\sum_{m=1}^{M(r_{\mathrm{cut}})}
d\!\left(X_{\mathcal D_m},\bar X^{\mathrm{ref}}_{k,\mathcal D_m}\right)^2
\right)^{1/2}.
\label{eq:s_final}
\end{equation}
Since the reference list grows during the simulation, $s(X, t)$ depends explicitly on $t$ through $N_{\mathrm{ref}}(t)$.
With $d_{\mathrm{cv}} = 1$ and time-averaged references $\bar X^{\mathrm{ref}}_j$ as hill centers, the metadynamics bias of Eqs.~\eqref{eq:metad_sum}--\eqref{eq:metad_gauss} becomes
\begin{equation}
V_{\mathrm{bias}}(X,t)
=
\sum_{t_j<t}
W(t_j)\,
\exp\!\left[-\frac{\left(s(X,t)-s(\bar X^{\mathrm{ref}}_j,t)\right)^2}{2\sigma^2}\right].
\label{eq:vbias_final}
\end{equation}

In this work, $r_\mathrm{cut}$ candidates are 
$\{1.2, 1.5, 2.0, 2.4, 3.0, 4.0, 10.0\}$~\AA{} for Stage 1, 
$\{1.2, 2.0,\allowbreak 3.0, 10.0\}$~\AA{} for Stages 2--3, and 
$\{2.1, 4.5, 10.0\}$~\AA{} for Stages 4--5.
The time-averaging window is 500~fs for Stages 1--3, and is drawn from $\{500, 300, 100\}$~fs (50\%/25\%/25\% probability) for Stages 4--5.
Hydrogen atoms are excluded from the CV metric for domains with $r_\mathrm{cut} >2.0$~\AA{} in Stage 3, and for all domains in Stages 4--5.

Taken together, the CV allows the bias energy to be relaxed across local domains (through the aggregated formalism), to be responsive to local modes within each domain (through the per-domain RMSD), and to couple selectively to slow modes (through the time-averaged reference).
Ablation experiments on spatial and temporal resolution of the CV are shown in Sec.~\ref{subsecS_ablation}, Fig.~\ref{figS_ablation_efficiency} and Fig.~\ref{figS_ablation_spatial_scale}.
Compared with the bare CV that ablates both resolutions, adding spatial resolution remarkably improves the sampling efficiency and the abundance in spatial extent of sampled reactions, achieving performance robust to increasing the molecule size.
The temporal resolution also brings slightly positive effect over the bare.
The combined spatiotemporally resolved CV is more than twice the efficiency of the bare on exploring random reactants from 5 molecular-size stages in Sec.~\ref{subsec:staged_reactant_selection}, and $\sim$10-fold the efficiency of the bare on large-size reactants from stage 5.
Besides, Table~\ref{tab:biased-segment-temperature} shows averaged temperatures below 400\,K in metadynamics simulations under a 300~K NVT ensemble, confirming that all CV formalisms maintain relatively mild sampling conditions.
In practice, random local-domain partitioning at multiple $r_\mathrm{cut}$ scales, explored in parallel, enables high-throughput sampling of reactions across diverse mechanisms and spatial scales.
Our design differs from Grimme's earlier CV\cite{Grimme2019} in three respects: it adopts an aggregated metric rather than separate RMSD for each reference, uses time-averaged rather than snapshot references, and randomized local-domain partitioning rather than whole-molecule or manually-assigned-subset RMSD.

\subsection{Training the production model}\label{subsec:production_model_training}

To assess whether OpenRxn26 enables reliable reactive MLIPs, we train a production-grade model (denoted DPA4\_rxn26) on the DPA-4\cite{dpa4} architecture; the resulting model provides the headline benchmarks of Sec.~\ref{sec3}.
The architecture uses four equivariant message-passing blocks, each with four SO(2) mixing layers and 64 feature channels.
The angular representations use $\ell_{\max}=4$ and $m_{\max}=1$, and the radial expansion contains 16 basis functions.
The neighbor cutoff is 6~\AA, with smooth polynomial cutoff envelopes applied to the radial basis functions and message-passing features.
The fitting network by default has a single layer of 192 neurons, followed by a scalar linear output layer.

The production model is trained from scratch for 810,000 steps, with the number of configurations per batch chosen so that the total atom count is $\sim$8000 (distributed across 4 GPU cards).
Druglike2025 and the curated reactive dataset produced by the concurrent learning workflow of Sec.~\ref{subsec:concurrent_learning} are mixed with sampling probabilities 1:3.
The learning rate is linearly warmed up from $6 \times 10^{-5}$ to $3 \times 10^{-4}$ over the first 0.3\% of training, held constant until 35\% of training, and then decayed to $1 \times 10^{-6}$ using a cosine schedule.
Throughout training, the loss combines the mean absolute error in energy per atom and the mean Euclidean norm of the atomic force errors, with constant energy and force prefactors of 10 and 5, respectively.
Optimization uses HybridMuon with a weight decay of $10^{-3}$ and a maximum gradient norm of 5.

For benchmarks in Sec.~\ref{sec3}, we additionally train two reference models, DPA3\_rxn26 and DPA3\_druglike, using the DPA-3\cite{DPA3zhang2025graph} architecture and the same training protocol below.
The architecture uses 24 layers, whose vertex, edge, and angle feature dimensions are set to 128, 64, and 32, respectively.
The vertex-edge and edge-angle graph cutoffs are 6 and 4~\AA, respectively, with smooth switching applied over 5--6~\AA{} and 3.5--4~\AA.
The fitting network has three layers with 240 neurons each.

The reference models are trained in two stages of 1,500,000 steps each, with the number of configurations per batch chosen so that the total atom count is $\sim$1024 (distributed across 4 GPU cards).
In the first stage, the learning rate decays exponentially from $1 \times 10^{-3}$ to $1 \times 10^{-5}$ under a mean-squared-error loss, and the energy and force loss prefactors change linearly with the learning rate from 0.2 and 100 to 20 and 20, respectively.
In the second stage, the learning rate decays exponentially from $5 \times 10^{-4}$ to $1 \times 10^{-5}$ under a Huber loss ($\delta = 0.01$), and the energy and force loss prefactors are held constant at 30 and 1, respectively.
For DPA3\_rxn26, the datasets of Druglike2025 and the curated reactive dataset by Sec.~\ref{subsec:concurrent_learning} are mixed with sampling probabilities 1:3, which is the same as DPA4\_rxn26.
DPA3\_druglike is trained only on Druglike2025.

\section{Results}\label{sec3}

The concurrent learning workflow (Sec.~\ref{subsec:concurrent_learning}) yields OpenRxn26, a dataset of 1.8~M configurations with DFT labels accumulated across its iterations.
It covers gas-phase unimolecular reactions of $\sim$0.6~M organic reactants drawn from two sources: $\sim$1K from Transition1x, and the rest down-sampled from $\sim$30~M PubChem records.
All reactants are neutral singlet organic molecules in the H/C/N/O chemical space with $N_\mathrm{heavy} \leq 30$.
OpenRxn26 excludes configurations with any absolute force component exceeding 15~eV/{\AA}.
We train a production model, DPA4\_rxn26, on OpenRxn26 combined with the non-reactive druglike dataset of Yang et al.\cite{Yang2025DPA2Drug} (Druglike2025); training details are described in Sec.~\ref{subsec:production_model_training}.
We characterize the novelty and coverage of OpenRxn26's atomic environments, assess the transferable accuracy of DPA4\_rxn26 on reaction properties, and examine its reliability in reactive dynamics trajectories.

\subsection{New atomic environments}\label{subsec3.1}

In an MLIP, energy is decomposed into per-atom contributions, each determined by the local atomic environment (the geometry of neighboring atoms within a cutoff radius).
The diversity of atomic environments in a training set therefore determines the range of configurations a model can reliably predict.
We find that OpenRxn26 contains atomic environments underrepresented in OMol25\cite{OMol25}, currently the most comprehensive general-purpose molecule dataset.
OMol25 spans the chemical and conformational space currently accessible through domain training sets, aggregating ANI-2x\cite{ANI2x}, SPICE2\cite{SPICE2023,SPICE2andNutMeg2024}, GEOM\cite{geom2022}, OrbNet Denali\cite{OrbNetDenali2021}, Transition1x\cite{Grambow2020,Transition1x}, RGD1\cite{RGD1}, ANI-1xBB\cite{ANI1xBB}, and MechDB curations\cite{rmechdb2023,pmechdb2024}, with additional coverage of biomolecules, electrolytes, and metal complexes.
Its conformations include both non-reactive and reactive structures, with abundant off-equilibrium structures sampled by moderate and elevated-temperature MD driven by empirical force field, semi-empirical method and MLIP on electrolytes, metal complexes and biomolecule clusters, making it a stringent reference for novelty.
We first qualitatively illustrate the novelty by comparing distributions of atomic-environment features through t-distributed stochastic neighbor embeddings (t-SNE) between samples from OpenRxn26 and an OMol25 subset.
We then quantify the extrapolation gap on the t-SNE-identified out-of-distribution (OOD) samples using a model trained on the full OMol25 set, finding prediction errors more than an order of magnitude larger than in-distribution (ID) baselines.

\begin{figure}[p]
    \centering
    \includegraphics[width=1.0\textwidth]{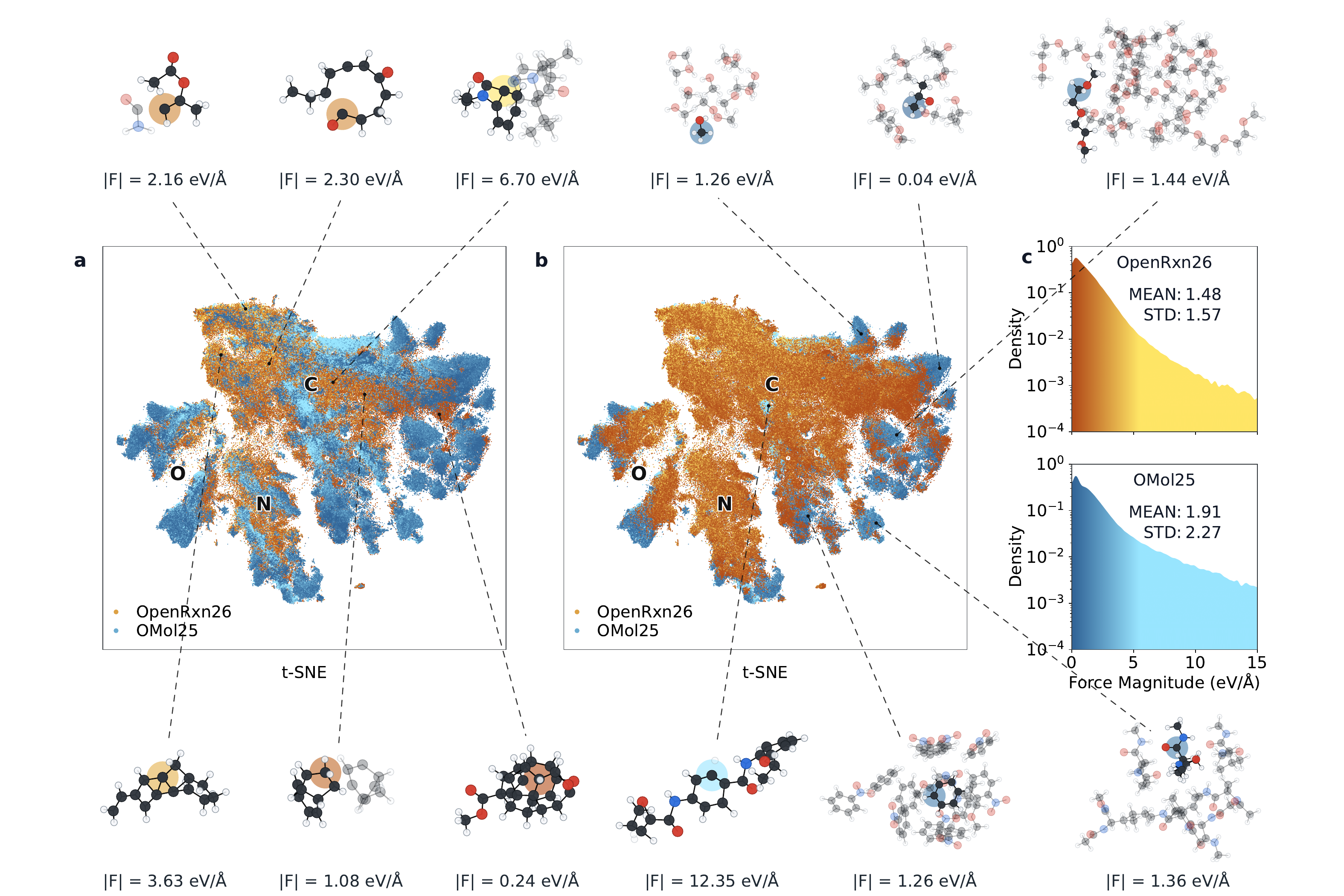}
    \caption{Atomic-environment coverage of OpenRxn26 compared with OMol25, visualized by t-SNE of DPA4\_rxn26 representations on heavy atoms (300,000 sampled from each set, $\geq$15~eV/{\AA} forces excluded; 600,000 atoms total).
    Orange points denote OpenRxn26 atomic environments and blue points OMol25; data-point brightness encodes the force magnitude on the target atom (mapping shown in c).
    a, t-SNE projection with OMol25 (blue) points overlaid on OpenRxn26 (orange) points, exposing OpenRxn26-dominated regions.
    Representative molecular structures carrying carbon atoms from these regions are shown next to the t-SNE; the target atom in each structure is marked by a halo matching the color and brightness of its scatter point, with the force magnitude labeled beneath.
    Fragments that do not contain the target atom are rendered transparently.
    Main elements of t-SNE clusters are labeled.
    b, As in a, but with OpenRxn26 points overlaid on OMol25 points, showing structures from OMol25-dominated regions.
    c, Density distributions of force magnitudes on the analyzed heavy atoms.
    }
    \label{fig2}  

\end{figure}

Fig.~\ref{fig2}a and b plot the t-SNE of atomic environment representations from DPA4\_rxn26 for heavy atoms sampled from OpenRxn26 and OMol25.
In these two panels, the overlay order of data points from the two sets is swapped to highlight the relative coverage and dominance of each set, and the brightness of the points encodes the force magnitude on the target atom (the atom whose environment representation gives the data point).
Example molecular structures carrying representative carbon atoms are connected to their t-SNE points by dashed lines.
Fig.~\ref{fig2}c plots the atomic force-magnitude distributions of sampled heavy atoms from OpenRxn26 and OMol25.
To plot the figure, an OMol25 subset is constructed by matching the conditions of OpenRxn26 (neutral singlet molecules in the H/C/N/O chemical space, excluding aqueous systems with $\geq3$ water molecules).
It comprises 34~M frames from 100~M total.
Recently, OMol25 is updated with extra 40~M frames beyond the scope of gas-phase neutral-singlet unimolecular organic reactions (denoted by version OMol-1, while the original 100~M by version OMol-0): systems with charged or higher spin states, element substitutions beyond H/C/N/O, as well as clusters from condensed-phase molecular crystals and MD boxes.
These samples would not align with the OpenRxn26 condition, and are not considered below.
Then 300~K heavy-atom representations are randomly down-sampled from each set, after excluding those with force magnitudes $\geq$15~eV/\AA.

In Fig.~\ref{fig2}a, blue OMol25 points are overlaid on top of orange OpenRxn26 points.
Regions where orange remains exposed reveal new atomic environments unique to OpenRxn26.
These regions diffuse among the separated OMol25 ridges and islands to form continuous continents in the t-SNE projection, especially for carbon atoms.
The distribution morphology supports the interpretation that the new environments are reactive and dynamic, blending features from distinct stable environments to fill the interior gaps left by OMol25.

For qualitatively perceiving the environment characteristics and locating representative samples for further quantitative test, we focus on the core regions of new C-atom environments, defined as smoothed t-SNE bins where the OpenRxn26 density is at least 20$\times$ the OMol25 density.
To ensure representativeness, bins where either the OpenRxn26 density or the total density falls below its 10th percentile are excluded (denoted OpenRxn26 20$\times$).
These regions host approximately 22\% of OpenRxn26 carbon atoms in the t-SNE plot.
Force magnitudes on these C atoms have mean/std of 1.77/1.82~eV/{\AA}, consistent with the full-sample OpenRxn26 distribution (1.48/1.57~eV/{\AA}, Fig.~\ref{fig2}c) and lower than the OMol25 distribution (1.91/2.27~eV/{\AA}).
This suggests that the new environments arise from reasonable but unseen molecular configurations, not from unphysical high-force outliers.

Geometric analysis of parent molecules shows that about 34\% of these C atoms lie in fragments cleaved from the heavy-atom scaffold of unimolecular reactants, with an average of 2.77 bonding neighbors (lower than the 3.16 of other C atoms in the same fragments), suggesting that the target atoms are likely the unsaturated bond-cleavage sites.
The parent molecules contain an average of 2.12 heavy-atom fragments.
An example molecule carrying a cleavage-site carbon is shown at the upper left of Fig.~\ref{fig2}a.
In addition, about 58\% of these C atoms are in ring structures, with each ring-membered target atom participating in 1.36 rings on average.
About half of the rings are 3- to 4-membered, the other half mainly 5- to 6-membered, and the largest reaches 20 members.
Polycyclic motifs strained by small rings, reactive toward strain relief, are 
exemplified by the lower-left molecule.
Small to medium-sized rings with target atoms near cleavage sites are shown by the lower-middle and upper-right molecules.
A large ring undergoing bond breaking is shown by the upper-middle molecule.
Stable environments with novel surroundings are also present, exemplified by the large cage-like molecule at the lower right; density profiles of the OpenRxn26 20$\times$ regions in t-SNE are shown in Fig.~\ref{figS_tsne_20x_OpenRxn26}.

In Fig.~\ref{fig2}b, the overlay is reversed: orange OpenRxn26 points are on top of blue OMol25 points, and OpenRxn26 nonetheless covers most of OMol25's distribution areas.
Within OpenRxn26's $N_\mathrm{heavy} \leq30$ size range, the coverage is even more complete (Fig.~\ref{figS_tsne_heavy30}).
This may seem counterintuitive given that OMol25 includes many non-reactive environments in addition to reactive ones.
However, it becomes reasonable from the local nature of bond-breaking and bond-forming events in organic reactions: atoms far from the reaction center contribute near-equilibrium local environments similar to those in non-reactive molecules.
Such atoms appear as the darker, lower-force points in Fig.~\ref{fig2}b.

In Fig.~\ref{fig2}b, several blue OMol25 clusters remain exposed at the exteriors of the continents.
An analogous analysis of the OMol25 regions having $\geq$20-fold the point density of OpenRxn26 shows that these atomic environments tend to be less reactive, high-force and less diversified.
Parts of them are not genuinely distinct from non-reactive atomic environments already covered by Druglike2025, which is the ground of OpenRxn26 and has been co-trained in DPA4\_rxn26.
These parts are excluded from the focused regions by simultaneously requiring $\geq$20-fold point density than the post-projected t-SNE of 300~K randomly sampled Druglike2025 heavy atoms to the existing embedding of Fig.~\ref{fig2}b.
The remaining regions (denoted OMol25 20$\times$) contain about 11\% of OMol25's carbon atoms, with 95\% of them from $N_\mathrm{heavy} >30$ systems (distribution profiles in Fig.~\ref{figS_tsne_20x_OMol25}).
Force magnitudes within show a pronounced $N_\mathrm{heavy}$-dependent split: atoms from $N_\mathrm{heavy} \leq30$ (within OpenRxn26's size range) systems have mean/std of 4.04/4.48~eV/{\AA}, while those from $N_\mathrm{heavy} >30$ systems are 1.47/1.19~eV/{\AA}, suggesting distinct atomic environments across the two size regimes.

Geometric analysis shows around 98\% of these C atoms lie in multi-fragment systems, in nearly saturated states with an average of 3.80 bonded neighbors (close to the sp$^3$ maximum of 4.00).
Other C atoms in the same fragment (3.79 neighbors) and the fragment average (3.78) are nearly equivalent, suggesting that the majority of target atoms are at random sites within intermolecular clusters of stable molecules, which are marginally relevant to elementary-step unimolecular reactions.
Such stable sites dominate the $N_\mathrm{heavy} >30$ regime, accounting for 99\% of its atoms, consistent with its low forces noted above.
These clusters are large, giving condensed-phase-like environments with an average of 9.57 non-hydrogen fragments; four examples (upper middle, upper right, lower middle, and lower right of Fig.~\ref{fig2}b) show the terminal carbon, middle carbon, aromatic carbon, and amide carbonyl carbon in stable monomers.

Focusing on the $N_\mathrm{heavy} \leq30$ regime, about 68\% of the remaining C atoms also lie in clusters of stable molecules (upper left of Fig.~\ref{fig2}b shows the carbon in a methanol monomer).
Ring structures are still common but in more regular scaffolds compared with OpenRxn26 20$\times$, hosting about 31\% of the C atoms, with each ring-membered target atom participating in only 1.02 rings on average and over 90\% of the rings being 6-membered.
These in-ring atoms exert high force magnitudes (mean/std 9.66/3.71~eV/{\AA}), accounting for the distinctive forces of this regime.
About 76\% of these high-force environments come from C-H bond stretching and ring distortions (lower-left molecule of Fig.~\ref{fig2}b).

Besides the qualitative distribution analysis, Table~\ref{tab:ID/OOD} quantifies the extrapolation gap from OMol25 to the t-SNE-identified OOD samples (OpenRxn26 20$\times$) by the prediction errors of MACE-OMol25\cite{MACE-MH2025crosslearning}, a MACE model trained on the full OMol25 set 
in OMol-0 version.
For comparison, DPA4\_rxn26 is evaluated on the inverse OOD samples (OMol25 20$\times$, split by frame origin into $N_\mathrm{heavy} \leq30$ and $N_\mathrm{heavy} >30$ regimes), and in contrast generalizes well to them.
The OOD test samples 300 carbon atoms from the t-SNE- and $N_\mathrm{heavy}$-identified regions, with DFT labels recomputed for each tested model using its training-data DFT method.
The ID baseline samples 100K random carbon atoms from each model's training distribution.
Samples with any absolute force component in the parent frame exceeding 15~eV/{\AA} are filtered out.
The OOD test further excludes compressed condensed-phase-like clusters from OMol25, which contain over-coordinated heavy atoms (C, N, O with more than 4, 4, and 2 covalent neighbors, respectively) and fall outside the scope of this work's gas-phase systems.
For each target atom, we examine force-component errors, together with force-component errors on all H, C, N, and O atoms in the parent frame, as well as the per-atom energy of that frame.

\begin{table}[tb]
\caption{Atomic-environment coverage gap between OpenRxn26 and OMol25, quantified by prediction MAEs on each set's ID samples and on cross-set t-SNE OOD samples (target-atom force; parent-frame force and energy).}
\label{tab:ID/OOD}
\centering
\fontsize{8}{8}\selectfont
\setlength{\tabcolsep}{2pt}
\renewcommand{\arraystretch}{1.2}

\resizebox{\linewidth}{!}{%
\begin{tabular}{@{}l l l r@{--}r@{\,}r r c c c@{}}
\toprule
\multirow{3}{*}{model}
& \multicolumn{6}{c}{evaluated atomic environments$^\S$}
& target atom
& \multicolumn{2}{c}{parent frame} \\
\cmidrule(lr){2-7}\cmidrule(lr){8-8}\cmidrule(l){9-10}
& \multirow{2}{*}{ID/OOD}
& \multicolumn{4}{c}{data source}
& \multirow{2}{*}{\#samples}
& force & force & energy \\
\cmidrule(lr){3-6}
& & name & \multicolumn{3}{c}{$N_\mathrm{heavy}$ (avg.)}
& & (kcal/mol/\AA) & (kcal/mol/\AA) & (kcal/mol/atom) \\
\midrule

\multirow{2}{*}{MACE-OMol25$^\dagger$}
& ID & OMol25 & 2 & 224 & (37.9) & 100,000
& 0.2 & 0.1 & 0.020 \\
& \mbox{t-SNE~OOD} & \mbox{OpenRxn26~20x$^\parallel$} & 4 & 30 & (8.7) & 300
& 6.8 & 3.2 & 0.288 \\

\addlinespace[0.6em]

\multirow{3}{*}{DPA4\_rxn26}
& ID & OpenRxn26 & 2 & 30 & (10.1) & 100,000
& 0.4 & 0.2 & 0.011 \\
& \mbox{t-SNE~OOD} & \mbox{OMol25~20x$^\P$} & 3 & 30 & (23.9) & 300
& 0.2 & 0.1 & 0.005 \\
& \mbox{t-SNE~OOD} & \mbox{OMol25~20x$^\P$} & 31 & 151 & (70.6) & 300
& 0.2 & 0.2 & 0.009 \\

\bottomrule
\end{tabular}
}

\begin{flushleft}
\footnotesize
\tabnote{$^\S$}{Carbon atoms sampled from the t-SNE in Fig.~\ref{fig2}, with all force components in the parent frame $\leq$15~eV/{\AA} ($\sim$346~kcal/mol/{\AA}). OOD samples exclude compressed condensed-phase 
clusters from OMol25 with over-coordinated heavy atoms (C, N, O with more than 4, 4, and 2 covalent neighbors, respectively).}
\tabnote{$^\dagger$}{A MACE model trained on the full set of OMol25 (version OMol-0)~\cite{MACE-MH2025crosslearning}. 
We were informed via email by the OMol25 team that the preprint of this manuscript led to an audit of open-shell configurations in OMol25.
The audit found that some configurations requiring broken-symmetry singlet treatment had been labeled as closed-shell states (planned to be addressed in a future OMol25 update).
It should be noted that MACE-OMol25's performance here and in the benchmarks below may be affected by this labeling discrepancy.
}
\tabnote{$^\parallel$}{t-SNE clusters where OpenRxn26 point density is $\geq$20$\times$ that of OMol25; sampled atoms are relabeled with the OMol25 DFT method. See Fig.~\ref{figS_tsne_20x_OpenRxn26} for the highlighted carbon regions.}
\tabnote{$^\P$}{t-SNE clusters where OMol25 point density is $\geq$20$\times$ that of OpenRxn26 and the post-projected Druglike2025; sampled atoms are relabeled with the OpenRxn26 DFT method. See Fig.~\ref{figS_tsne_20x_OMol25} for the highlighted carbon regions.}
\end{flushleft}

\end{table}

Remarkably, MACE-OMol25's force MAE on target carbon atoms inflates from an ID baseline of 0.2~kcal/mol/\AA{} to 6.8~kcal/mol/\AA{} on OpenRxn26 20$\times$ OOD samples, a $\sim$30-fold gap.
The gap persists at the parent-frame level: force MAE rises from 0.1 to 3.2~kcal/mol/\AA, and per-atom energy MAE from 0.020 to 0.288~kcal/mol.
These results imply that an OMol25-trained model suffers severe accuracy degradation on systems containing OpenRxn26's new environments, supporting their distinguishability from OMol25's environments.
DPA4\_rxn26's MAEs on OMol25 20$\times$ OOD samples match or undercut the ID baselines (0.1--0.2~kcal/mol/\AA{} for forces, 0.005--0.009~kcal/mol for per-atom energy).
DPA4\_rxn26's ability to handle OMol25's OOD environments suggests that OpenRxn26 (together with Druglike2025) covers a broader, more challenging atomic-environment space than OMol25 under the same condition.

OpenRxn26 20$\times$ samples have milder force magnitudes than the $N_\mathrm{heavy} \leq30$ OMol25 20$\times$ samples in DFT labels (Table~\ref{tabS:Force-statistics}).
Despite the milder forces, they are more challenging to predict: DPA4\_rxn26 shows MAEs on them (Table~\ref{tab:ood_id_reference_mae}) larger than on $N_\mathrm{heavy} \leq30$ OMol25 20$\times$ samples, even though they are within its training distribution.
This is consistent with the highly reactive and diverse nature of their atomic environments established above.

\subsection{Transferable accuracy in reaction properties}\label{subsec3.2}

The basic utility of a reactive MLIP is benchmarked by its accuracy and transferability in predicting barrier heights and reaction energies.
We perform a two-phase examination.
Phase 1 combines a data ablation testing the sampling strategy and a demonstration of the final production model: we compare DPA3\_rxn26 (trained from scratch on Druglike2025 + OpenRxn26) against its baseline DPA3\_druglike (only on Druglike2025), on broad reaction collections (Transition1x and RGD1), to isolate the effect of adding OpenRxn26.
Both models inherit the DPA3 architecture adopted by the model ensemble at the beginning of the data generation workflow (Sec.~\ref{subsec:concurrent_learning} and Sec.~\ref{subsec:production_model_training}).
Then, adopting a newer DPA4 model architecture, DPA4\_rxn26 trained on the same two datasets is compared.
Phase 2 benchmarks the production DPA4\_rxn26 against other reactive MLIPs on mechanism-classified reaction benchmarks (samples from BH9, Textbook181, cyclo[3+2]).
Throughout, barrier heights and reaction energies are reported as potential-energy differences.

\FloatBarrier

\begin{figure}[p]
    \centering
    \includegraphics[width=1.0\textwidth]{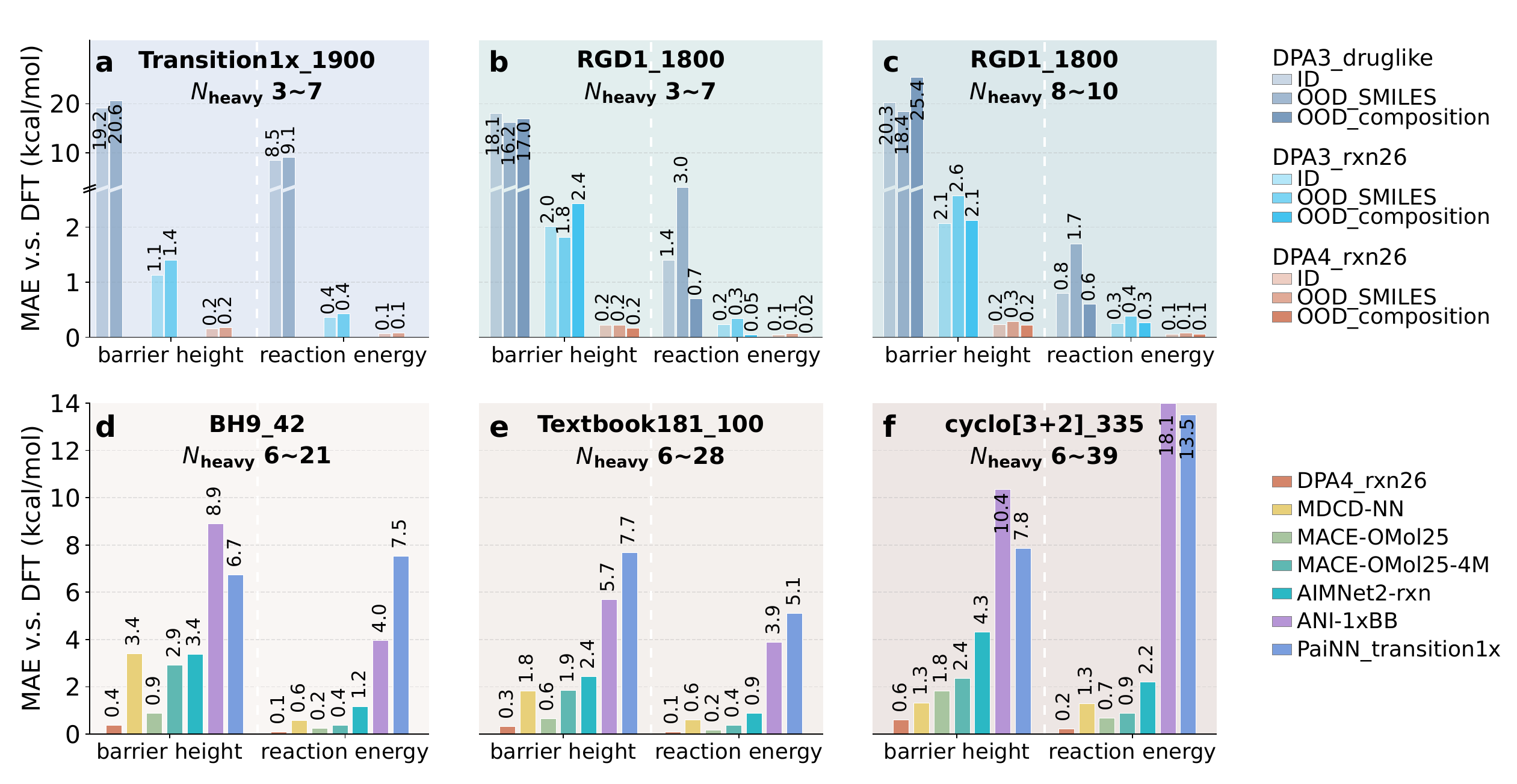}
    \caption{Benchmarks of model accuracy and generalization on barrier heights and reaction energies (reported as potential-energy differences).
    a--c, Effect of adding OpenRxn26 to training: DPA3\_rxn26 (trained on Druglike2025 + OpenRxn26) v.s. its baseline DPA3\_druglike (trained on Druglike2025 only), and the performance of the production model DPA4\_rxn26 (Druglike2025 + OpenRxn26), tested on subsets of Transition1x and RGD1.
    a, Transition1x\_1900 ($\sim$1900 reactions sampled from Transition1x).
    b, c, RGD1\_1800 partitioned by $N_\mathrm{heavy}$ 3--7 (b) and 8--10 (c), $\sim$900 reactions each.
    Reactions are categorized by whether the reactant's SMILES is in OpenRxn26's reactant pool (ID), only its elemental composition matches (OOD\_SMILES), or neither (OOD\_composition).
    d--f, Comparison of reactive MLIPs on three domain-specific reaction collections, filtered to H/C/N/O neutral-singlet reactions.
    d, BH9\_42 (42 reactions from BH9).
    e, Textbook181\_100 (100 reactions from Textbook181).
    f, cyclo[3+2]\_335 (335 reactions sampled from cyclo[3+2]).
    DFT references are recalculated using each tested MLIP's training-data DFT method.}
    \label{fig3} 

\end{figure}

In Phase 1, test sets are constructed from two reaction databases: Transition1x\cite{Grambow2020,Transition1x} ($\sim$10K reactions) and RGD1\cite{RGD1} ($\sim$177K reactions).
Respectively, around 1900 and 1800 reactions are sampled and their reactant, TS, and product geometries are relabeled at OpenRxn26's DFT level (also used for Druglike2025), forming Transition1x\_1900 and RGD1\_1800.
Each test set is partitioned into ID, OOD\_SMILES, and OOD\_composition subsets according to whether the reactant's SMILES is in OpenRxn26's reactant pool (ID), only its elemental composition matches (OOD\_SMILES), or neither (OOD\_composition).

RGD1\_1800 is further partitioned by molecule size into $N_\mathrm{heavy}$ 3--7 and $N_\mathrm{heavy}$ 8--10 groups.
The 3--7 group, in the same size range as Transition1x\_1900, tests sensitivity to chemical diversity at fixed system size: RGD1's reactants are derived from PubChem with tailoring allowing multi-molecules while Transition1x's reactants are unimolecules from GDB-7\cite{Ruddigkeit2012}.
The 8--10 group, compared with the 3--7 group, tests scaling with system size.
Subset sizes are: Transition1x\_1900 ID/OOD\_SMILES ($\sim$1000/900, Fig.~\ref{fig3}a); RGD1\_1800 $N_\mathrm{heavy}$ 3--7 ID/OOD\_SMILES/OOD\_composition ($\sim$400/400/80, Fig.~\ref{fig3}b); RGD1\_1800 $N_\mathrm{heavy}$ 8--10 ID/OOD\_SMILES/OOD\_composition ($\sim$300 in each, Fig.~\ref{fig3}c).

All Transition1x compositions are covered in OpenRxn26, so Transition1x\_1900 has no OOD\_composition subset.
RGD1's reactants undergo tailoring and re-saturation around reaction centers, which can produce small systems without carbon atoms; these inorganic systems are excluded from RGD1\_1800.

As shown in Fig.~\ref{fig3}a--c, training on OpenRxn26 substantially improves DPA3's accuracy on reaction properties, verifying the effectiveness of the sampling strategy and the value of OpenRxn26.
Across all subsets, barrier-height MAEs drop from $\sim$20~kcal/mol to $\leq$2.6~kcal/mol, and reaction-energy MAEs from up to 9.1~kcal/mol to $\leq$0.4~kcal/mol (within chemical accuracy with respect to the DFT reference).
Importantly, DPA3\_rxn26 shows a small ID/OOD generalization gap: the largest is $\sim$0.5~kcal/mol in barrier height (ID $\to$ OOD\_SMILES for RGD1\_1800 $N_\mathrm{heavy}$ 8--10), and the reaction-energy gap is hardly observable ($\sim$0.1~kcal/mol).
Performance is also stable across both axes of variation: across chemical diversity, barrier-height MAE rises from 1.1--1.4~kcal/mol on Transition1x\_1900 to 1.8--2.4~kcal/mol on the more diverse RGD1\_1800 (Fig.~\ref{fig3}a vs.\ b, a fluctuation of $\sim$1~kcal/mol); across system size, MAE is 1.8--2.4~kcal/mol for $N_\mathrm{heavy}$ 3--7 and 2.1--2.6~kcal/mol for 8--10 (Fig.~\ref{fig3}b vs.\ c).
Reaction-energy accuracy is stable across both axes.
Moreover, DPA4\_rxn26 further suppresses MAEs across all ID/OOD test sets to sub-kcal/mol with respect to the DFT labels, giving 0.2--0.3 on barriers and around 0.1 on reaction energies.
This promising result indicates that the sampling strategy could support the DPA4 architecture to obtain a reactive PES model, satisfying the basic demands of accuracy and transferability in predicting gas-phase organic reactions.
Accordingly, Phase 2 and all subsequent tests will further evaluate the performance and the reliability of this data-model pipeline.

In Phase 2, we benchmark DPA4\_rxn26 against several representative reactive MLIPs specialized for gas-phase organic reactions.
MDCD-NN\cite{MDCD20} is a sister DPA3 model trained on the MDCD20 reactive dataset (1.08~M structures of H/C/N/O molecules with $N_\mathrm{heavy}\leq20$).
MACE-OMol25\cite{MACE-MH2025crosslearning} and MACE-OMol25-4M\cite{mace_omol_4m} are MACE models trained on the full 100 M-frame OMol25 set (version OMol-0) and on a uniformly sampled 4~M subset\cite{OMol25}, respectively.
AIMNet2-rxn\cite{AIMNet2rxn} is an AIMNet2-family model specialized for organic reactions, trained on 4.7~M H/C/N/O geometries around MEPs of RGD1\cite{RGD1} records and $\sim$400\,k extra reactions via graph enumeration, as well as non-reactive conformations from AIMNet2\cite{AIMNet2} training sets.
ANI-1xBB\cite{ANI1xBB} is an ANI-family model specialized for elementary-step organic reactions, trained on 13.1~M bond-breaking structures from PubChem ($N_\mathrm{heavy}\leq7$) plus 5.5~M non-reactive ANI-1x configurations.
PaiNN\_transition1x\cite{NeuralNEB} is a PaiNN model trained on 9.6~M Transition1x configurations.

Test sets are selected from three domain-specific reaction collections, BH9\cite{BH9,BH9Correction}, Textbook181\cite{MDCD20}, and cyclo[3+2]\cite{Cyclo3+2}, filtered to H/C/N/O neutral-singlet reactions.
BH9\_42 contains 42 reactions with $N_\mathrm{heavy}$ 6--21: 40 unimolecular pericyclic reactions and 2 proton-transfer reactions.
Unimolecular pericyclic reactions demand concerted bonding of multiple sites and, in flexible molecules, large-amplitude motions to bring distant sites together, which makes training-set construction challenging.
Textbook181\_100 contains the singlet subset of 100 classical organic-chemistry textbook reactions (collected by Li and co-workers\cite{MDCD20}) with $N_\mathrm{heavy}$ 6--28, spanning a broad range of mechanisms (cycloaddition, electrocyclization, sigmatropic rearrangement, aldol reaction, Baeyer--Villiger oxidation, Cope elimination, Eschweiler--Clarke, Kiliani--Fischer synthesis, Huang--Minlon modified Wolff--Kishner reduction, Michael addition, Prilezhaev epoxidation, Arndt--Eistert homologation, Alder--ene reaction, Hantzsch reaction, Hofmann elimination, and Leuckart--Wallach reaction).
For cyclo[3+2], we randomly sample 335 reactions to form cyclo[3+2]\_335 ($N_\mathrm{heavy}$ 6--39); these are 3+2 cycloadditions, which require concerted bonding of 
two reactants and therefore lie outside OpenRxn26's unimolecular sampling scope, providing a strict generalization test.
For a fair comparison, single-point energies on these geometries are relabeled at each benchmarked model's training DFT level.

Importantly, all reactions in the three test sets are OOD to DPA4\_rxn26.
OpenRxn26 was constructed to exclude any configuration whose reactant, TS, or product SMILES appears in these test sets, making all of them at least OOD\_SMILES; reactions with $N_\mathrm{heavy} >30$ in cyclo[3+2]\_335 further fall outside OpenRxn26's scope and are OOD\_composition.
For MDCD-NN, Textbook181\_100 and cyclo[3+2]\_335 are partially ID, since about half of Textbook181\_100's reactants and $\sim$1/10 of cyclo[3+2]'s are used to initialize the MDCD20 training set\cite{MDCD20}.
MACE-OMol25 is expected to deliver excellent performance given the scale of its training set, while MACE-OMol25-4M serves as a matched-scale reference (4~M frames, comparable to OpenRxn26 + Druglike2025 at $\sim$1.8~M + $\sim$1.4~M).

Across all three test sets (Fig.~\ref{fig3}d--f), DPA4\_rxn26 ranks first in both barrier heights and reaction energies, and is the only model that maintains sub-kcal/mol accuracy on all sets (0.3--0.6~kcal/mol for barrier heights and 0.1--0.2~kcal/mol for reaction energies), demonstrating satisfactory accuracy and transferability across diverse gas-phase reactions.
On BH9\_42 (Fig.~\ref{fig3}d), DPA4\_rxn26's barrier MAE is 0.4~kcal/mol (0.5 ahead of the second-best MACE-OMol25 and 2.5 ahead of the third-best MACE-OMol25-4M).
On Textbook181\_100 (Fig.~\ref{fig3}e), DPA4\_rxn26's barrier MAE is 0.3~kcal/mol (0.3 ahead of the second-best MACE-OMol25 and 1.5 ahead of the third-best MDCD-NN).
On both sets, DPA4\_rxn26's reaction-energy MAEs remain as low as 0.1~kcal/mol, while the second- and third-best models give 0.2--0.4~kcal/mol.
On cyclo[3+2]\_335 (Fig.~\ref{fig3}f), prediction becomes more challenging for most models; only the partially-ID MDCD-NN improves its barrier MAE (1.3~kcal/mol, second best).
A slight error increase of DPA4\_rxn26 is expected, since bimolecular cycloadditions lie outside OpenRxn26's sampling scope of unimolecular reactants; yet DPA4\_rxn26 still leads, with a barrier MAE of 0.6~kcal/mol (0.7 ahead of MDCD-NN and 1.2 ahead of MACE-OMol25) and a reaction-energy MAE of 0.2~kcal/mol (0.5 ahead of MACE-OMol25 and 1.1 ahead of MDCD-NN).

PaiNN\_transition1x and ANI-1xBB struggle on all three test sets, while AIMNet2-rxn ranks outside the top-3.
Despite their larger training sets (9.6~M frames for PaiNN\_transition1x, 13.1~M for ANI-1xBB, 4.7~M for AIMNet2-rxn), they appear under-sampled in the regions tested, supporting the data efficiency of OpenRxn26's (1.8~M frames) sampling strategy on reactive chemical and configurational space.

\subsection{Reliability in reactive dynamic trajectories}\label{subsec3.3}

\begin{figure}[p]
    \centering
    \includegraphics[width=0.99\textwidth]{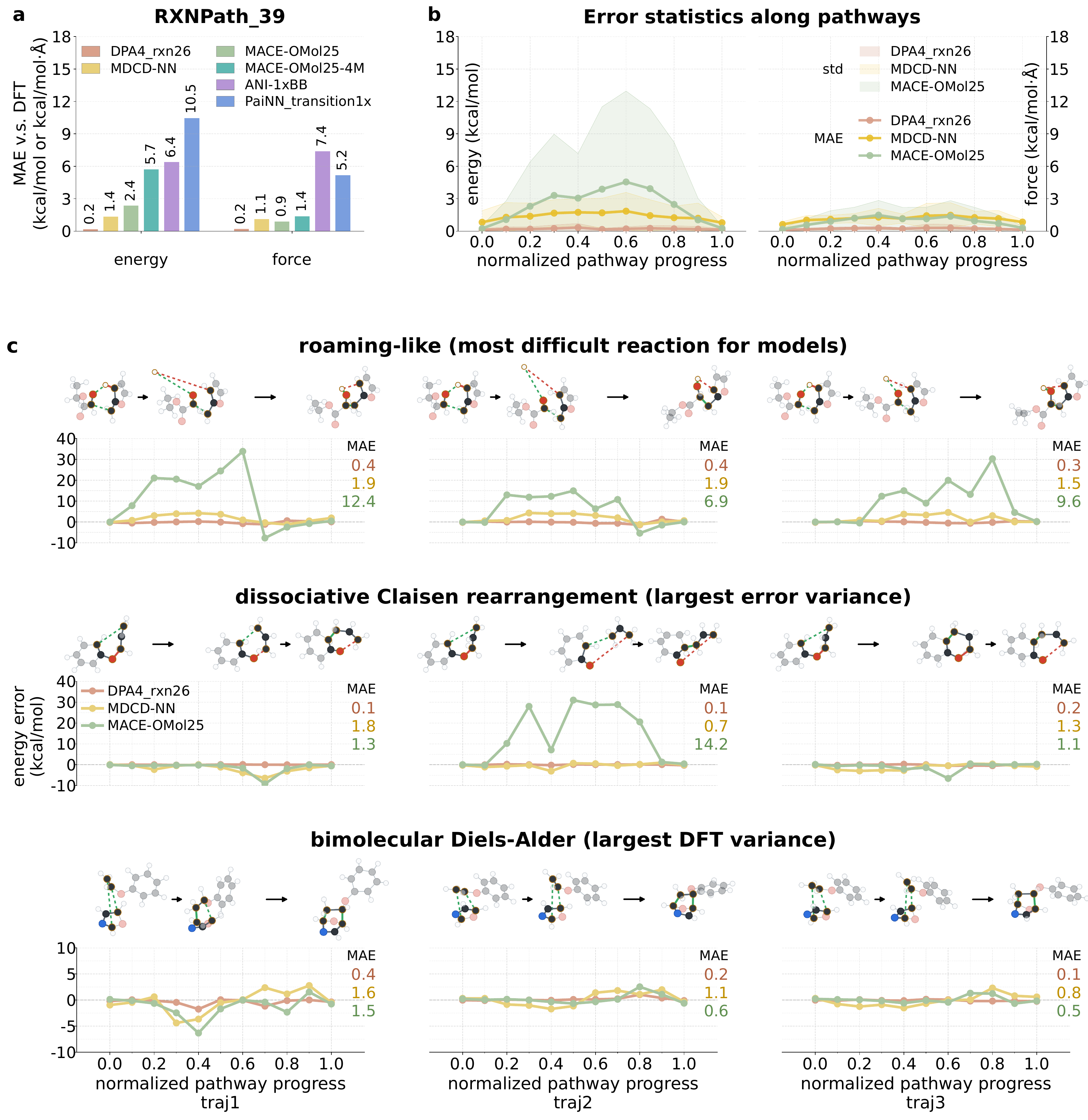}

    \caption{Benchmark of model accuracy on RXNPath\_39 dynamic reactive configurations.
    a, Overall energy and force MAEs.
    b, Per-progress energy and force error statistics for the top-three models in a. Trajectories are projected onto the reaction space defined by distances of bond-changing atom pairs; pathway progress is the normalized cumulative arc length on this projection,
    and the progress of each selected snapshot is rounded to the nearest 0.1 interval.
    c, Per-trajectory errors for the top-three models on three reactions: a carbonyl-ene reaction with spurious roaming-like H transfer (Case 1; largest error among all models);
    a dissociative Claisen rearrangement (Case 2; largest model-error variance across its 3 trajectories);
    a bimolecular Diels--Alder cycloaddition (Case 3; largest DFT energy variance across its 3 trajectories).
    Structures at three progress values for each trajectory are shown with the reaction center and breaking (red)/forming (green) bonds highlighted (details in Fig.~\ref{figS_trj_roam}--\ref{figS_trj_DA}).
    DFT references are recalculated using each model's training-data DFT method.
    }
    \label{fig4}  

\end{figure}

Finally, we probe whether reactive training datasets support reliable MLIP predictions on the dynamic process of reactions, which is the foundation for harnessing MLIPs towards in-situ mechanism mining, rate and selectivity prediction, and process optimization.
To support reliable MD propagation, the training data must cover broader off-MEP configurations encountered in diverse trajectories, including spurious high-energy pathways\cite{HeadGordon2023}.
This goes beyond the static accuracy tested in Sec.~\ref{subsec3.2} (barrier heights and reaction energies on MEP-anchored geometries).
Gaps in this coverage manifest as model errors that bias the relative population of reaction channels or, more severely, pull the trajectory into unphysical regions with cascading bond breaking and simulation collapse\cite{HeadGordon2023}.
Accurate prediction on dynamics-derived reactive configurations is therefore the necessary condition we use below to assess each reactive dataset's MD-readiness, measured via the MLIP trained on it.

Test conformations are sampled from a spatially continuous trajectory connecting reactant and product, constructed by a relay of biased shooting MD segments driven by DPA4\_rxn26 and labeled at the DFT reference level.
A reference path is defined by interpolated images along the documented reactant-TS-product triad.
Each MD segment then starts from the farthest-progressed snapshot of the preceding segment, with initial velocities of reaction-center atoms modulated along the bond-changing direction.
The bond-changing direction points toward the pair distances in the nearest next reference image if the current distance is within the interpolation range, or toward those in the final product otherwise.
Consequently, trajectories between reactant and product do not necessarily traverse the documented TS, can vary substantially from one another, and may include spurious pathways with high-energy configurations.
Finally, the trajectory is projected onto a low-dimensional reaction space defined by the distances of bond-changing atom pairs, and sparse configurations are selected as snapshots nearest to equal arc-length intervals along the projected arc; each selected snapshot retains its full atomic geometry.
This sampling method is distinct from the metadynamics used to curate OpenRxn26, providing a complementary viewpoint on dataset coverage and an OOD-style test for DPA4\_rxn26.

Specifically, we construct the RXNPath\_39 test set from 39 dynamic reactive trajectories,
sampled on 13 reactions randomly selected from Textbook181\_100.
For each reactant-product pair, 3 trajectories are sampled, each contributing 11 sparse configurations at evenly spaced pathway progresses (from 0.0 to 1.0 with an interval of 0.1
after normalization), giving 429 data points in total.
Configurations from high-energy spurious pathways including roaming-like ones (see below) are collected as expected, providing all MLIPs strict tests for reliabilities on broad off-MEP configurational space.
DFT labels of configurations are further recalculated for each model using its training-data DFT method to provide a fair test.
We use Textbook181\_100 as the reference to (1) leverage its broad coverage of organic reaction mechanisms and (2) enable direct comparison with Sec.~\ref{subsec3.2} in assessing performance beyond MEPs.

As shown in Fig.~\ref{fig4}a, DPA4\_rxn26 achieves the lowest energy MAE (0.2~kcal/mol, being the only model reaching sub-kcal/mol accuracy against its training-level DFT reference), followed by MDCD-NN at 1.4~kcal/mol, MACE-OMol25 at 2.4~kcal/mol, and the remaining models at 5.7--10.5~kcal/mol.
DPA4\_rxn26's energy errors on RXNPath\_39 are comparable to its Textbook181\_100 errors in Sec.~\ref{subsec3.2} (0.3~kcal/mol for barrier heights and 0.1~kcal/mol for reaction energies), indicating its uniform accuracy across the conformational space covered by OpenRxn26.
Recall that Textbook181\_100 (and hence RXNPath\_39) is OOD to DPA4\_rxn26; the best performance of DPA4\_rxn26 reaffirms the data efficiency of OpenRxn26's sampling strategy.
In contrast, MACE-OMol25 degrades from Textbook181\_100 to RXNPath\_39,
dropping from second to third place, with its MAE increasing around fourfold from 0.6~kcal/mol on barrier heights to 2.4~kcal/mol on dynamic configurations.
It should be noted again that the performance of MACE-OMol25 may be affected by the known label noise in OMol25 (version OMol-0 and OMol-1), as noted in the footnote of Table~\ref{tab:ID/OOD}.
Besides, ANI-1xBB and PaiNN\_transition1x also give increased energy MAEs respectively from 5.7 and 7.7 kcal/mol on Textbook181\_100 barrier heights to 6.4 and 10.5 kcal/mol on RXNPath\_39.

These results suggest that MEP-anchored accuracy for MLIPs does not naturally imply their accuracy on dynamically sampled off-MEP reactive configurations.
On forces, DPA4\_rxn26 also ranks first (0.2~kcal/mol/\AA{}), followed by MACE-OMol25 and MDCD-NN (0.9--1.1~kcal/mol/\AA{}).

Fig.~\ref{fig4}b tracks the error distribution of all 39 trajectories with respect to normalized pathway progress, resolving the per-model performance along reaction propagation for the top-three models in Fig.~\ref{fig4}a.
Energy MAE of DPA4\_rxn26 rises from 0.1~kcal/mol at the ends of the error curve to a maximum of 0.3~kcal/mol at progress 0.4 (std $\sim$0.4~kcal/mol), and its force MAE rises from 0.1~kcal/mol/\AA{} at the ends to 0.3~kcal/mol/\AA{} at the same progress (std $\sim$0.2~kcal/mol/\AA).
MDCD-NN gives an arched error curve: its energy MAE rises from 0.8~kcal/mol at the ends to a maximum of 1.9~kcal/mol at progress 0.6 (std $\sim$1.7~kcal/mol), and its force MAE rises from 0.6--0.8~kcal/mol/\AA{} at the ends to 1.4~kcal/mol/\AA{} at the same progress (std $\sim$1.1~kcal/mol/\AA).
MACE-OMol25 is relatively accurate near both ends (0.2--0.3~kcal/mol) but degrades severely at mid-progress: its energy MAE exceeds 4~kcal/mol at progress 0.6 with std over 8~kcal/mol, suggesting that some tested pathways may fall outside its training distribution.
Its force errors give an MAE $\sim$1.5~kcal/mol/\AA{} and a std $\sim$1.4~kcal/mol/\AA{} at progress 0.4.
Plots for the remaining (lower-ranked) models are shown in Fig.~\ref{figS_traj_b}.

Fig.~\ref{fig4}c illustrates the performance differences among the top-three models by typical reactions and specific pathways.
Three reactions are selected to expose trajectories that are (1) hardest for all benchmarked models, (2) most divergent in model errors across their three test trajectories, and (3) most divergent in DFT energies across their three test trajectories.
Case (1) is identified by the largest averaged prediction RMSE of snapshot energies across the six models in Fig.~\ref{fig4}a: a carbonyl-ene reaction proceeding through high-energy spurious roaming-like pathways (top row), in which the H atom first dissociates from a C--H bond, drifts around, and recombines (detailed geometry evolution in Fig.~\ref{figS_trj_roam}).
Case (2) is identified by the largest standard deviation of model errors across the three test trajectories at each progress, averaged over the six models: a Claisen rearrangement in dissociative pathways, where the extent of O--C cleavage before C--C bond formation varies among trajectories (Fig.~\ref{figS_trj_dissociative}).
Case (3) is identified by the largest standard deviation of DFT energies across the three test trajectories at each progress: a bimolecular Diels--Alder cycloaddition, where the relative motion of two reactants and the alignment timing of two C--C bonding sites differ among the three trajectories (Fig.~\ref{figS_trj_DA}).

Case (1) shows that roaming-like conformations from spurious pathways are difficult for MACE-OMol25, with energy MAE on these trajectories ranging 7--12~kcal/mol and peaks up to $\sim$34~kcal/mol.
Even more dangerously, some non-top-three models artificially lower the energy by 30--40~kcal/mol at the roaming stages (Fig.~\ref{figS_trj_c},~\ref{figS_trj_Aimnet2-rxn}), a tendency that makes unphysical regions artificially more accessible to trajectories.
In contrast, DPA4\_rxn26 handles these configurations best with MAEs of 0.3--0.4~kcal/mol and no observable fluctuation during roaming, while MDCD-NN falls in between (MAE 1.5--1.9~kcal/mol).

Case (2) shows substantial trajectory-to-trajectory variation in MACE-OMol25's reliability on the dissociative Claisen rearrangement.
On trajectory 2, MACE-OMol25 has the largest MAE among top-three models (14.2~kcal/mol, with a peak up to 31~kcal/mol), yet on trajectories 1 and 3, its MAEs drop to 1.1--1.3~kcal/mol.
This could be correlated to the early and complete O--C cleavage in trajectory 2, which leaves a loose fragment pair from progress around 0.3, until the new C--C bond forms at the end.
In contrast, DPA4\_rxn26 maintains the smallest MAEs of 0.1--0.2~kcal/mol across all three trajectories, including the mid-phase geometries of trajectory 2 where both rearranging bonds are outside the bonding range.

Case (3) shows that DPA4\_rxn26 is the only model maintaining sub-kcal/mol accuracy across all three trajectories of this demanding and delicate Diels--Alder reaction.
Its MAEs across the three trajectories are 0.1--0.4~kcal/mol, while the other models give MAEs exceeding 1~kcal/mol (0.5--1.5 by MACE-OMol25 and 0.8--1.6~kcal/mol by MDCD-NN).
This is notable because the selectivity prediction and substrate screening of Diels--Alder reactions could depend on kcal/mol-level barrier differences. 
Given that its wide configurational space of intermolecular approaching conformers is intrinsically difficult to cover, and bimolecular reactions are beyond the sampling starting point of OpenRxn26 as the unimolecular reactant, the excellent performance of DPA4\_rxn26 further demonstrates the transferability of the model, as well as the efficiency of the sampling strategy.

DPA4\_rxn26 is the most consistent across the nine trajectories of Fig.~\ref{fig4}c: its per-trajectory MAE varies by 0.4~kcal/mol, compared with 1.1~kcal/mol for MDCD-NN and 13.8~kcal/mol for MACE-OMol25.
Taken together, the RXNPath\_39 benchmark shows that DPA4\_rxn26 is the only model in this benchmark suite combining sub-kcal/mol overall accuracy with respect to the training-DFT reference on dynamics-derived configurations, with consistency across diverse reaction trajectories.
OpenRxn26 is thereby demonstrated as the MD-ready reactive training data. 

The superior reliability of DPA4\_rxn26 in the RXNPath\_39 benchmark is robust to models sampling the test-set geometries. 
Verifying benchmarks on trajectories respectively sampled by MDCD-NN and MACE-OMol25 are shown in Fig.~\ref{figS_trj_MDCD_geom} and Fig.~\ref{figS_trj_MACE_geom}. 
They preserve the same accuracy ranking as that in Fig.~\ref{fig4}, where the overall MAEs in energy and force for DPA4\_rxn26 are still 0.2~kcal/mol and 0.2~kcal/mol/\AA{}.
Energy MAEs of MDCD-NN are 1.5~kcal/mol in both verifying benchmarks, while those of MACE-OMol25 are respectively 3.2 (MDCD-NN geometries) and 1.8 (MACE-OMol25 geometries)~kcal/mol.
Benchmarks for AIMNet2-rxn are shown in Fig.~\ref{figS_trj_Aimnet2-rxn}, whose overall relative-energy (with respect to the first frame for each trajectory) MAE on RXNPath\_39 is 5.5~kcal/mol and the force MAE is 4.0~kcal/mol/\AA{} (5.7/5.0~kcal/mol and 4.1/4.1~kcal/mol/\AA{} on the MDCD-NN/MACE-OMol25 trajectories). 
AIMNet2-rxn outputs energy shifts relative to the composition sum of unreported per-element references, thus cannot be directly compared with the total energies in Fig.~\ref{fig4}.

\FloatBarrier

\section{Conclusion}\label{sec:conclusion}

We developed a metadynamics-driven sampling methodology that produces reactive training data with sufficient off-MEP coverage to make general MLIPs MD-ready for gas-phase organic reactions.
The core methodological contribution is a system-independent collective variable built from Cartesian RMSD against a dynamically expanding library of time-averaged references, with local-domain partitioning for tunable spatial resolution.
Applied within a DPGEN2 concurrent-learning workflow, this produces OpenRxn26, a dataset of 1.8~M DFT-labeled configurations covering reactions from $\sim$0.6~M unimolecular reactant molecules in the H/C/N/O chemical space ($N_\mathrm{heavy} \leq30$).
A DPA4 model trained on OpenRxn26 (denoted DPA4\_rxn26) serves as the production-grade demonstration to evaluate the data-model pipeline.

Benchmark results establish three points.
First, OpenRxn26 covers atomic environments that general-purpose datasets such as OMol25 do not, with a model trained on OMol25 showing more than an order of magnitude force-MAE inflation on the OpenRxn26-distinguished regions.
Second, OpenRxn26-trained DPA4 (DPA4\_rxn26) is the only model in the benchmark suite to deliver transferable sub-kcal/mol accuracy on barrier heights and reaction energies, with respect to labeling methods of training data.
Third, DPA4\_rxn26 is the only model in the benchmark suite to deliver sub-kcal/mol overall MAE on dynamics-derived reactive configurations and the smallest trajectory-to-trajectory variation across diverse reactions, with respect to labeling methods of training data.
The sampling strategy is demonstrated to be capable of supporting the DPA4 architecture in developing reliable general-purpose reactive MLIPs for gas-phase organic reactions, and OpenRxn26 is verified as MD-ready training data for dynamics simulation on gas-phase neutral singlet unimolecular organic reactions.

High accuracy on static properties along MEP does not naturally imply reliability on off-MEP configurations from reactive dynamics trajectories, as exemplified by the change of accuracy ranking of MACE-OMol25 from Textbook181\_100 barrier heights to energies of reactive MD snapshots of RXNPath\_39.
Evaluating reactive MLIPs at the trajectory level is therefore necessary to assess their readiness for direct or enhanced-sampling MD.
The current OpenRxn26 sampling targets gas-phase neutral singlet reactants in the H/C/N/O chemical space with $N_\mathrm{heavy} \leq30$; natural extensions include broader chemical space (additional elements, larger molecules, transition-metal catalysis), charged and higher-spin species, and condensed-phase environments.

\section*{Supporting Information}
Supporting Information is available online and includes details of ablation experiments on spatiotemporal resolution of CV for sampling performance; SMARTS templates for down-sampling reactants; t-SNE analysis for OpenRxn26 and OMol25; label statistics and test on t-SNE ID/OOD samples; benchmarks on RXNPath\_39 for lower-ranked models; verifying RXNPath\_39 benchmarks on trajectories sampled by MDCD-NN and MACE-OMol25; reaction center evolutions in exemplified trajectories of RXNPath\_39; Tables S1--S4 and Figures S1--S13. 

\section*{Data Availability}
The training dataset OpenRxn26 is available online from AIS Square repository at \url{https://www.aissquare.com/datasets/detail?pageType=datasets&name=OpenRxn26&id=425}.
The production model DPA4\_rxn26 and corresponding training scripts are available online from AIS Square repository at
\url{https://www.aissquare.com/models/detail?pageType=models&name=DPA4_rxn26&id=447}.
The reference model DPA3\_rxn26 and corresponding training scripts are available online from AIS Square repository at
\url{https://www.aissquare.com/models/detail?pageType=models&name=DPA3_rxn&id=441}.
The benchmark reference data and scripts of the sampling strategy are available from \url{https://github.com/Vibsteamer/OpenRxn26_benchmark-ref_sampling-script}.

\section*{Acknowledgements}
The authors thank Mr. Duo Zhang for support with the DPA3 model.
The authors thank Mr. Bowen Li for discussions on model training.
The authors thank Mr. Zhaojia Dong for discussions on the domain test set.
The authors thank Mr. Guoao Li for discussions on the domain test set.
The authors thank Dr. Xinzijian Liu for support with DPGEN2.
Computational resources are supported by the Bohrium Platform of DP Technology.
The work of Han Wang is supported by the National Key R\&D Program of China (Grant No. 2022YFA1004300), and the National Natural Science Foundation of China (Grants No.~12525113 and No.~12561160120). 
The work of Tong Zhu is supported by the National Natural Science Foundation of China (Grant Nos. 92570205, 92461313, and 22361132538), and the AI for Science Program of the Shanghai Municipal Commission of Economy and Informatization (Grant No. 2025-GZL-RGZN-BTBX-01004). Tong Zhu also acknowledges the support from the Open Research Fund of Suzhou Laboratory (No. SZLAB-1508-2024-TS017).

\section*{Declarations}
The authors declare no conflict of interest.

\bibliography{sn-bibliography}

\clearpage
\setcounter{section}{0}
\renewcommand{\thesection}{S\arabic{section}}

\renewcommand{\theHsection}{S\arabic{section}}
\renewcommand{\theHsubsection}{\theHsection.\arabic{subsection}}

\section*{Supporting Information}

\renewcommand{\thefigure}{S\arabic{figure}}
\renewcommand{\theHfigure}{S\arabic{figure}}   
\setcounter{figure}{0}

\renewcommand{\thetable}{S\arabic{table}}
\renewcommand{\theHtable}{S\arabic{table}}     
\setcounter{table}{0}

\section{Supporting Contexts}\label{secS1}
\subsection{Ablation experiment on spatiotemporal resolution of CV\label{subsecS_ablation}}

We evaluated the performance in sampling reactive pathways of four CV formalisms.
With respect to the full (spatial+temporal) CV, ablated formalisms respectively retain the spatial, temporal, or neither (bare) resolution.
The compared performance aspects include efficiency, the abundance of spatial scales of sampled pathways, especially focusing on the scale-up ability towards large molecules. 
The averaged temperature in metadynamics is further checked for guaranteeing a relatively mild condition. 

A fixed set of 20 preselected reactant molecules was prepared, with four molecules randomly taken from each of 5 molecular-size stages from Sec.~\ref{subsec:staged_reactant_selection} to exhibit the variance of the sampling performance with respect to the increase of molecular size. 
6 individual experiment batches are conducted to evaluate the averaged result, with the velocity initialization seeds being different among batches.
In each of the 6 batches, each of the 4 CV formalisms conducts 6 metadynamics runs for each of the 20 molecules, giving 2880 metadynamics runs in total.

For 1) spatiotemporally resolved formalism (spatial+temporal), a simulation scheme close to the OpenRxn26 production scheme is adopted:
$r_\mathrm{cut}$ candidates are \{1.2, 2.1, 3.0, 4.0, 4.5, 10.0\}~\AA{} and each candidate value is used once to give 6 sampling runs; 
the time-averaging window for hill-center reference is 500 fs; 
the height candidates for bias hill are \{1, 2, 4\} eV and hill width candidates are \{0.20, 0.35\}~\AA, with each $r_\mathrm{cut}$-specific run randomly adopting one height-width combination from the candidates; 
hydrogen atoms are excluded from the CV metric if $r_\mathrm{cut}$ $\geq$ 2.0~\AA{} (1/6 runs adopting $r_\mathrm{cut}$ = 1.2~\AA{} consider H in metrics).

For 2) spatial formalism, no time-averaging of the reference is conducted, with the other settings the same as in 1).

For 3) temporal formalism, no local-partition of the molecule in CV metric is conducted (still 6 runs per molecule adopting random height-width combinations for bias hill), and only the random 1/6 of runs consider hydrogen in the CV metric to match the sampling condition of 1);
other settings are the same as in 1).

For 4) the bare formalism, neither the time-averaging nor the local partition is conducted (still 6 runs per molecule adopting random height-width combinations for bias hill), and only the random 1/6 of runs include hydrogen in the CV metric to match the sampling condition of 1);
other settings are the same as in 1).

For all runs, one bias hill is deposited every 500 fs and MD time step is 1 fs, using the 300~K NVT ensemble with a Nosé--Hoover thermostat adopting a 100~$\mathrm{fs}$ damping time.
The simulation length of each run is self-adaptive from 3.5 ps to 9.5 ps according to the hill height (the lower the hill the longer the simulation), and in total giving 17.75 ns for 2880 runs.
The simulation adopts the DPA3 model used in the DPGEN2 workflow for generating OpenRxn26.

For analysis of sampling efficiency and spatial extent of sampled events, distinct reactive pathways within each experiment batch for each CV are recognized.
On each trajectory, bond-change events in one trajectory linking distinct stable molecules are merged into one committed pathway if they are separated by less than 100~fs.
Each committed pathway was assigned a unique identifier, consisting of the connectivity of the pre- and post- stable states of the pathway, and the ordered sequence of committed bond-change pairs (atom pairs with altered connectivity between pre- and post- stable states).
Among runs within the same batch for each CV, de-duplication is conducted according to the identifier, and the batch-level distinct pathways are used for evaluating the final batch-averaged results.

\begin{figure}[htbp]
    \centering
    \includegraphics[width=1.0\textwidth]
    {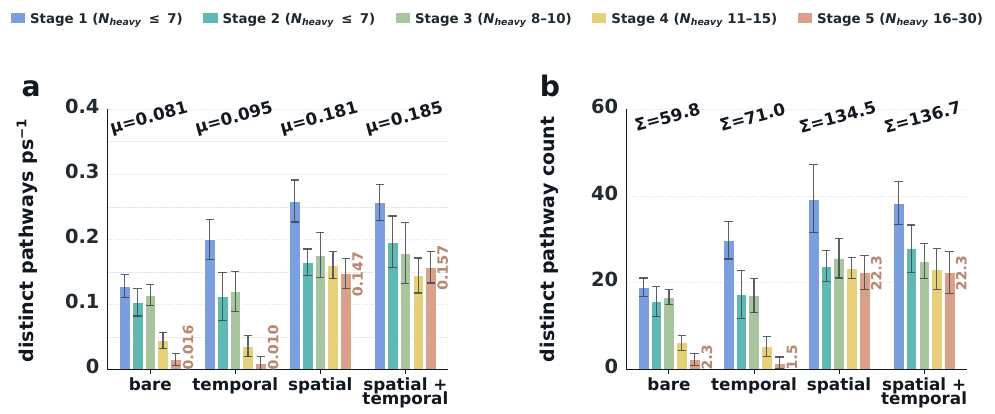}
  
    \caption{Ablation on spatiotemporal resolution of CVs for sampling efficiency.
    Experimental settings and the de-duplication of reactive pathways are described in Sec.~\ref{subsecS_ablation}.
    a, averaged sampling efficiency of distinct reactive pathways using 4 CV formalisms, respectively retaining the spatial, temporal, or neither (bare) resolution of the full (spatial+temporal) CV. 
    $\mu$ values denote the all-stage averaged efficiencies, and standard deviations of each stage across experiment batches are depicted.
    The averaged efficiencies on stage 5 molecules are particularly labeled.
    b, averaged number of sampled distinct reactive pathways, with the standard deviations of each stage across experiment batches also depicted.
    $\Sigma$ values denote the all-stage sum of distinct reactive pathways.
    The averaged sum on stage 5 molecules is particularly labeled.
    The averaged simulation time of each batch for each CV on 20 molecules from 5 stages is around 0.7 ns.
    The spatial+temporal CV has the highest sampling efficiency, with relative advantages being mainly attributed to the spatial resolution that gives similarly good efficiency.
    Temporal resolution also shows positive improvement to the bare. 
    Both spatial+temporal and spatial CVs show relatively stable efficiency facing larger molecules thus are capable of scaling up to large systems.
    }
    \label{figS_ablation_efficiency}  

\end{figure}

\begin{figure}[htbp]
    \centering
    \includegraphics[width=1.0\textwidth]
    {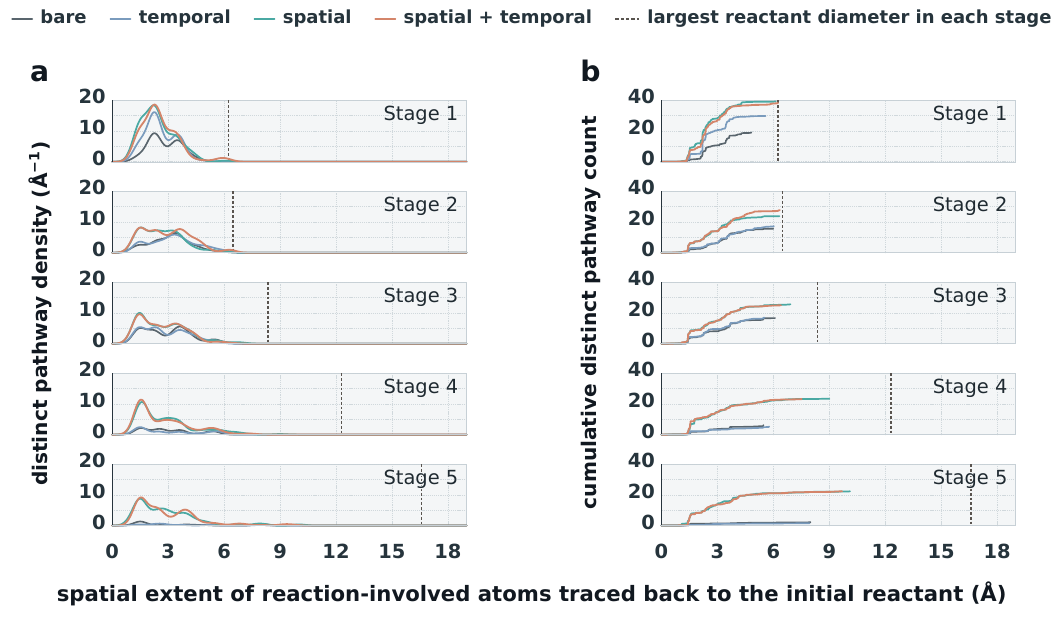}
    
    \caption{Ablation on the spatiotemporal resolution of CV for spatial extent of sampled reactive process.
    Experimental settings are described in Sec.~\ref{subsecS_ablation}.   
    For each pathway, the pairwise distances between all reaction-involved atoms in the initial reactant are examined, and the maximum of these distances defines the spatial extent a reactive process can reach on the initial reactant, i.e., the initial separation between the two most distant atoms that participate in the reaction.
    The distances are taken as time-averaged values over the unbiased MD segment that initializes each exploration run.
    a, the averaged distribution density of distinct pathways with respect to its spatial extent on initial reactant. b, the averaged cumulative distinct pathways with respect to its spatial extent on initial reactant. The curve stops after the reactive event with the largest spatial extent is accumulated, and the endpoint marks the largest explored extent.
    Dashed lines indicate the maximum extent allowed by each stage (maximum atomic pair-distance of selected reactants).
    The CVs with spatial resolution obtain remarkably more abundant pathways on the full spatial extents accessible by 4 CVs, particularly for the larger molecules.
    Spatial resolution is the principal contributor to the increased abundance in spatial extent.
    }
    \label{figS_ablation_spatial_scale}  

\end{figure}

\begin{figure}[htbp]
    \centering
    \includegraphics[width=0.80\textwidth]
    {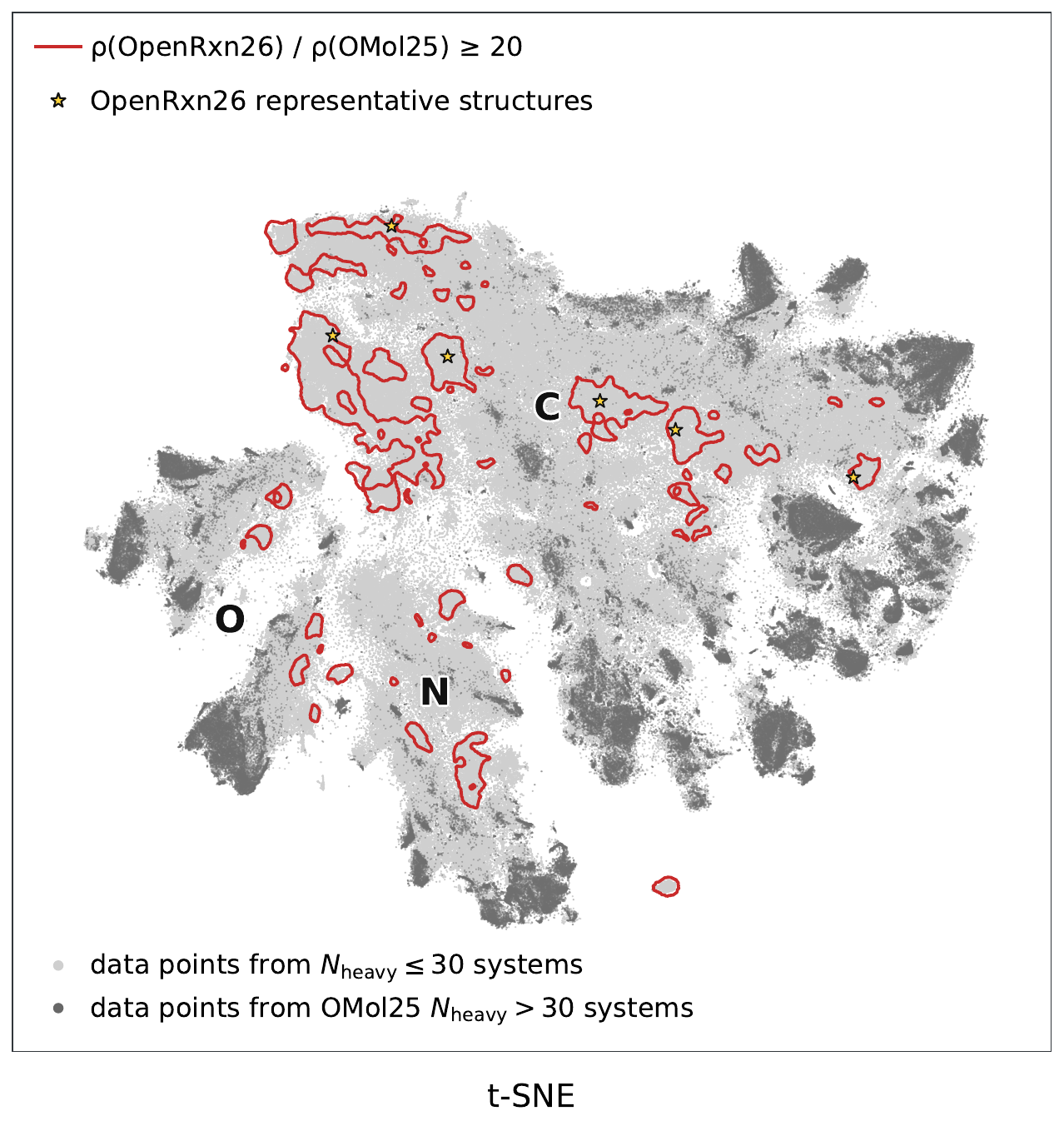}
    
    \caption{Highlighted t-SNE regions where OpenRxn26 covers atomic environments beyond OMol25, on the same projection as Fig.~\ref{fig2}.
    Red contour lines enclose regions with OpenRxn26-to-OMol25 point-density ratios of $\geq$20;
    the 20$\times$ regions provide the cross-set OOD samples used in Table~\ref{tab:ID/OOD}.
    Regions where either the OpenRxn26 density or the total density is below its 10th percentile are excluded.
    Yellow stars mark positions of OpenRxn26 representative structures shown in Fig.~\ref{fig2}a.
    Data points from OpenRxn26 and OMol25 $N_\mathrm{heavy} \leq$ 30 systems are in light gray, while those from OMol25 $N_\mathrm{heavy} >$ 30 systems are in dark gray.
    Element-dominant continents are labeled (C, N, O).
    }
    \label{figS_tsne_20x_OpenRxn26}  

\end{figure}

\begin{figure}[htbp]
    \centering
    \includegraphics[width=0.80\textwidth]
    {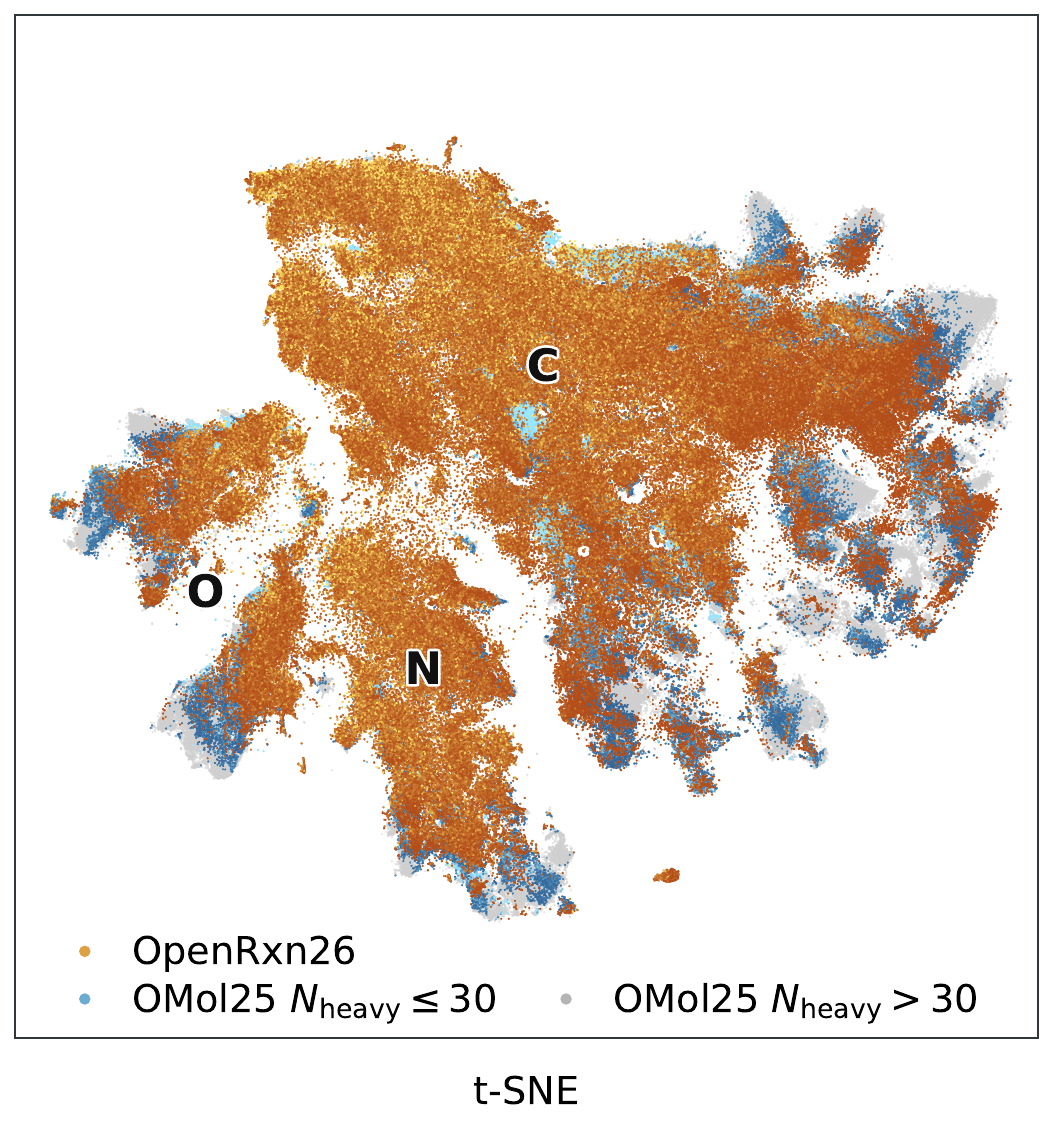}

    \caption{OpenRxn26 well covers the t-SNE distribution of heavy atoms from OMol25 $N_\mathrm{heavy} \leq$ 30 systems. 
    t-SNE projection is the same as Fig.~\ref{fig2}b, with data points from OMol25 $N_\mathrm{heavy} >30$ being plotted in gray.}
    \label{figS_tsne_heavy30} 
\end{figure}

\begin{figure}[htbp]
    \centering
    \includegraphics[width=0.80\textwidth]
    {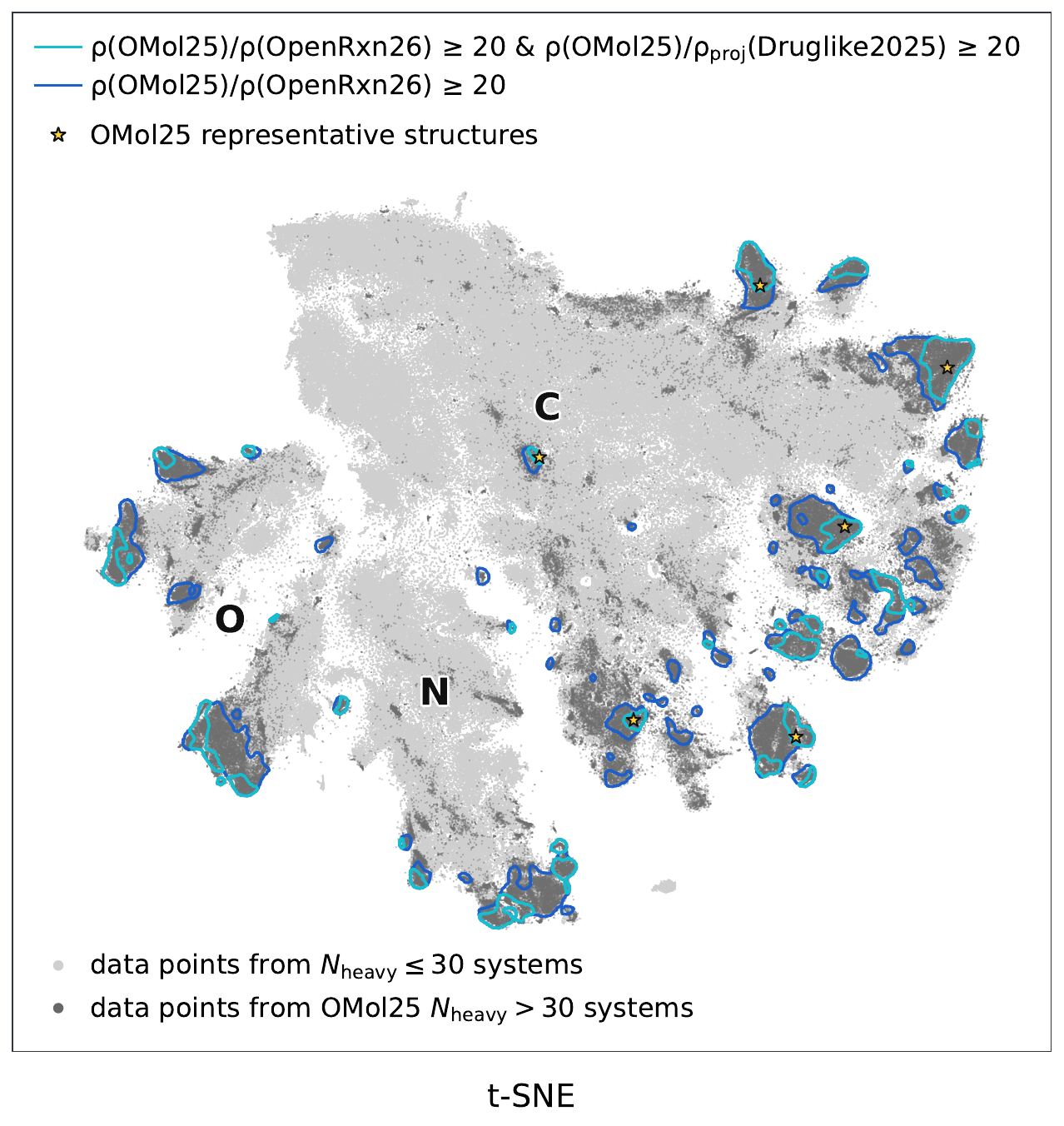}
    
    \caption{Highlighted t-SNE regions where OMol25 point density exceeds OpenRxn26 and the post-projected Druglike2025, on the same projection as Fig.~\ref{fig2}.
    Blue contour lines enclose regions with OMol25-to-OpenRxn26 point-density ratios of $\geq$20, and cyan contour lines enclose the subset simultaneously with $\geq$20-fold the point density of 300K randomly sampled Druglike2025 heavy atoms post-projected onto the existing OpenRxn26-OMol25 two-set t-SNE embedding;
    the cyan regions provide the cross-set OOD samples used in Table~\ref{tab:ID/OOD}.
    Regions where either the OMol25 density or the total (OpenRxn26 + OMol25) density is below its 10th percentile are excluded.
    Yellow stars mark positions of OMol25 representative structures shown in Fig.~\ref{fig2}b.
    Data points from OpenRxn26 and OMol25 $N_\mathrm{heavy} \leq$ 30 systems are in light gray, while those from OMol25 $N_\mathrm{heavy} >30$ systems are in dark gray.
    Element-dominant continents are labeled (C, N, O).
    }
    \label{figS_tsne_20x_OMol25} 
\end{figure}

\begin{figure}[htbp]
    \centering
    \includegraphics[width=1.0\textwidth]
    {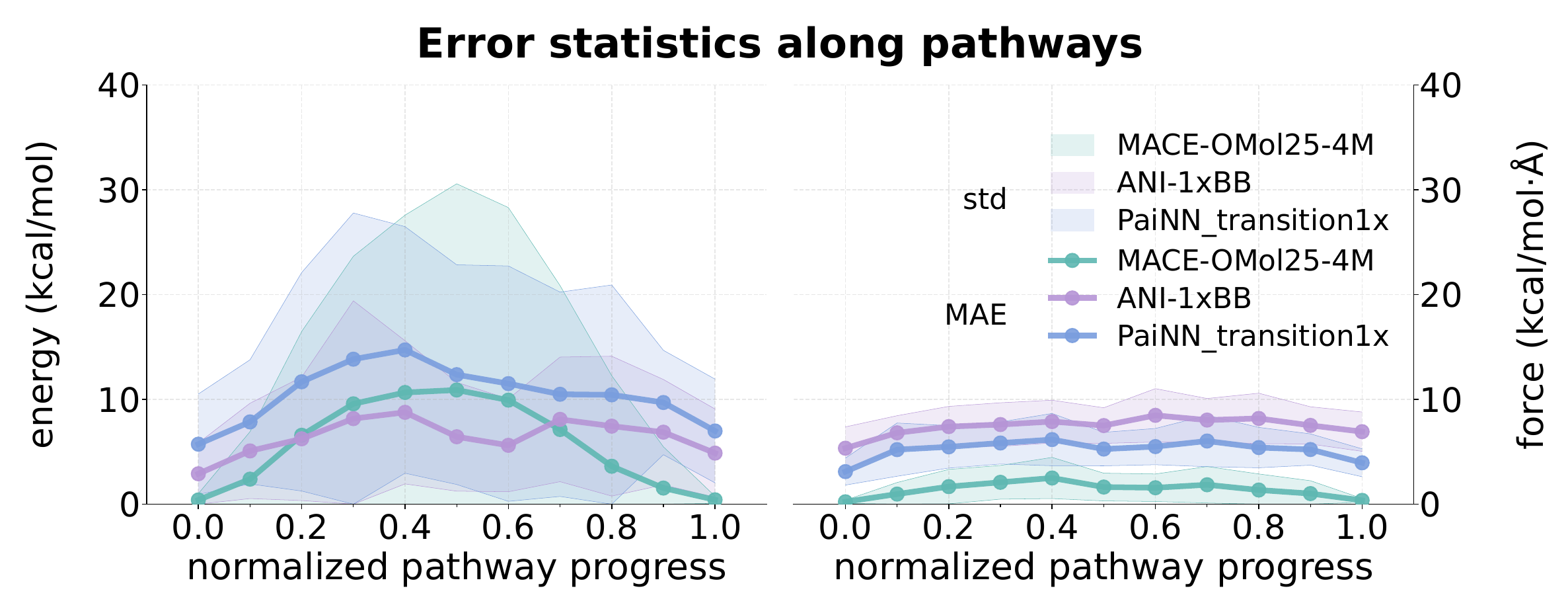}
    
    \caption{Per-progress energy and force error statistics on RXNPath\_39 for the lower-ranked models (MACE-OMol25-4M, ANI-1xBB, PaiNN\_transition1x), complementing Fig.~\ref{fig4}b.
    Solid lines show MAE; shaded bands show one standard deviation.
    Trajectory projection and pathway-progress definition follow Fig.~\ref{fig4}.
    DFT references are recalculated using each model's training-data DFT method.}
    \label{figS_traj_b} 

\end{figure}

\begin{figure}[htbp]
    \centering
    \includegraphics[width=1.0\textwidth]
    {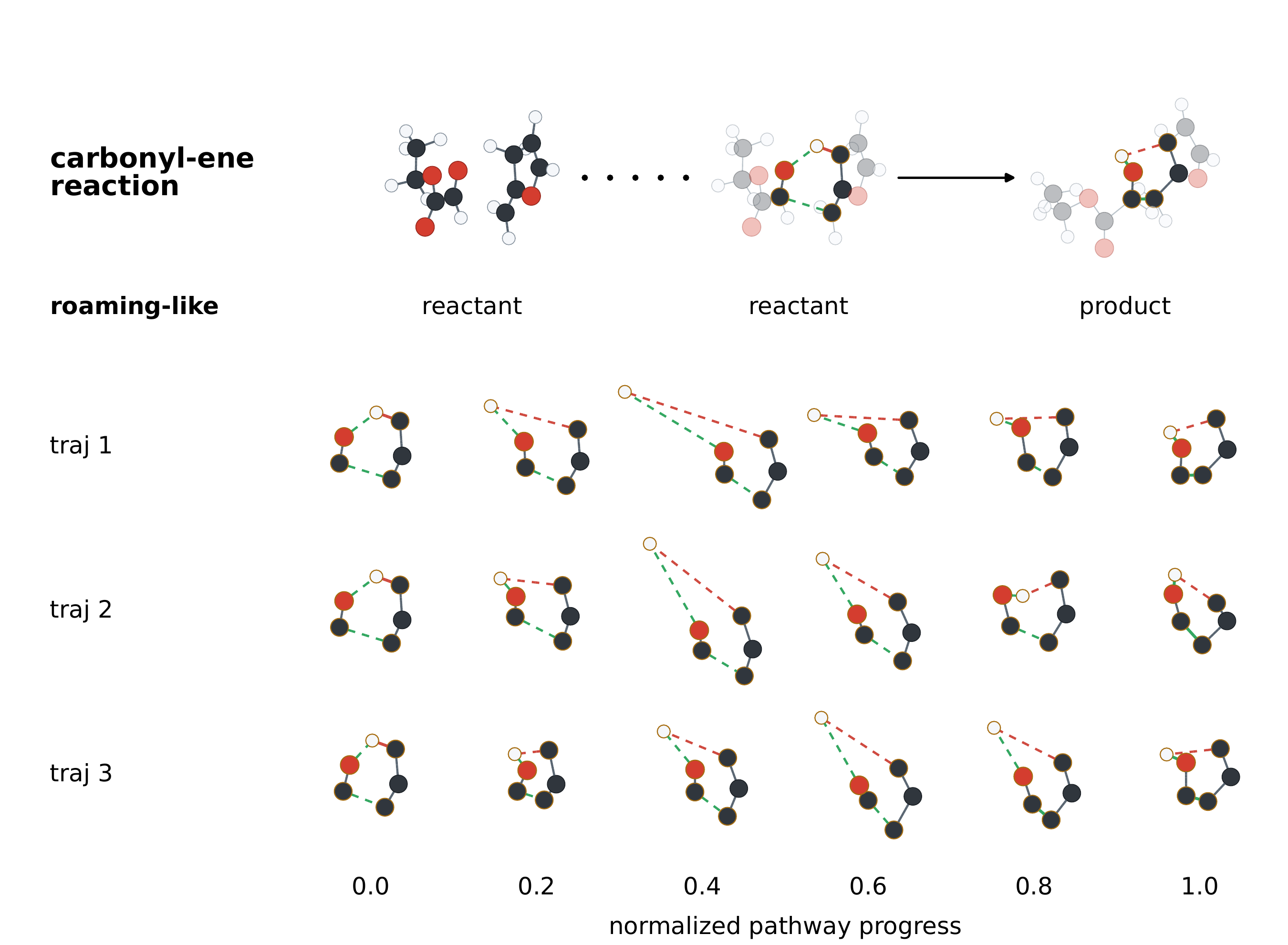}
    
    \caption{Carbonyl-ene reaction proceeding through high-energy spurious roaming-like pathways (Case 1 in Fig.~\ref{fig4}c).
    Top row: 
    Highlights of the reaction center in reactant and product.
    Rows 2--4: reaction-center evolution along the 3 trajectories with roaming-like H atoms from RXNPath\_39 (labeled traj 1, 2, 3), sampled at evenly spaced pathway progresses 0.0--1.0.
    Red and green lines mark breaking and forming bonds; solid lines are within bonding distance (1.3$\times$ sum of covalent radii), dashed lines beyond.
    }
    \label{figS_trj_roam} 

\end{figure}

\begin{figure}[htbp]
    \centering
    \includegraphics[width=1.0\textwidth]
    {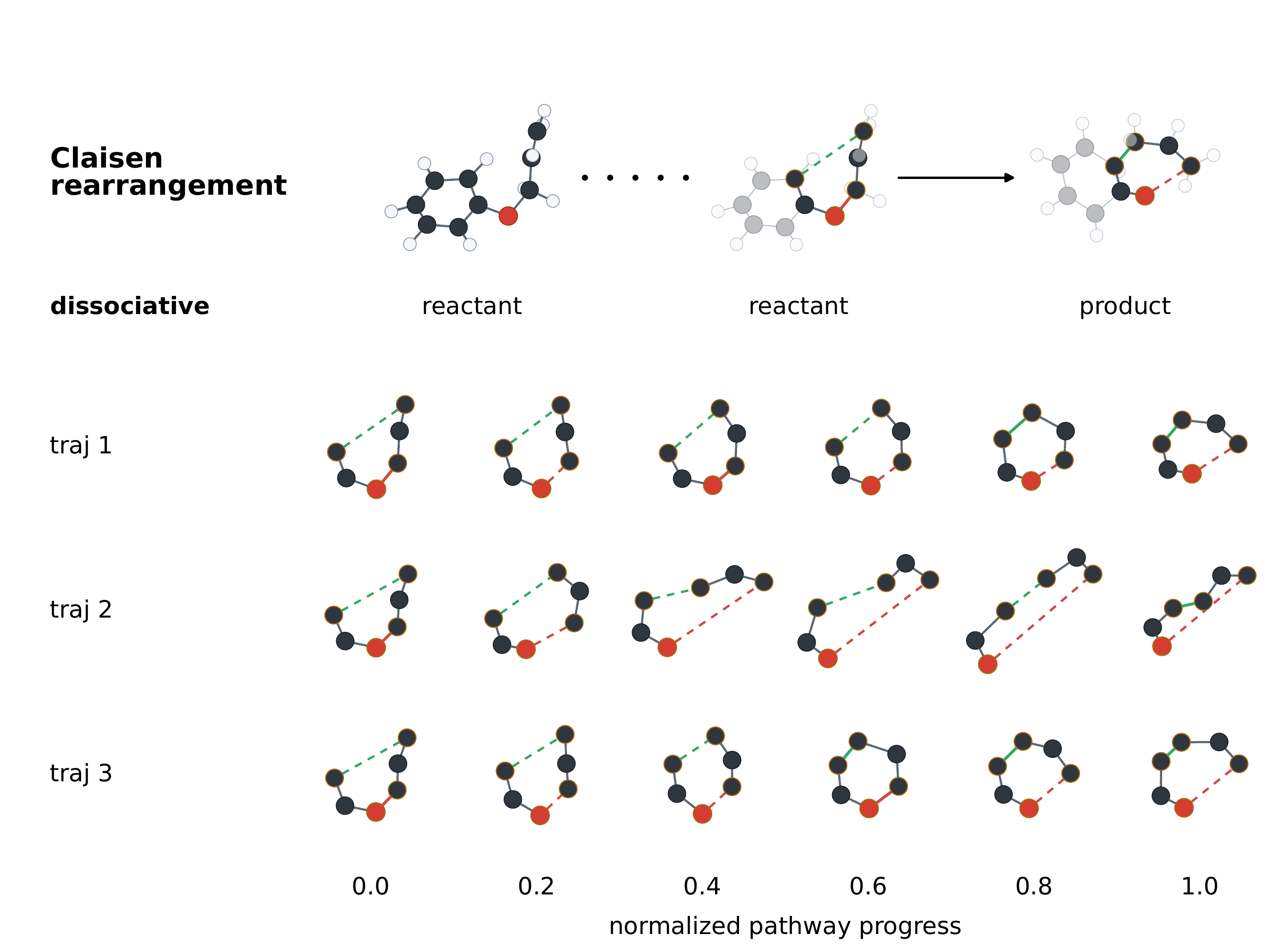}
    
    \caption{Claisen rearrangement in dissociative pathways (Case 2 in Fig.~\ref{fig4}c).
    Top row: 
    Highlights of the reaction center in reactant and product.
    Rows 2--4: reaction-center evolution along the 3 trajectories from RXNPath\_39 (labeled traj 1, 2, 3), sampled at evenly spaced pathway progresses 0.0--1.0.
    Red and green lines mark breaking and forming bonds; solid lines are within bonding distance (1.3$\times$ sum of covalent radii), dashed lines beyond.
    }
    \label{figS_trj_dissociative} 

\end{figure}

\begin{figure}[htbp]
    \centering
    \includegraphics[width=1.0\textwidth]
    {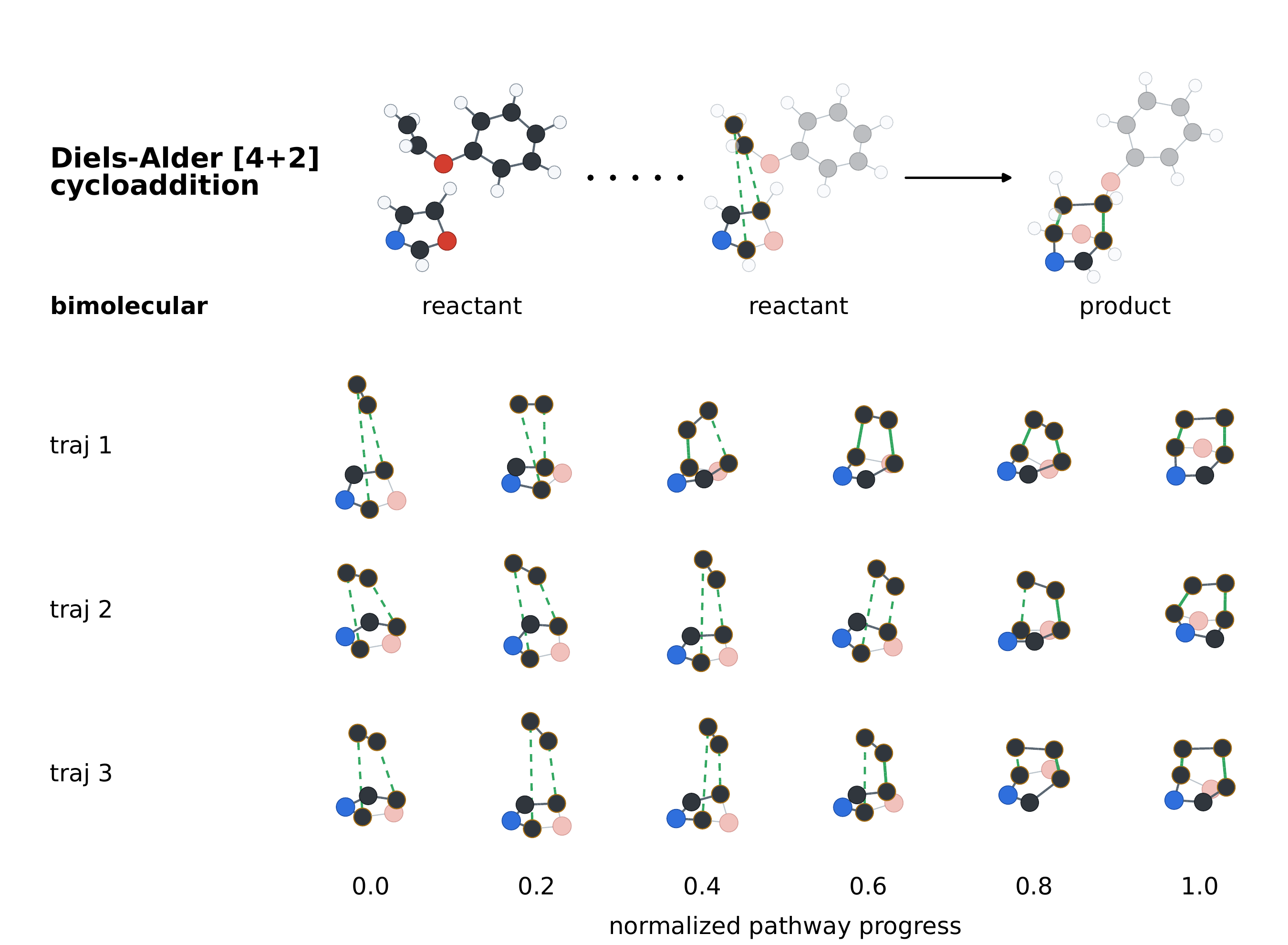}
    
    \caption{Bimolecular Diels--Alder cycloaddition (Case 3 in Fig.~\ref{fig4}c).
    Top row: Highlights of the reaction center in reactant and product.
    Rows 2--4: reaction-center evolution along the 3 trajectories from RXNPath\_39 (labeled traj 1, 2, 3), sampled at evenly spaced pathway progresses 0.0--1.0.
    Red and green lines mark breaking and forming bonds; solid lines are within bonding distance (1.3$\times$ sum of covalent radii), dashed lines beyond.
    }
    \label{figS_trj_DA}

\end{figure}

\begin{figure}[htbp]
    \centering
    \includegraphics[width=1.0\textwidth]
    {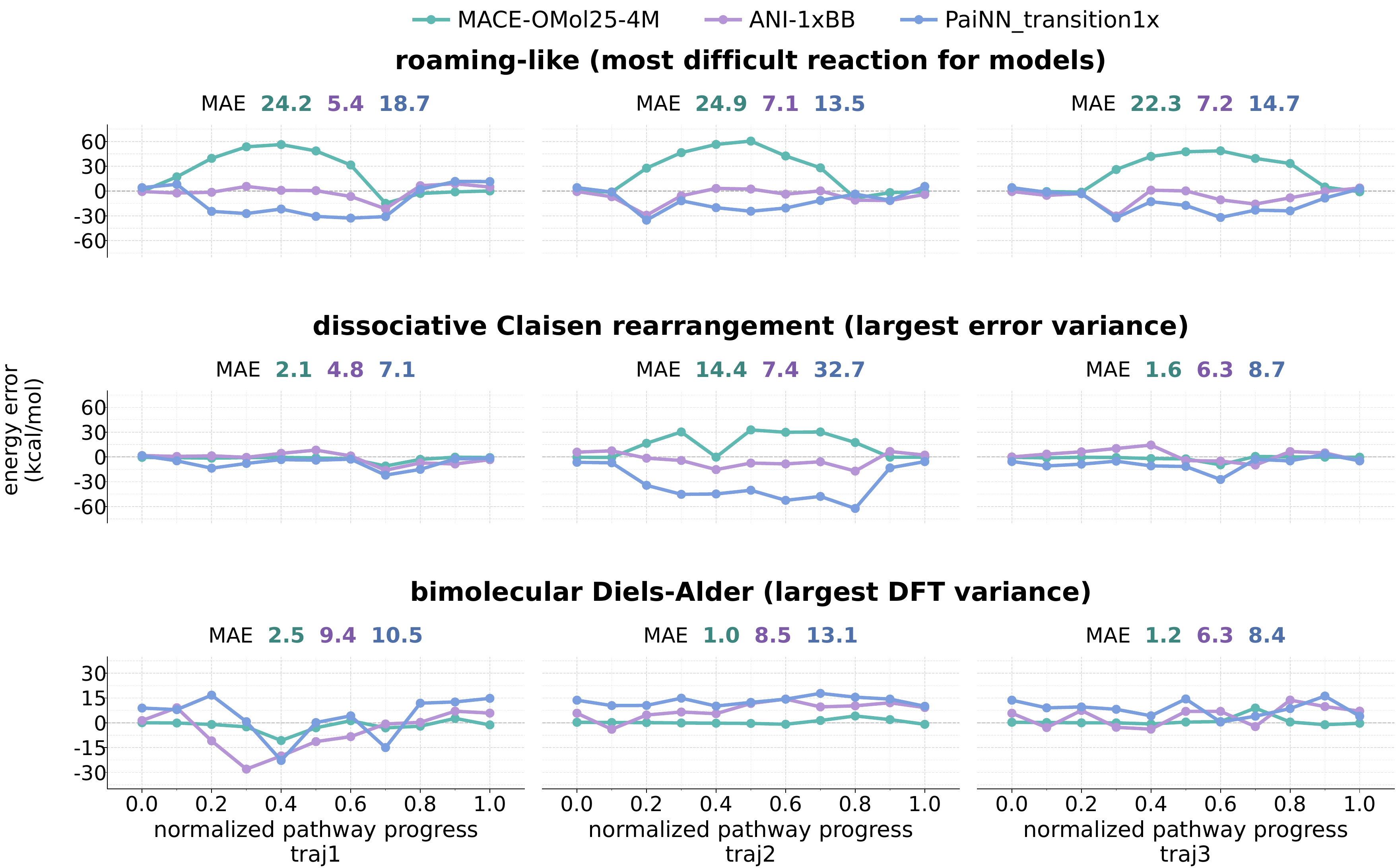}
    
    \caption{Per-trajectory energy errors for the lower-ranked models (MACE-OMol25-4M, ANI-1xBB and PaiNN\_transition1x) on the three reactions in Fig.~\ref{fig4}c.
    Rows correspond to the three cases: roaming-like carbonyl-ene reaction (Case 1, Fig.~\ref{figS_trj_roam}), dissociative Claisen rearrangement (Case 2, Fig.~\ref{figS_trj_dissociative}), bimolecular Diels--Alder cycloaddition (Case 3, Fig.~\ref{figS_trj_DA}); columns show the 3 trajectories per reaction (traj 1--3).
    Each subplot annotates the per-model MAE (kcal/mol).
    DFT references are recalculated using each model's training-data DFT method.
    }
    \label{figS_trj_c} 

\end{figure}

\begin{figure}[htbp]
    \centering
    \includegraphics[width=1.0\textwidth]
    {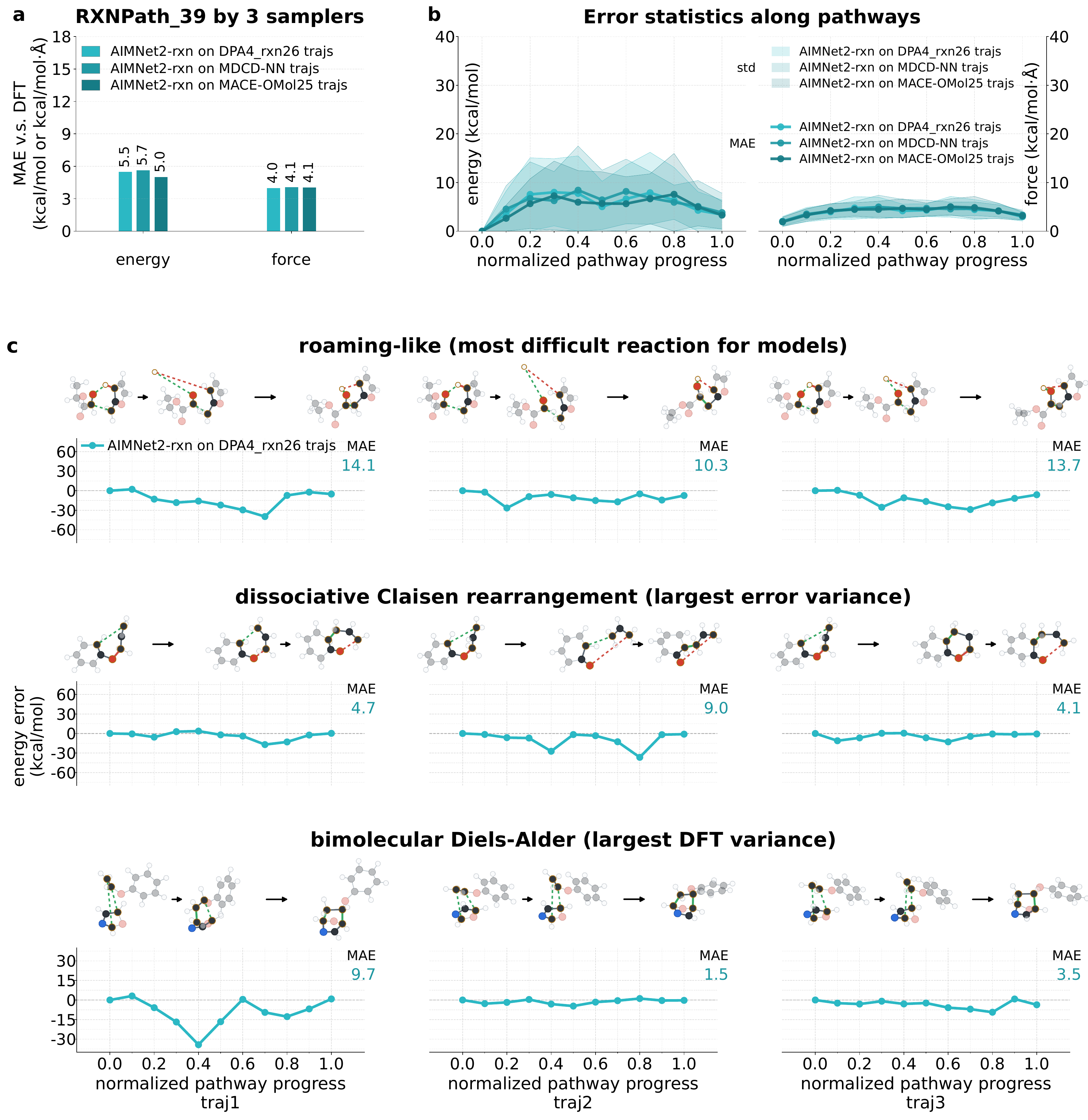}
    
    \caption{Benchmark of AIMNet2-rxn accuracy on RXNPath\_39 dynamic reactive configurations (sampled by DPA4\_rxn26), as well as on configurations sampled by MDCD-NN and MACE-OMol25.
    For each trajectory, the relative energies with respect to the first frame are compared.
    AIMNet2-rxn outputs energy shifts relative to the composition sum of unreported per-element references, which cannot be directly compared to total energies (also not to the formal formation energies). Forces are not shifted.
    a, Overall energy and force MAEs.
    b, Per-progress energy and force error statistics. The normalized pathway progress is the same as that described in Fig.~\ref{fig4}b.
    The range of y axis is aligned with that in Fig.~\ref{figS_traj_b}.
    c, Per-trajectory errors on three representative reactions from RXNPath\_39 (sampled by DPA4\_rxn26). Three reactions are the same as those in Fig.~\ref{fig4}c. The y-axis ranges are aligned with those in Fig.~\ref{figS_trj_c}.
    DFT references are recalculated using AIMNet2-rxn's training-data DFT method.}
    \label{figS_trj_Aimnet2-rxn} 

\end{figure}

\begin{figure}[htbp]
    \centering
    \includegraphics[width=1.0\textwidth]
    {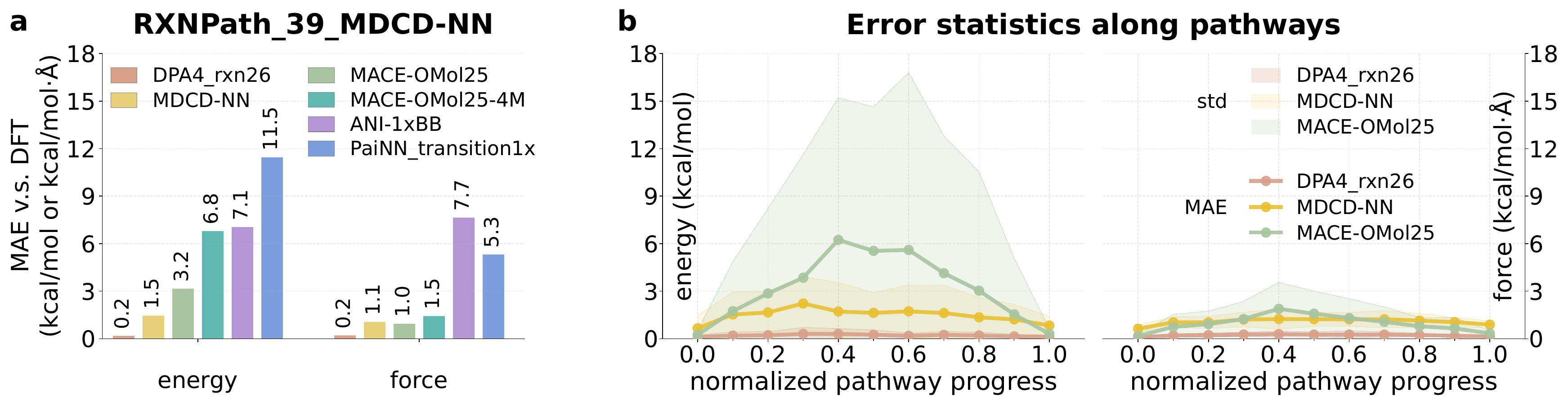}
    
    \caption{Verifying benchmark of model accuracy on dynamic reactive configurations sampled by MDCD-NN. The same sampling strategy for RXNPath\_39 in Fig.~\ref{fig4} is used, and the test set is denoted as RXNPath\_39\_MDCD-NN.
    a, Overall energy and force MAEs.
    b, Per-progress energy and force error statistics for the top-three models in a. Trajectories are projected onto the reaction space defined by distances of bond-changing atom pairs; pathway progress is the normalized cumulative arc length on this projection,
    and selected snapshots are rounded to nearest 0.1 intervals.}
    \label{figS_trj_MDCD_geom} 

\end{figure}

\begin{figure}[htbp]
    \centering
    \includegraphics[width=1.0\textwidth]
    {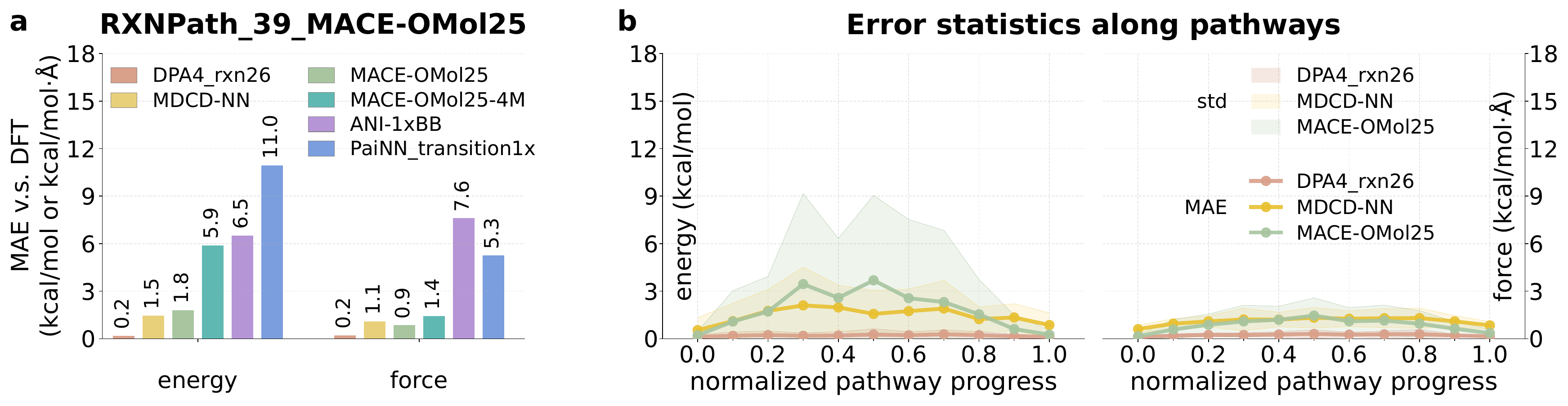}
    
    \caption{Verifying benchmark of model accuracy on dynamic reactive configurations sampled by MACE-OMol25. The same sampling strategy for RXNPath\_39 in Fig.~\ref{fig4} is used, and the test set is denoted as RXNPath\_39\_MACE-OMol25.
    a, Overall energy and force MAEs.
    b, Per-progress energy and force error statistics for the top-three models in a. Trajectories are projected onto the reaction space defined by distances of bond-changing atom pairs; pathway progress is the normalized cumulative arc length on this projection,
    and selected snapshots are rounded to nearest 0.1 intervals.}
    \label{figS_trj_MACE_geom} 

\end{figure}

Results are shown in Fig.~\ref{figS_ablation_efficiency}--Fig.~\ref{figS_ablation_spatial_scale} and Table~\ref{tab:biased-segment-temperature}.

\newpage

\section{Supporting Figures}\label{secS2}

\FloatBarrier

\newpage

\newpage

\FloatBarrier

\section{Supporting Tables}\label{secS3}
\input{SI_table_SMARTS.tex}

\begin{table}[htbp]
\caption{Averaged temperature (K) of metadynamics simulations in the ablation experiment using 4 formalisms of the CV. With respect to the full (spatial+temporal) CV, ablated formalisms respectively retain the spatial, temporal, or neither (bare) resolution.}
\label{tab:biased-segment-temperature}
\centering
\fontsize{8}{8}\selectfont
\setlength{\tabcolsep}{4pt}
\renewcommand{\arraystretch}{1.2}
\begin{tabular}{lcccccc}
\toprule
CV & Stage 1 & Stage 2 & Stage 3 & Stage 4 & Stage 5 & All stages \\
\midrule
bare & $379.7 \pm 51.0$ & $375.2 \pm 44.1$ & $387.8 \pm 55.4$ & $359.8 \pm 33.7$ & $344.8 \pm 24.9$ & $369.5 \pm 45.9$ \\
temporal & $386.4 \pm 49.3$ & $381.8 \pm 43.4$ & $383.3 \pm 51.0$ & $348.4 \pm 26.1$ & $327.9 \pm 17.9$ & $365.5 \pm 46.2$ \\
spatial & $385.1 \pm 53.1$ & $383.6 \pm 53.5$ & $377.4 \pm 56.4$ & $349.3 \pm 30.0$ & $333.7 \pm 25.4$ & $365.9 \pm 50.1$ \\
spatial + temporal & $386.6 \pm 44.5$ & $389.2 \pm 49.6$ & $376.4 \pm 47.4$ & $338.8 \pm 24.8$ & $331.2 \pm 19.5$ & $364.4 \pm 46.2$ \\
\bottomrule
\end{tabular}

\begin{flushleft}
\footnotesize
The simulation is conducted under the 300~K NVT ensemble using a Nosé--Hoover thermostat adopting a 100~$\mathrm{fs}$ damping time.
The unbiased MD segments initializing metadynamics runs are excluded in statistics.
\end{flushleft}
\end{table}

\begin{table}[p]
\caption{DFT force-magnitude statistics on target C atoms. 
OpenRxn26 20$\times$ samples carry lower force magnitudes than OMol25 20$\times$ $N_\mathrm{heavy}\leq30$ samples, despite both from cross-set t-SNE OOD regions.}
\label{tabS:Force-statistics}
\centering
\fontsize{8}{8}\selectfont
\setlength{\tabcolsep}{2pt}
\renewcommand{\arraystretch}{1.2}

\begin{tabular}{@{}l l r@{--}r@{\,}r r c c@{}}
\toprule
\multirow{3}{*}{label method}
& \multicolumn{5}{c}{selected atomic environments}
& \multicolumn{2}{c}{$|\mathbf{F}|$ label mean/std} \\
\cmidrule(lr){2-6}\cmidrule(l){7-8}
& \multicolumn{4}{c}{data source}
& \multirow{2}{*}{\#samples}
& (kcal/mol/\AA)
& (eV/\AA) \\
\cmidrule(lr){2-5}
& name & \multicolumn{3}{c}{$N_\mathrm{heavy}$ (avg.)}
& & & \\
\midrule

\multirow{4}{*}{OMol25}
& OMol25
& 2 & 224 & (37.9)
& 100,000
& 33.9 / 35.4
& 1.47 / 1.54 \\

& \mbox{OpenRxn26~20x}
& 4 & 30 & (\phantom{0}8.7)
& 300
& 40.4 / 36.7
& 1.75 / 1.59 \\

& \mbox{OMol25~20x}
& 3 & 30 & (23.9)
& 300
& 47.3 / 60.3
& 2.05 / 2.62 \\

& \mbox{OMol25~20x}
& 31 & 151 & (70.6)
& 300
& 32.1 / 20.7
& 1.39 / 0.90 \\

\addlinespace[0.6em]

\multirow{4}{*}{OpenRxn26}
& OpenRxn26
& 2 & 30 & (10.1)
& 100,000
& 34.2 / 37.2
& 1.48 / 1.61 \\

& \mbox{OpenRxn26~20x}
& 4 & 30 & (\phantom{0}8.7)
& 300
& 39.2 / 36.7
& 1.70 / 1.59 \\

& \mbox{OMol25~20x}
& 3 & 30 & (23.9)
& 300
& 50.0 / 63.6
& 2.17 / 2.76 \\

& \mbox{OMol25~20x}
& 31 & 151 & (70.6)
& 300
& 33.6 / 22.0
& 1.45 / 0.95 \\

\bottomrule
\end{tabular}

\begin{flushleft}
\footnotesize
Samples are the same as in Table~\ref{tab:ID/OOD}. For cross-set comparison, the OpenRxn26 20$\times$ and OMol25 20$\times$ samples each have two sets of DFT reference labels, computed using the OMol25 and OpenRxn26 label methods, respectively.
\end{flushleft}
\end{table}

\begin{table}[htbp]
\caption{MAEs on cross-set 20$\times$ t-SNE regions as ID samples (MACE-OMol25 on two sets from OMol25 20$\times$; DPA4\_rxn26 on samples from OpenRxn26 20$\times$). Compared with DPA4\_rxn26's performance in Table~\ref{tab:ID/OOD}, OpenRxn26 20$\times$ regions are harder to predict than $N_\mathrm{heavy} \leq30$ OMol25 20$\times$ samples.}
\label{tab:ood_id_reference_mae}
\centering
\fontsize{8}{8}\selectfont
\setlength{\tabcolsep}{2pt}
\renewcommand{\arraystretch}{1.2}

\begin{tabular}{@{}l l l r@{--}r@{\,}r r c c c@{}}
\toprule
\multirow{3}{*}{model}
& \multicolumn{6}{c}{evaluated atomic environments}
& target atom
& \multicolumn{2}{c}{parent frame} \\
\cmidrule(lr){2-7}\cmidrule(lr){8-8}\cmidrule(l){9-10}
& \multirow{2}{*}{ID/OOD}
& \multicolumn{4}{c}{data source}
& \multirow{2}{*}{\#samples}
& force & force & energy \\
\cmidrule(lr){3-6}
& & name & \multicolumn{3}{c}{$N_\mathrm{heavy}$ (avg.)}
& & (kcal/mol/\AA) & (kcal/mol/\AA) & (kcal/mol/atom)\\
\midrule

\multirow{2}{*}{MACE-OMol25}
& ID & \mbox{OMol25~20x}
& 3 & 30 & (23.9)
& 300
& 0.1 & 0.1 & 0.020\\

& ID & \mbox{OMol25~20x}
& 31 & 151 & (70.6)
& 300
& 0.1 & 0.1 & 0.027\\

\addlinespace[0.6em]

DPA4\_rxn26
& ID & \mbox{OpenRxn26~20x}
& 4 & 30 & (\phantom{0}8.7)
& 300
& 0.4 & 0.3 & 0.011\\

\bottomrule
\end{tabular}

\begin{flushleft}
\footnotesize
\end{flushleft}
\end{table}

\end{document}

%% file: SI_table_SMARTS.tex
%
\begingroup
\fontsize{9}{7.5}\selectfont
\setlength{\tabcolsep}{3pt}
\renewcommand{\arraystretch}{1.5}
\setlength{\LTcapwidth}{\textwidth}
\providecommand{\catrule}{\midrule[0.4pt]}
\begin{longtable}{@{}r@{\hspace{0.7em}}>{\raggedright\arraybackslash}p{0.19\linewidth}@{\hspace{0.7em}}>{\raggedright\arraybackslash}p{0.21\linewidth}@{\hspace{0.7em}}>{\raggedright\arraybackslash}p{0.52\linewidth}@{}}
\caption{The 26 SMARTS reaction templates used in the staged reactant-selection strategy (Sec.~\ref{subsec:staged_reactant_selection}) to infer possible product SMILES from each candidate reactant.}
\label{tabS:SMARTS}\\
\toprule
\textbf{\#} & \textbf{Rule name} & \textbf{Description} & \textbf{Reaction SMARTS} \\
\midrule
\endfirsthead
\multicolumn{4}{l}{\textit{Table~\ref{tabS:SMARTS} (continued)}} \\
\toprule
\textbf{\#} & \textbf{Rule name} & \textbf{Description} & \textbf{Reaction SMARTS} \\
\midrule
\endhead
\midrule
\multicolumn{4}{r}{\textit{continued on next page}} \\
\endfoot
\bottomrule
\endlastfoot
\multicolumn{4}{@{}l}{\textbf{Sigmatropic H-shifts}} \\*
1 & Allylic H shift & H transfers across C--C=C--C; the C=C shifts one position & \hyphenpenalty=10000\exhyphenpenalty=10000 {\ttfamily [C:1]\allowbreak{}-[C:2]\allowbreak{}=[C:3]\allowbreak{}-[CH:4]\allowbreak{}}~\ensuremath{>>}\newline {\ttfamily [CH:1]\allowbreak{}=[C:2]\allowbreak{}-[C:3]\allowbreak{}-[C:4]\allowbreak{}} \\
2 & 1,5-hydrogen shift & Allylic methine H relocates across a 1,3-diene & \hyphenpenalty=10000\exhyphenpenalty=10000 {\ttfamily [C:1]\allowbreak{}=[C:2]\allowbreak{}-[C:3]\allowbreak{}=[C:4]\allowbreak{}-[CH:5]\allowbreak{}}~\ensuremath{>>}\newline {\ttfamily [CH:1]\allowbreak{}=[C:2]\allowbreak{}-[C:3]\allowbreak{}=[C:4]\allowbreak{}=[C:5]\allowbreak{}} \\
\catrule
\multicolumn{4}{@{}l}{\textbf{Sigmatropic C--C shifts}} \\*
3 & Allylic ester $[3,3]$ shift & $[3,3]$-shift of an allylic ester motif (including Carroll rearrangement) & \hyphenpenalty=10000\exhyphenpenalty=10000 {\ttfamily [C:1]\allowbreak{}=[C:2]\allowbreak{}[C:3]\allowbreak{}[O:4]\allowbreak{}[C:5]\allowbreak{}(=[O:6]\allowbreak{})\allowbreak{}[C:7]\allowbreak{}}~\ensuremath{>>}\newline {\ttfamily [C:1]\allowbreak{}[C:2]\allowbreak{}=[C:3]\allowbreak{}[C:7]\allowbreak{}([OH:4]\allowbreak{})\allowbreak{}[C:5]\allowbreak{}=[O:6]\allowbreak{}} \\
4 & Carbonyl$\to$dienol shift & Breaks C$_\alpha$--C$_\beta$, bonds the carbonyl C to the alkene terminus $\to$ dienol & \hyphenpenalty=10000\exhyphenpenalty=10000 {\ttfamily [O:1]\allowbreak{}=[C:2]\allowbreak{}-[C:3]\allowbreak{}-[C:4]\allowbreak{}-[C:5]\allowbreak{}=[C:6]\allowbreak{}}~\ensuremath{>>}\newline {\ttfamily [C:4]\allowbreak{}-[C:5]\allowbreak{}=[C:6]\allowbreak{}-[C:2]\allowbreak{}([O:1]\allowbreak{})\allowbreak{}=[C:3]\allowbreak{}} \\
\catrule
\multicolumn{4}{@{}l}{\textbf{Electrocyclizations}} \\*
5 & 4$\pi$ electrocyclization & Diene closes to a cyclobutene & \hyphenpenalty=10000\exhyphenpenalty=10000 {\ttfamily [C:1]\allowbreak{}=[C:2]\allowbreak{}-[C:3]\allowbreak{}=[C:4]\allowbreak{}}~\ensuremath{>>}\newline {\ttfamily [C:1]\allowbreak{}1-[C:2]\allowbreak{}=[C:3]\allowbreak{}-[C:4]\allowbreak{}-1} \\
6 & 6$\pi$ electrocyclization & Triene closes to a 1,3-cyclohexadiene & \hyphenpenalty=10000\exhyphenpenalty=10000 {\ttfamily [C:1]\allowbreak{}=[C:2]\allowbreak{}-[C:3]\allowbreak{}=[C:4]\allowbreak{}-[C:5]\allowbreak{}=[C:6]\allowbreak{}}~\ensuremath{>>}\newline {\ttfamily [C:1]\allowbreak{}1=[C:2]\allowbreak{}-[C:3]\allowbreak{}=[C:4]\allowbreak{}-[C:5]\allowbreak{}-[C:6]\allowbreak{}-1} \\
7 & 8$\pi$ electrocyclization & Tetraene closes to a cyclooctatriene & \hyphenpenalty=10000\exhyphenpenalty=10000 {\ttfamily [C:1]\allowbreak{}=[C:2]\allowbreak{}-[C:3]\allowbreak{}=[C:4]\allowbreak{}-[C:5]\allowbreak{}=[C:6]\allowbreak{}-[C:7]\allowbreak{}=[C:8]\allowbreak{}}~\ensuremath{>>}\newline {\ttfamily [C:1]\allowbreak{}1=[C:2]\allowbreak{}-[C:3]\allowbreak{}=[C:4]\allowbreak{}-[C:5]\allowbreak{}=[C:6]\allowbreak{}-[C:7]\allowbreak{}-[C:8]\allowbreak{}-1} \\
\catrule
\multicolumn{4}{@{}l}{\textbf{Ring transformations}} \\*
8 & Ring expansion & 4-membered ring $\to$ 5-membered ring & \hyphenpenalty=10000\exhyphenpenalty=10000 {\ttfamily [C:1]\allowbreak{}1-[C:2]\allowbreak{}-[C:3]\allowbreak{}-[C:4]\allowbreak{}-1}~\ensuremath{>>}\newline {\ttfamily [C:1]\allowbreak{}1-[C:2]\allowbreak{}-[C:3]\allowbreak{}-[C:4]\allowbreak{}-[C:5]\allowbreak{}-1} \\
9 & Ring contraction & 5-membered ring $\to$ 4-membered ring & \hyphenpenalty=10000\exhyphenpenalty=10000 {\ttfamily [C:1]\allowbreak{}1-[C:2]\allowbreak{}-[C:3]\allowbreak{}-[C:4]\allowbreak{}-[C:5]\allowbreak{}-1}~\ensuremath{>>}\newline {\ttfamily [C:1]\allowbreak{}1-[C:2]\allowbreak{}-[C:3]\allowbreak{}-[C:4]\allowbreak{}-1} \\
10 & Fused bicyclization & Chain closes to a 3-membered ring fused to a 5-membered ring & \hyphenpenalty=10000\exhyphenpenalty=10000 {\ttfamily [C:1]\allowbreak{}([C:5]\allowbreak{})\allowbreak{}([C:6]\allowbreak{})\allowbreak{}-[C:2]\allowbreak{}=[C:3]\allowbreak{}-[C:4]\allowbreak{}}~\ensuremath{>>}\newline {\ttfamily [C:1]\allowbreak{}12-[C:5]\allowbreak{}-[C:6]\allowbreak{}-1-[C:2]\allowbreak{}-[C:3]\allowbreak{}-[C:4]\allowbreak{}-2} \\
11 & Aziridine ring closure & C--N--C closes to a three-membered aziridine ring & \hyphenpenalty=10000\exhyphenpenalty=10000 {\ttfamily [C:1]\allowbreak{}[N:2]\allowbreak{}[C:3]\allowbreak{}}~\ensuremath{>>}\newline {\ttfamily [C:1]\allowbreak{}1-[N:2]\allowbreak{}-[C:3]\allowbreak{}-1} \\
12 & Hydroxycyclopentene motif fragmentation & Hydroxycyclopentene motif cleaves; groups on C$_6$ could fragment together & \hyphenpenalty=10000\exhyphenpenalty=10000 {\ttfamily [C:1]\allowbreak{}1-[C:2]\allowbreak{}=[C:3]\allowbreak{}-[C:4]\allowbreak{}([OH:5]\allowbreak{})\allowbreak{}-[C:6]\allowbreak{}-1}~\ensuremath{>>}\newline {\ttfamily [C:1]\allowbreak{}=[C:2]\allowbreak{}-[C:3]\allowbreak{}=[C:4]\allowbreak{}=[O:5]\allowbreak{}.[C:6]\allowbreak{}} \\
13 & Meinwald rearrangement & Epoxide rearranges to a carbonyl & \hyphenpenalty=10000\exhyphenpenalty=10000 {\ttfamily [C:1]\allowbreak{}1-[O:2]\allowbreak{}-[C:3]\allowbreak{}-1}~\ensuremath{>>}\newline {\ttfamily [C:1]\allowbreak{}(=[O:2]\allowbreak{})\allowbreak{}-[C:3]\allowbreak{}} \\
14 & Epoxide to cyclopropanol & Epoxide O leaves the ring; adjacent C closes a cyclopropanol & \hyphenpenalty=10000\exhyphenpenalty=10000 {\ttfamily [C:1]\allowbreak{}1-[O:2]\allowbreak{}-[C:3]\allowbreak{}([C:4]\allowbreak{})\allowbreak{}-1}~\ensuremath{>>}\newline {\ttfamily [C:1]\allowbreak{}1-[C:3]\allowbreak{}([O:2]\allowbreak{})\allowbreak{}-[C:4]\allowbreak{}-1} \\
\catrule
\multicolumn{4}{@{}l}{\textbf{Heteroatom migrations}} \\*
15 & Nitrogen walk & N moves to the far end of an allyl unit & \hyphenpenalty=10000\exhyphenpenalty=10000 {\ttfamily [N:1]\allowbreak{}-[C:2]\allowbreak{}=[C:3]\allowbreak{}-[C:4]\allowbreak{}}~\ensuremath{>>}\newline {\ttfamily [C:2]\allowbreak{}=[C:3]\allowbreak{}-[C:4]\allowbreak{}-[N:1]\allowbreak{}} \\
16 & Oxygen walk & O moves to the far end of an allyl unit & \hyphenpenalty=10000\exhyphenpenalty=10000 {\ttfamily [O:1]\allowbreak{}-[C:2]\allowbreak{}=[C:3]\allowbreak{}-[C:4]\allowbreak{}}~\ensuremath{>>}\newline {\ttfamily [C:2]\allowbreak{}=[C:3]\allowbreak{}-[C:4]\allowbreak{}-[O:1]\allowbreak{}} \\
\catrule
\multicolumn{4}{@{}l}{\textbf{Tautomerizations \& $\pi$-bond isomerizations}} \\*
17 & Diene isomerization & C=C migrates to give a conjugated 1,3-diene & \hyphenpenalty=10000\exhyphenpenalty=10000 {\ttfamily [C:1]\allowbreak{}=[C:2]\allowbreak{}-[C:3]\allowbreak{}-[C:4]\allowbreak{}=[C:5]\allowbreak{}-[C:6]\allowbreak{}}~\ensuremath{>>}\newline {\ttfamily [C:1]\allowbreak{}-[C:2]\allowbreak{}=[C:3]\allowbreak{}-[C:4]\allowbreak{}=[C:5]\allowbreak{}-[C:6]\allowbreak{}} \\
18 & Imine--enamine tautom. & Imine $\to$ conjugated dienamine (aza-diene) & \hyphenpenalty=10000\exhyphenpenalty=10000 {\ttfamily [N:1]\allowbreak{}=[C:2]\allowbreak{}-[C:3]\allowbreak{}-[C:4]\allowbreak{}=[C:5]\allowbreak{}-[C:6]\allowbreak{}}~\ensuremath{>>}\newline {\ttfamily [N:1]\allowbreak{}-[C:2]\allowbreak{}=[C:3]\allowbreak{}-[C:4]\allowbreak{}=[C:5]\allowbreak{}-[C:6]\allowbreak{}} \\
19 & Oxime--nitroso tautom. & Oxime (C=N--OH) $\to$ C--nitroso (C--N=O) & \hyphenpenalty=10000\exhyphenpenalty=10000 {\ttfamily [C:1]\allowbreak{}=[N:2]\allowbreak{}[OH:3]\allowbreak{}}~\ensuremath{>>}\newline {\ttfamily [C:1]\allowbreak{}[N:2]\allowbreak{}=[O:3]\allowbreak{}} \\
20 & Dienone$\to$phenol aromatization & Cyclohexa-2,5-dienone aromatizes to phenol & \hyphenpenalty=10000\exhyphenpenalty=10000 {\ttfamily [C:1]\allowbreak{}1=[C:2]\allowbreak{}[C:3]\allowbreak{}(=[O:4]\allowbreak{})\allowbreak{}[C:5]\allowbreak{}=[C:6]\allowbreak{}[C:7]\allowbreak{}-1}~\ensuremath{>>}\newline {\ttfamily [c:1]\allowbreak{}1[c:2]\allowbreak{}[c:3]\allowbreak{}([OH:4]\allowbreak{})\allowbreak{}[c:5]\allowbreak{}[c:6]\allowbreak{}[c:7]\allowbreak{}1} \\
21 & Keto-enol tautomerism & C=O $\rightleftharpoons$ C=C--OH & \hyphenpenalty=10000\exhyphenpenalty=10000 {\ttfamily [C:1]\allowbreak{}-[C:2]\allowbreak{}=[O:3]\allowbreak{}}~\ensuremath{>>}\newline {\ttfamily [C:1]\allowbreak{}=[C:2]\allowbreak{}-[OH:3]\allowbreak{}} \\
\catrule
\multicolumn{4}{@{}l}{\textbf{Acyl / nitrogen rearrangements}} \\*
22 & Benzilic acid rearrangement & 1,2-diaryl diketone $\to$ $\alpha$-hydroxy acid (aryl migration) & \hyphenpenalty=10000\exhyphenpenalty=10000 {\ttfamily [c:1]\allowbreak{}[C:2]\allowbreak{}(=[O:3]\allowbreak{})\allowbreak{}[C:4]\allowbreak{}(=[O:5]\allowbreak{})\allowbreak{}[c:6]\allowbreak{}}~\ensuremath{>>}\newline {\ttfamily [c:1]\allowbreak{}[C:2]\allowbreak{}([OH:3]\allowbreak{})\allowbreak{}([c:6]\allowbreak{})\allowbreak{}[C:4]\allowbreak{}(=[O:5]\allowbreak{})\allowbreak{}[OH]\allowbreak{}} \\
23 & Wolff rearrangement & $\alpha$-diazoketone $\to$ ketene ($-$N$_2$) & \hyphenpenalty=10000\exhyphenpenalty=10000 {\ttfamily [C:1]\allowbreak{}(=[O:2]\allowbreak{})\allowbreak{}[C:3]\allowbreak{}=[N+:4]\allowbreak{}=[N-:5]\allowbreak{}}~\ensuremath{>>}\newline {\ttfamily [C:1]\allowbreak{}=[C:3]\allowbreak{}=[O:2]\allowbreak{}} \\
24 & Curtius rearrangement & Acyl azide $\to$ isocyanate ($-$N$_2$) & \hyphenpenalty=10000\exhyphenpenalty=10000 {\ttfamily [C:1]\allowbreak{}(=[O:2]\allowbreak{})\allowbreak{}[N:3]\allowbreak{}=[N+:4]\allowbreak{}=[N-:5]\allowbreak{}}~\ensuremath{>>}\newline {\ttfamily [N:3]\allowbreak{}=[C:1]\allowbreak{}=[O:2]\allowbreak{}} \\
25 & Lossen rearrangement & Hydroxamic acid $\to$ isocyanate & \hyphenpenalty=10000\exhyphenpenalty=10000 {\ttfamily [C:1]\allowbreak{}(=[O:2]\allowbreak{})\allowbreak{}[N:3]\allowbreak{}[OH:4]\allowbreak{}}~\ensuremath{>>}\newline {\ttfamily [N:3]\allowbreak{}=[C:1]\allowbreak{}=[O:2]\allowbreak{}} \\
26 & Schmidt rearrangement & Ketone $\to$ amide via formal N insertion & \hyphenpenalty=10000\exhyphenpenalty=10000 {\ttfamily [C:1]\allowbreak{}(=[O:2]\allowbreak{})\allowbreak{}[C:3]\allowbreak{}([C:4]\allowbreak{})\allowbreak{}([C:5]\allowbreak{})\allowbreak{}}~\ensuremath{>>}\newline {\ttfamily [C:3]\allowbreak{}([C:4]\allowbreak{})\allowbreak{}([C:5]\allowbreak{})\allowbreak{}[N:6]\allowbreak{}[C:1]\allowbreak{}(=[O:2]\allowbreak{})\allowbreak{}} \\
\end{longtable}
\endgroup

%% file: sn-bibliography.bib
@article{Marcelin1915,
  title={Contribution {\`a} l'{\'e}tude de la cin{\'e}tique physico-chimique},
  author={Marcelin, Ren{\'e}},
  journal={Annales de Physique},
  volume={9},
  number={3},
  pages={120--231},
  year={1915},
  publisher={EDP Sciences}
}

@article{Eyring1935,
  author  = {Eyring, Henry},
  title   = {The Activated Complex in Chemical Reactions},
  journal = {The Journal of Chemical Physics},
  volume  = {3},
  number  = {2},
  pages   = {107--115},
  year    = {1935},
  doi     = {10.1063/1.1749604}
}

@article{Wigner1938,
  author  = {Wigner, E.},
  title   = {The Transition State Method},
  journal = {Transactions of the Faraday Society},
  volume  = {34},
  pages   = {29--41},
  year    = {1938},
  doi     = {10.1039/TF9383400029}
}

@inbook{Keck1967,
author = {Keck, James C.},
publisher = {John Wiley \& Sons, Ltd},
isbn = {9780470140154},
title = {Variational Theory of Reaction Rates},
booktitle = {Advances in Chemical Physics},
chapter = {},
pages = {85-121},
doi = {https://doi.org/10.1002/9780470140154.ch5},
url = {https://onlinelibrary.wiley.com/doi/abs/10.1002/9780470140154.ch5},
eprint = {https://onlinelibrary.wiley.com/doi/pdf/10.1002/9780470140154.ch5},
year = {1967}
}

@article{Truhlar1980,
  author  = {Truhlar, Donald G. and Garrett, Bruce C.},
  title   = {Variational Transition-State Theory},
  journal = {Accounts of Chemical Research},
  volume  = {13},
  number  = {12},
  pages   = {440--448},
  year    = {1980},
  doi     = {10.1021/ar50156a002}
}

@article{Rice1927,
  author  = {Rice, Oscar Knefler and Ramsperger, Herman C.},
  title   = {Theories of Unimolecular Gas Reactions at Low Pressures},
  journal = {Journal of the American Chemical Society},
  volume  = {49},
  number  = {7},
  pages   = {1617--1629},
  year    = {1927},
  doi     = {10.1021/ja01406a001}
}

@article{Kassel1928,
  author  = {Kassel, Louis S.},
  title   = {Studies in Homogeneous Gas Reactions. I},
  journal = {The Journal of Physical Chemistry},
  volume  = {32},
  number  = {2},
  pages   = {225--242},
  year    = {1928},
  doi     = {10.1021/j150284a007}
}

@article{Marcus1952,
  author  = {Marcus, R. A.},
  title   = {Unimolecular Dissociations and Free Radical Recombination Reactions},
  journal = {The Journal of Chemical Physics},
  volume  = {20},
  number  = {3},
  pages   = {359--364},
  year    = {1952},
  doi     = {10.1063/1.1700424}
}

@article{Carpenter1992,
  author  = {Carpenter, Barry K.},
  title   = {Intramolecular Dynamics for the Organic Chemist},
  journal = {Accounts of Chemical Research},
  volume  = {25},
  number  = {11},
  pages   = {520--528},
  year    = {1992},
  doi     = {10.1021/ar00023a006}
}

@article{Carpenter2005,
  author  = {Carpenter, Barry K.},
  title   = {Nonstatistical Dynamics in Thermal Reactions of Polyatomic Molecules},
  journal = {Annual Review of Physical Chemistry},
  volume  = {56},
  pages   = {57--89},
  year    = {2005},
  doi     = {10.1146/annurev.physchem.56.092503.141240}
}

@article{Ess2008,
  author  = {Ess, Daniel H. and Wheeler, Steven E. and Iafe, Robert G. and Xu, Lai and {\c{C}}elebi-{\"O}l{\c{c}}{\"u}m, Nihan and Houk, Kendall N.},
  title   = {Bifurcations on Potential Energy Surfaces of Organic Reactions},
  journal = {Angewandte Chemie International Edition},
  volume  = {47},
  number  = {40},
  pages   = {7592--7601},
  year    = {2008},
  doi     = {10.1002/anie.200800918}
}

@article{Suits2020,
  author  = {Suits, Arthur G.},
  title   = {Roaming Reactions and Dynamics in the van der Waals Region},
  journal = {Annual Review of Physical Chemistry},
  volume  = {71},
  pages   = {77--100},
  year    = {2020},
  doi     = {10.1146/annurev-physchem-050317-020929}
}

@incollection{Tantillo2021,
title = {Chapter One - Beyond transition state theory—Non-statistical dynamic effects for organic reactions},
editor = {Ian H. Williams and Nicholas H. Williams},
booktitle = {Advances in Physical Organic Chemistry},
publisher = {Academic Press},
volume = {55},
pages = {1-16},
year = {2021},
issn = {0065-3160},
doi = {https://doi.org/10.1016/bs.apoc.2021.06.001},
url = {https://www.sciencedirect.com/science/article/pii/S0065316021000010},
author = {Dean J. Tantillo}
}

@article{IUPAC2022,
url = {https://doi.org/10.1515/pac-2018-1010},
title = {Glossary of terms used in physical organic chemistry (IUPAC Recommendations 2021)},
author = {Charles L. Perrin and Israel Agranat and Alessandro Bagno and Silvia E. Braslavsky and Pedro Alexandrino Fernandes and Jean-François Gal and Guy C. Lloyd-Jones and Herbert Mayr and Joseph R. Murdoch and Norma Sbarbati Nudelman and Leo Radom and Zvi Rappoport and Marie-Françoise Ruasse and Hans-Ullrich Siehl and Yoshito Takeuchi and Thomas T. Tidwell and Einar Uggerud and Ian H. Williams},
pages = {353--534},
volume = {94},
number = {4},
journal = {Pure and Applied Chemistry},
doi = {10.1515/pac-2018-1010},
year = {2022},
lastchecked = {2026-03-20}
}

@article{Carpenter1995,
  author  = {Carpenter, Barry K.},
  title   = {Dynamic Matching: The Cause of Inversion of Configuration in the [1,3] Sigmatropic Migration?},
  journal = {Journal of the American Chemical Society},
  volume  = {117},
  number  = {23},
  pages   = {6336--6344},
  year    = {1995},
  doi     = {10.1021/ja00128a024}
}

@article{Townsend2004,
  author  = {Townsend, D. and Lahankar, S. A. and Lee, S. K. and Chambreau, S. D. and Suits, A. G. and Zhang, X. and Rheinecker, J. and Harding, L. B. and Bowman, J. M.},
  title   = {The Roaming Atom: Straying from the Reaction Path in Formaldehyde Decomposition},
  journal = {Science},
  volume  = {306},
  number  = {5699},
  pages   = {1158--1161},
  year    = {2004},
  doi     = {10.1126/science.1104386}
}

@article{Lahankar2008,
title = {Further aspects of the roaming mechanism in formaldehyde dissociation},
journal = {Chemical Physics},
volume = {347},
number = {1},
pages = {288-299},
year = {2008},
note = {Ultrafast Photoinduced Processes in Polyatomic Molecules},
issn = {0301-0104},
doi = {10.1016/j.chemphys.2007.11.007},
author = {S.A. Lahankar and V. Goncharov and F. Suits and J.D. Farnum and J.M. Bowman and Arthur G. Suits},

}

@article{Caramella2002,
  author  = {Caramella, Pierluigi and Quadrelli, Paolo and Toma, Lucio},
  title   = {An Unexpected Bispericyclic Transition Structure Leading to 4+2 and 2+4 Cycloadducts in the Endo Dimerization of Cyclopentadiene},
  journal = {Journal of the American Chemical Society},
  volume  = {124},
  number  = {7},
  pages   = {1130--1131},
  year    = {2002},
  doi     = {10.1021/ja016622h}
}

@article{Wang2009,
author = {Wang, Zhihong and Hirschi, Jennifer S. and Singleton, Daniel A.},
title = {Recrossing and Dynamic Matching Effects on Selectivity in a Diels-Alder Reaction},
journal = {Angewandte Chemie International Edition},
volume = {48},
number = {48},
pages = {9156-9159},
doi = {10.1002/anie.200903293},
eprint = {https://onlinelibrary.wiley.com/doi/pdf/10.1002/anie.200903293},
year = {2009}
}

@article{Hou2024,
  author  = {Hou, Yi-Fan and Zhang, Quanhao and Dral, Pavlo O.},
  title   = {Surprising Dynamics Phenomena in the {Diels-Alder} Reaction of {C$_{60}$} Uncovered with {AI}},
  journal = {The Journal of Organic Chemistry},
  volume  = {89},
  number  = {20},
  pages   = {15041--15047},
  year    = {2024},
  doi     = {10.1021/acs.joc.4c01763}
}

@article{Berson1968,
  author  = {Berson, Jerome A.},
  title   = {The Stereochemistry of Sigmatropic Rearrangements. Tests of the Predictive Power of Orbital Symmetry Rules},
  journal = {Accounts of Chemical Research},
  volume  = {1},
  number  = {5},
  pages   = {152--160},
  year    = {1968},
  doi     = {10.1021/ar50005a004}
}

@article{Baldwin2008,
  author  = {Baldwin, John E. and Leber, Phyllis A.},
  title   = {Molecular Rearrangements through Thermal [1,3] Carbon Shifts},
  journal = {Organic \& Biomolecular Chemistry},
  volume  = {6},
  number  = {1},
  pages   = {36--47},
  year    = {2008},
  doi     = {10.1039/B711494J}
}

@article{Mardirossian2017,
author = {Narbe Mardirossian and Martin Head-Gordon},
title = {Thirty years of density functional theory in computational chemistry: an overview and extensive assessment of 200 density functionals},
journal = {Molecular Physics},
volume = {115},
number = {19},
pages = {2315--2372},
year = {2017},
publisher = {Taylor \& Francis},
doi = {10.1080/00268976.2017.1333644},
}

@article{Truhlar1996,
  author  = {Truhlar, Donald G. and Garrett, Bruce C. and Klippenstein, Stephen J.},
  title   = {Current Status of Transition-State Theory},
  journal = {The Journal of Physical Chemistry},
  volume  = {100},
  number  = {31},
  pages   = {12771--12800},
  year    = {1996},
  doi     = {10.1021/jp953748q}
}

@article{vanDuin2001,
  author  = {van Duin, Adri C. T. and Dasgupta, Siddharth and Lorant, Francois and Goddard, William A.},
  title   = {ReaxFF: A Reactive Force Field for Hydrocarbons},
  journal = {The Journal of Physical Chemistry A},
  volume  = {105},
  number  = {41},
  pages   = {9396--9409},
  year    = {2001},
  doi     = {10.1021/jp004368u}
}

@article{Behler2007,
title = {Generalized Neural-Network Representation of High-Dimensional Potential-Energy Surfaces},
  author = {Behler, J\"org and Parrinello, Michele},
  journal = {Physical Review Letters},
  year = {2007},
  volume = {98},
  number = {14},
  pages = {146401},
  doi = {10.1103/PhysRevLett.98.146401}
}

@article{MLIPReview2025,
  author  = {Xia, Junfan and Zhang, Yaolong and Jiang, Bin},
  title   = {The Evolution of Machine Learning Potentials for Molecules, Reactions and Materials},
  journal = {Chemical Society Reviews},
  volume  = {54},
  number  = {10},
  pages   = {4790--4821},
  year    = {2025},
  doi     = {10.1039/D5CS00104H}
}

@article{Transition1x,
  author  = {Schreiner, Mathias and Bhowmik, Arghya and Vegge, Tejs and Busk, Jonas and Winther, Ole},
  title   = {Transition1x - a Dataset for Building Generalizable Reactive Machine Learning Potentials},
  journal = {Scientific Data},
  volume  = {9},
  pages   = {779},
  year    = {2022},
  doi     = {10.1038/s41597-022-01870-w}
}

@article{RGD1,
  author  = {Zhao, Qiyuan and Vaddadi, Sai Mahit and Woulfe, Michael and Ogunfowora, Lawal A. and Garimella, Sanjay S. and Isayev, Olexandr and Savoie, Brett M.},
  title   = {Comprehensive Exploration of Graphically Defined Reaction Spaces},
  journal = {Scientific Data},
  volume  = {10},
  pages   = {145},
  year    = {2023},
  doi     = {10.1038/s41597-023-02043-z}
}

@article{
AIMNet2rxn,
author = {Dylan M. Anstine  and Qiyuan Zhao  and Roman Zubatyuk  and Shuhao Zhang  and Veerupaksh Singla  and Filipp Nikitin  and Brett M. Savoie  and Olexandr Isayev },
title = {AIMNet2-rxn: A machine-learned potential for generalized reaction modeling on a millions-of-pathways scale},
journal = {Science Advances},
volume = {12},
number = {41},
pages = {eaea1557},
year = {2026},
doi = {10.1126/sciadv.aea1557},
URL = {https://www.science.org/doi/abs/10.1126/sciadv.aea1557},
eprint = {https://www.science.org/doi/pdf/10.1126/sciadv.aea1557},
}

@article{ANI1xBB,
  author  = {Zhang, Shuhao and Zubatyuk, Roman I. and Yang, Yinuo and Roitberg, Adrian and Isayev, Olexandr},
  title   = {{ANI-1xBB}: An {ANI}-Based Reactive Potential for Small Organic Molecules},
  journal = {Journal of Chemical Theory and Computation},
  volume  = {21},
  number  = {9},
  pages   = {4365--4374},
  year    = {2025},
  doi     = {10.1021/acs.jctc.5c00347}
}

@article{MDCD20,
author = {Guoao Li  and Haobo Ling  and Chaoxu Su  and Zhengxuan Liu  and Guoqiang Wang  and Manyi Yang  and Shuhua Li },
title = {A Data-Efficient Reactive Machine Learning Potential to Accelerate Automated Exploration of Complex Reaction Networks},
journal = {CCS Chemistry},
year = {2026},
note    = {https://doi.org/10.31635/ccschem.026.202607339},
doi = {10.31635/ccschem.026.202607339},

URL = {https://www.chinesechemsoc.org/doi/abs/10.31635/ccschem.026.202607339},
eprint = {https://www.chinesechemsoc.org/doi/pdf/10.31635/ccschem.026.202607339}

}

@misc{OMol25,
  author       = {Levine, Daniel S. and Shuaibi, Muhammed and Spotte-Smith, Evan Walter Clark and Taylor, Michael G. and Hasyim, Muhammad R. and Michel, Kyle and Batatia, Ilyes and Cs{\'a}nyi, G{\'a}bor and Dzamba, Misko and Eastman, Peter and Frey, Nathan C. and Fu, Xiang and Gharakhanyan, Vahe and Krishnapriyan, Aditi S. and Rackers, Joshua A. and Raja, Sanjeev and Rizvi, Ammar and Rosen, Andrew S. and Ulissi, Zachary and Vargas, Santiago and Zitnick, C. Lawrence and Blau, Samuel M. and Wood, Brandon M.},
  title        = {The Open Molecules 2025 (OMol25) Dataset, Evaluations, and Models},
  howpublished = {arXiv},
  year = {2025},
  eprint = {2505.08762},
  archivePrefix = {arXiv},
  primaryClass = {physics.chem-ph},
  doi = {10.48550/arXiv.2505.08762},
  note = {DOI: \href{https://doi.org/10.48550/arXiv.2505.08762}
                   {10.48550/arXiv.2505.08762}}
}

@article{Huber1994,
  title={Local elevation: a method for improving the searching properties of molecular dynamics simulation},
  author={Huber, Thomas and Torda, Andrew E and van Gunsteren, Wilfred F},
  journal={Journal of Computer-Aided Molecular Design},
  volume={8},
  number={6},
  pages={695--708},
  year={1994},
  publisher={Springer},
  doi={10.1007/BF00124016}
}

@article{Laio2002,
  author  = {Laio, Alessandro and Parrinello, Michele},
  title   = {Escaping Free-Energy Minima},
  journal = {Proceedings of the National Academy of Sciences of the United States of America},
  volume  = {99},
  number  = {20},
  pages   = {12562--12566},
  year    = {2002},
  doi     = {10.1073/pnas.202427399}
}

@article{Yang2022,
  author  = {Yang, Manyi and Bonati, Luigi and Polino, Daniela and Parrinello, Michele},
  title   = {Using Metadynamics to Build Neural Network Potentials for Reactive Events: The Case of Urea Decomposition in Water},
  journal = {Catalysis Today},
  volume  = {387},
  pages   = {143--149},
  year    = {2022},
  doi     = {10.1016/j.cattod.2021.03.018}
}

@article{Vitartas2025,
title = {Active learning meets metadynamics: automated workflow for reactive machine learning interatomic potentials},
journal = {Digital Discovery},
volume = {5},
number = {1},
pages = {108-122},
year = {2026},
issn = {2635-098X},
doi = {10.1039/d5dd00261c},
author = {Valdas Vitartas and Hanwen Zhang and Veronika Juraskova and Tristan Johnston-Wood and Fernanda Duarte},
}

@article{HeadGordon2023,
  author  = {Guan, Xingyi and Heindel, Joseph P. and Ko, Taehee and Yang, Chao and Head-Gordon, Teresa},
  title   = {Using Machine Learning to Go Beyond Potential Energy Surface Benchmarking for Chemical Reactivity},
  journal = {Nature Computational Science},
  volume  = {3},
  pages   = {965--974},
  year    = {2023},
  doi     = {10.1038/s43588-023-00549-5}
}

@article{Grimme2019,
  author  = {Grimme, Stefan},
  title   = {Exploration of Chemical Compound, Conformer, and Reaction Space with Meta-Dynamics Simulations Based on Tight-Binding Quantum Chemical Calculations},
  journal = {Journal of Chemical Theory and Computation},
  volume  = {15},
  number  = {5},
  pages   = {2847--2862},
  year    = {2019},
  doi     = {10.1021/acs.jctc.9b00143}
}

@misc{RXNxTBAL,
author = {Bowen Li  and Sixuan Mi  and Jin Xiao  and Duo Zhang  and Shuwen Zhang  and John Zhang  and Han Wang  and Tong Zhu },
title = {General reactive machine learning potentials for CHON elements},
howpublished = {ChemRxiv},
year = {2025},
doi = {10.26434/chemrxiv-2025-1d293-v2},
note = {DOI: \href{https://doi.org/10.26434/chemrxiv-2025-1d293-v2}
               {10.26434/chemrxiv-2025-1d293-v2}}
}

@article{SSW2013,
  author  = {Shang, Cheng and Liu, Zhi-Pan},
  title   = {Stochastic Surface Walking Method for Structure Prediction and Pathway Searching},
  journal = {Journal of Chemical Theory and Computation},
  year    = {2013},
  volume  = {9},
  number  = {3},
  pages   = {1838--1845},
  doi     = {10.1021/ct301010b}
}

@article{Ruddigkeit2012,
  author  = {Ruddigkeit, Lars and van Deursen, Ruud and Blum, Lorenz C. and Reymond, Jean-Louis},
  title   = {Enumeration of 166 Billion Organic Small Molecules in the Chemical Universe Database GDB-17},
  journal = {Journal of Chemical Information and Modeling},
  volume  = {52},
  number  = {11},
  pages   = {2864--2875},
  year    = {2012},
  doi     = {10.1021/ci300415d}
}

@article{Fukui1970,
  author  = {Fukui, Kenichi},
  title   = {Formulation of the reaction coordinate},
  journal = {The Journal of Physical Chemistry},
  volume  = {74},
  number  = {23},
  pages   = {4161--4163},
  year    = {1970},
  doi     = {10.1021/j100717a029}
}

@article{GonzalezJames2012,
  author  = {Gonzalez-James, Ollie M. and Kwan, Eugene E. and Singleton, Daniel A.},
  title   = {Entropic Intermediates and Hidden Rate-Limiting Steps in Seemingly Concerted Cycloadditions. Observation, Prediction, and Origin of an Isotope Effect on Recrossing},
  journal = {Journal of the American Chemical Society},
  volume  = {134},
  number  = {4},
  pages   = {1914--1917},
  year    = {2012},
  doi     = {10.1021/ja208779k},
  url     = {https://doi.org/10.1021/ja208779k}
}

@article{Stuart2000,
  author  = {Stuart, Steven J. and Tutein, Alan B. and Harrison, Judith A.},
  title   = {A reactive potential for hydrocarbons with intermolecular interactions},
  journal = {The Journal of Chemical Physics},
  volume  = {112},
  number  = {14},
  pages   = {6472--6486},
  year    = {2000},
  doi     = {10.1063/1.481208}
}

@article{GSM2013,
  author  = {Zimmerman, Paul M.},
  title   = {Reliable transition state searches integrated with the growing string method},
  journal = {Journal of Chemical Theory and Computation},
  volume  = {9},
  number  = {7},
  pages   = {3043--3050},
  year    = {2013},
  doi     = {10.1021/ct400319w}
}

@article{AFIR,
author = {Sameera, W. M. C. and Kumar Sharma, Akhilesh and Maeda, Satoshi and Morokuma, Keiji},
title = {Artificial Force Induced Reaction Method for Systematic Determination of Complex Reaction Mechanisms},
journal = {The Chemical Record},
volume = {16},
number = {5},
pages = {2349-2363},
year = {2016},
doi     = {10.1002/tcr.201600052}
}

@article{Gonzalez1989,
  author = {Gonzalez, Carlos and Schlegel, H. Bernhard},
  title = {An improved algorithm for reaction path following},
  journal = {The Journal of Chemical Physics},
  volume = {90},
  number = {4},
  pages = {2154-2161},
  year = {1989},
  doi = {10.1063/1.456010}
}

@article{Bussi2020Metadynamics,
  author  = {Bussi, Giovanni and Laio, Alessandro},
  title   = {Using metadynamics to explore complex free-energy landscapes},
  journal = {Nature Reviews Physics},
  year    = {2020},
  volume  = {2},
  pages   = {200--212},
  doi     = {10.1038/s42254-020-0153-0}
}

@article{EndExploreNetwork2020,
  author  = {Unsleber, Jan P. and Reiher, Markus},
  title   = {The Exploration of Chemical Reaction Networks},
  journal = {Annual Review of Physical Chemistry},
  year    = {2020},
  volume  = {71},
  number  = {1},
  pages   = {121--142},
  doi     = {10.1146/annurev-physchem-071119-040123}
}

@article{DiscriminantsCV2018,
  author  = {Piccini, GiovanniMaria and Mendels, Dan and Parrinello, Michele},
  title   = {Metadynamics with Discriminants: A Tool for Understanding Chemistry},
  journal = {Journal of Chemical Theory and Computation},
  year    = {2018},
  volume  = {14},
  number  = {10},
  pages   = {5040--5044},
  doi     = {10.1021/acs.jctc.8b00634}
}

@article{HLDA2019,
  title={Blind Search for Complex Chemical Pathways Using Harmonic Linear Discriminant Analysis},
  author={Rizzi, Valerio and Mendels, Dan and Sicilia, Emilia and Parrinello, Michele},
  journal={Journal of Chemical Theory and Computation},
  volume={15},
  number={8},
  pages={4507--4515},
  year={2019},
  publisher={American Chemical Society},
  doi={10.1021/acs.jctc.9b00358}
}

@article{GNNCV2024,
  author  = {Zhang, Jintu and Bonati, Luigi and Trizio, Enrico and Zhang, Odin and Kang, Yu and Hou, TingJun and Parrinello, Michele},
  title   = {Descriptor-Free Collective Variables from Geometric Graph Neural Networks},
  journal = {Journal of Chemical Theory and Computation},
  year    = {2024},
  volume  = {20},
  number  = {24},
  pages   = {10787--10797},
  doi     = {10.1021/acs.jctc.4c01197}
}

@article{Grambow2020,
  author  = {Grambow, Colin A. and Pattanaik, Lagnajit and Green, William H.},
  title   = {Reactants, products, and transition states of elementary chemical reactions based on quantum chemistry},
  journal = {Scientific Data},
  year    = {2020},
  volume  = {7},
  number  = {1},
  pages   = {137},
  doi     = {10.1038/s41597-020-0460-4}
}

@article{YARP2021,
  author  = {Zhao, Qiyuan and Savoie, Brett M.},
  title   = {Simultaneously improving reaction coverage and computational cost in automated reaction prediction tasks},
  journal = {Nature Computational Science},
  year    = {2021},
  volume  = {1},
  pages   = {479--490},
  doi     = {10.1038/s43588-021-00101-3}
}

@inbook{NEB1998,
  author    = {J{\'o}nsson, Hannes and Mills, Greg and Jacobsen, Karsten W.},
  title     = {Nudged Elastic Band Method for Finding Minimum Energy Paths of Transitions},
booktitle = {Classical and Quantum Dynamics in Condensed Phase Simulations},
chapter = {},
pages = {385-404},
editor = {Berne, Bruce J. and
        Ciccotti, Giovanni and
        Coker, David F.},
publisher = {World Scientific},
address = {Singapore},
year = {1998},
doi = {10.1142/9789812839664_0016},
URL = {https://www.worldscientific.com/doi/abs/10.1142/9789812839664_0016},
eprint = {https://www.worldscientific.com/doi/pdf/10.1142/9789812839664_0016}
}

@article{enhancesamplingReview2019,
    author = {Yang, Yi Isaac and Shao, Qiang and Zhang, Jun and Yang, Lijiang and Gao, Yi Qin},
    title = {Enhanced sampling in molecular dynamics},
    journal = {The Journal of Chemical Physics},
    volume = {151},
    number = {7},
    pages = {070902},
    year = {2019},
    month = {08},
    issn = {0021-9606},
    doi = {10.1063/1.5109531},
    url = {https://doi.org/10.1063/1.5109531},
    eprint = {https://pubs.aip.org/aip/jcp/article-pdf/doi/10.1063/1.5109531/19987375/070902_1_1.5109531.pdf},
}

@article{MDCD2018,
author = {Yang, Manyi and Yang, Lijiang and Wang, Guoqiang and Zhou, Yanzi and Xie, Daiqian and Li, Shuhua},
title = {Combined Molecular Dynamics and Coordinate Driving Method for Automatic Reaction Pathway Search of Reactions in Solution},
journal = {Journal of Chemical Theory and Computation},
volume = {14},
number = {11},
pages = {5787-5796},
year = {2018},
doi = {10.1021/acs.jctc.8b00799},
note ={PMID: 30351922}
}

@article{ANI2x,
  title={{Extending the Applicability of the ANI Deep Learning Molecular Potential to Sulfur and Halogens}},
  author={Devereux, Christian and Smith, Justin S. and Huddleston, Kate K. and Barros, Kipton and Zubatyuk, Roman and Isayev, Olexandr and Roitberg, Adrian E.},
  journal={Journal of Chemical Theory and Computation},
  volume={16},
  number={7},
  pages={4192--4202},
  year={2020},
  publisher={ACS Publications},
  doi={10.1021/acs.jctc.0c00121}
}

@article{SPICE2023,
  title={{SPICE, A Dataset of Drug-like Molecules and Peptides for Training Machine Learning Potentials}},
  author={Eastman, Peter and Behara, Pavan Kumar and Dotson, David L. and Galvelis, Raimondas and Herr, John E. and Horton, Josh T. and Mao, Yuezhi and Chodera, John D. and Pritchard, Benjamin P. and Wang, Yuanqing and De Fabritiis, Gianni and Markland, Thomas E.},
  journal={Scientific Data},
  volume={10},
  number={1},
  pages={11},
  year={2023},
  publisher={Nature Publishing Group},
  doi={10.1038/s41597-022-01882-6}
}

@article{SPICE2andNutMeg2024,
  title={{Nutmeg and SPICE}: Models and Data for Biomolecular Machine Learning},
  author={Eastman, Peter and Pritchard, Benjamin P. and Chodera, John D. and Markland, Thomas E.},
  journal={Journal of Chemical Theory and Computation},
  volume={20},
  number={19},
  pages={8583--8593},
  year={2024},
  publisher={ACS Publications},
  doi={10.1021/acs.jctc.4c00794}
}

@article{geom2022,
  title={{GEOM, energy-annotated molecular conformations for property prediction and molecular generation}},
  author={Axelrod, Simon and G{\'o}mez-Bombarelli, Rafael},
  journal={Scientific Data},
  volume={9},
  number={1},
  pages={185},
  year={2022},
  publisher={Nature Publishing Group},
  doi={10.1038/s41597-022-01288-4}
}

@article{OrbNetDenali2021,
  title={{OrbNet Denali}: A machine learning potential for biological and organic chemistry with semi-empirical cost and {DFT} accuracy},
  author={Christensen, Anders S. and Sirumalla, Sai Krishna and Qiao, Zhuoran and O'Connor, Michael B. and Smith, Daniel G. A. and Ding, Feizhi and Bygrave, Peter J. and Anandkumar, Animashree and Welborn, Matthew and Manby, Frederick R. and Miller, III, Thomas F.},
  journal={The Journal of Chemical Physics},
  volume={155},
  number={20},
  pages={204103},
  year={2021},
  publisher={AIP Publishing},
  doi={10.1063/5.0061990}
}

@article{pmechdb2024,
  title={{PMechDB}: A Public Database of Elementary Polar Reaction Steps},
  author={Tavakoli, Mohammadamin and Miller, Ryan J. and Angel, Mirana Claire and Pfeiffer, Michael A. and Gutman, Eugene S. and Mood, Aaron D. and Van Vranken, David and Baldi, Pierre},
  journal={Journal of Chemical Information and Modeling},
  volume={64},
  number={6},
  pages={1975--1983},
  year={2024},
  publisher={ACS Publications},
  doi={10.1021/acs.jcim.3c01810}
}

@article{rmechdb2023,
  title={{RMechDB}: A Public Database of Elementary Radical Reaction Steps},
  author={Tavakoli, Mohammadamin and Chiu, Yin Ting T. and Baldi, Pierre and Carlton, Ann Marie and Van Vranken, David},
  journal={Journal of Chemical Information and Modeling},
  volume={63},
  number={4},
  pages={1114--1123},
  year={2023},
  publisher={ACS Publications},
  doi={10.1021/acs.jcim.2c01359}
}

@article{MACE-MH2025crosslearning,
  author        = {Batatia, Ilyes and Lin, Chen and Hart, Joseph and Kasoar, Elliott and Elena, Alin M. and Norwood, Sam Walton and Wolf, Thomas and Cs{\'a}nyi, G{\'a}bor},
  title         = {Cross Learning between Electronic Structure Theories for Unifying Molecular, Surface, and Inorganic Crystal Foundation Force Fields},
  journal   = {arXiv},
  year          = {2025},
  eprint        = {2510.25380},
  archivePrefix = {arXiv},
  primaryClass  = {physics.chem-ph},
  note           = {10.48550/arXiv.2510.25380}
}

@article{NeuralNEB,
  author    = {Mathias Schreiner and Arghya Bhowmik and Tejs Vegge and Peter Bj{\o}rn J{\o}rgensen and Ole Winther},
  title     = {{NeuralNEB}---neural networks can find reaction paths fast},
  journal   = {Machine Learning: Science and Technology},
  volume    = {3},
  number    = {4},
  pages     = {045022},
  year      = {2022},
  publisher = {IOP Publishing},
  doi       = {10.1088/2632-2153/aca23e},
  url       = {https://doi.org/10.1088/2632-2153/aca23e}
}

@article{BH9,
  author  = {Prasad, Viki Kumar and Pei, Zhipeng and Edelmann, Simon and Otero-de-la-Roza, Alberto and DiLabio, Gino A.},
  title   = {{BH9}, a New Comprehensive Benchmark Data Set for Barrier Heights and Reaction Energies: Assessment of Density Functional Approximations and Basis Set Incompleteness Potentials},
  journal = {Journal of Chemical Theory and Computation},
  volume  = {18},
  number  = {1},
  pages   = {151--166},
  year    = {2022},
  doi     = {10.1021/acs.jctc.1c00694}
}

@article{BH9Correction,
  author  = {Prasad, Viki Kumar and Pei, Zhipeng and Edelmann, Simon and Otero-de-la-Roza, Alberto and DiLabio, Gino A.},
  title   = {Correction to ``{BH9}, a New Comprehensive Benchmark Data Set for Barrier Heights and Reaction Energies: Assessment of Density Functional Approximations and Basis Set Incompleteness Potentials''},
  journal = {Journal of Chemical Theory and Computation},
  volume  = {18},
  number  = {6},
  pages   = {4041--4044},
  year    = {2022},
  doi     = {10.1021/acs.jctc.2c00362}
}

@article{Cyclo3+2,
  author  = {Stuyver, Thijs and Jorner, Kjell and Coley, Connor W.},
  title   = {Reaction profiles for quantum chemistry-computed {[3 + 2]} cycloaddition reactions},
  journal = {Scientific Data},
  volume  = {10},
  number  = {1},
  pages   = {66},
  year    = {2023},
  doi     = {10.1038/s41597-023-01977-8}
}

@article{Yang2025DPA2Drug,
  author  = {Yang, Manyi and Zhang, Duo and Wang, Xinyan and Li, BoWen and Zhang, Linfeng and E, Weinan and Zhu, Tong and Wang, Han},
  title   = {Ab Initio Accuracy Neural Network Potential for Drug-Like Molecules},
  journal = {Research},
  volume  = {8},
  pages   = {0837},
  year    = {2025},
  doi     = {10.34133/research.0837}
}

@article{DPA2,
  author  = {Zhang, Duo and Liu, Xinzijian and Zhang, Xiangyu and Zhang, Chengqian and Cai, Chun and Bi, Hangrui and Du, Yiming and Qin, Xuejian and Peng, Anyang and Huang, Jiameng and Li, Bowen and Shan, Yifan and Zeng, Jinzhe and Zhang, Yuzhi and Liu, Siyuan and Li, Yifan and Chang, Junhan and Wang, Xinyan and Zhou, Shuo and Liu, Jianchuan and Luo, Xiaoshan and Wang, Zhenyu and Jiang, Wanrun and Wu, Jing and Yang, Yudi and Yang, Jiyuan and Yang, Manyi and Gong, Fu-Qiang and Zhang, Linshuang and Shi, Mengchao and Dai, Fu-Zhi and York, Darrin M. and Liu, Shi and Zhu, Tong and Zhong, Zhicheng and Lv, Jian and Cheng, Jun and Jia, Weile and Chen, Mohan and Ke, Guolin and E, Weinan and Zhang, Linfeng and Wang, Han},
  title   = {{DPA-2}: a large atomic model as a multi-task learner},
  journal = {npj Computational Materials},
  volume  = {10},
  number  = {1},
  pages   = {293},
  year    = {2024},
  doi     = {10.1038/s41524-024-01493-2}
}

@article{BRICS,
  author  = {Degen, J{\"o}rg and Wegscheid-Gerlach, Christof and Zaliani, Andrea and Rarey, Matthias},
  title   = {On the Art of Compiling and Using 'Drug-Like' Chemical Fragment Spaces},
  journal = {ChemMedChem},
  volume  = {3},
  number  = {10},
  pages   = {1503--1507},
  year    = {2008},
  doi     = {10.1002/cmdc.200800178}
}

@article{RECAP,
  author  = {Lewell, Xiao Qing and Judd, Duncan B. and Watson, Stephen P. and Hann, Michael M.},
  title   = {{RECAP}---Retrosynthetic Combinatorial Analysis Procedure: A Powerful New Technique for Identifying Privileged Molecular Fragments with Useful Applications in Combinatorial Chemistry},
  journal = {Journal of Chemical Information and Computer Sciences},
  volume  = {38},
  number  = {3},
  pages   = {511--522},
  year    = {1998},
  doi     = {10.1021/ci970429i}
}

@article{Murcko,
  author  = {Bemis, Guy W. and Murcko, Mark A.},
  title   = {The Properties of Known Drugs. 1. Molecular Frameworks},
  journal = {Journal of Medicinal Chemistry},
  volume  = {39},
  number  = {15},
  pages   = {2887--2893},
  year    = {1996},
  doi     = {10.1021/jm9602928}
}

@article{Dimer1999,
  author  = {Henkelman, Graeme and J{\'o}nsson, Hannes},
  title   = {A dimer method for finding saddle points on high dimensional potential surfaces using only first derivatives},
  journal = {The Journal of Chemical Physics},
  volume  = {111},
  number  = {15},
  pages   = {7010--7022},
  year    = {1999},
  doi     = {10.1063/1.480097}
}

@article{wB97XD2008,
  author  = {Chai, Jeng-Da and Head-Gordon, Martin},
  title   = {Long-range corrected hybrid density functionals with damped atom-atom dispersion corrections},
  journal = {Physical Chemistry Chemical Physics},
  volume  = {10},
  number  = {44},
  pages   = {6615--6620},
  year    = {2008},
  doi     = {10.1039/B810189B}
}

@misc{g16,
author={M. J. Frisch and G. W. Trucks and H. B. Schlegel and G. E. Scuseria and M. A. Robb and J. R. Cheeseman and G. Scalmani and V. Barone and G. A. Petersson and H. Nakatsuji and X. Li and M. Caricato and A. V. Marenich and J. Bloino and B. G. Janesko and R. Gomperts and B. Mennucci and H. P. Hratchian and J. V. Ortiz and A. F. Izmaylov and J. L. Sonnenberg and D. Williams-Young and F. Ding and F. Lipparini and F. Egidi and J. Goings and B. Peng and A. Petrone and T. Henderson and D. Ranasinghe and V. G. Zakrzewski and J. Gao and N. Rega and G. Zheng and W. Liang and M. Hada and M. Ehara and K. Toyota and R. Fukuda and J. Hasegawa and M. Ishida and T. Nakajima and Y. Honda and O. Kitao and H. Nakai and T. Vreven and K. Throssell and Montgomery, {Jr.}, J. A. and J. E. Peralta and F. Ogliaro and M. J. Bearpark and J. J. Heyd and E. N. Brothers and K. N. Kudin and V. N. Staroverov and T. A. Keith and R. Kobayashi and J. Normand and K. Raghavachari and A. P. Rendell and J. C. Burant and S. S. Iyengar and J. Tomasi and M. Cossi and J. M. Millam and M. Klene and C. Adamo and R. Cammi and J. W. Ochterski and R. L. Martin and K. Morokuma and O. Farkas and J. B. Foresman and D. J. Fox},
title={Gaussian~16 {R}evision {C}.01},
year={2016},
note={Gaussian Inc. Wallingford CT}
}

@article{Stable_WF_Seeger1977,
  author  = {Seeger, Rolf and Pople, John A.},
  title   = {Self-consistent molecular orbital methods. XVIII. Constraints and stability in Hartree--Fock theory},
  journal = {The Journal of Chemical Physics},
  volume  = {66},
  number  = {7},
  pages   = {3045--3050},
  year    = {1977},
  doi     = {10.1063/1.434318}
}

@article{StabLe_DFT_Bauernschmitt1996,
  author  = {Bauernschmitt, R{\"u}diger and Ahlrichs, Reinhart},
  title   = {Stability analysis for solutions of the closed shell Kohn--Sham equation},
  journal = {The Journal of Chemical Physics},
  volume  = {104},
  number  = {22},
  pages   = {9047--9052},
  year    = {1996},
  doi     = {10.1063/1.471637}
}

@article{BS_Isobe2003DielsAlder,
  author  = {Isobe, Hiroshi and Takano, Yu and Kitagawa, Yasutaka and Kawakami, Takashi and Yamanaka, Syusuke and Yamaguchi, Kizashi and Houk, K. N.},
  title   = {Systematic Comparisons between Broken Symmetry and Symmetry-Adapted Approaches to Transition States by Chemical Indices: A Case Study of the Diels--Alder Reactions},
  journal = {The Journal of Physical Chemistry A},
  volume  = {107},
  number  = {5},
  pages   = {682--694},
  year    = {2003},
  doi     = {10.1021/jp021125o}
}

@article{BS_good_approxi_MR_but_kink,
  author  = {Kedziora, Gary S. and Barr, Stephen A. and Berry, Rajiv and Moller, James C. and Breitzman, Timothy D.},
  title   = {Bond breaking in stretched molecules: multi-reference methods versus density functional theory},
  journal = {Theoretical Chemistry Accounts},
  volume  = {135},
  number  = {3},
  pages   = {79},
  year    = {2016},
  doi     = {10.1007/s00214-016-1822-z}
}

@article{BS_rxn_mechanism_plannar_2015,
  author  = {Skraba-Joiner, Sarah L. and Johnson, Richard P. and Agarwal, Jay},
  title   = {Dehydropericyclic Reactions: Symmetry-Controlled Routes to Strained Reactive Intermediates},
  journal = {The Journal of Organic Chemistry},
  volume  = {80},
  number  = {23},
  pages   = {11779--11787},
  year    = {2015},
  doi     = {10.1021/acs.joc.5b01488}
}

@article{BS_MD_nonstatic,
  author  = {Hamaguchi, Masashi and Nakaishi, Masahiro and Nagai, Toshikazu and Nakamura, Takeshi and Abe, Manabu},
  title   = {Notable Effect of an Electron-Withdrawing Group at {C3} on the Selective Formation of Alkylidenecyclobutanes in the Thermal Denitrogenation of 4-Spirocyclopropane-1-pyrazolines. Nonstatistical Dynamics Effects in the Denitrogenation Reactions},
  journal = {Journal of the American Chemical Society},
  volume  = {129},
  number  = {43},
  pages   = {12981--12988},
  year    = {2007},
  doi     = {10.1021/ja068513e}
}

@article{LAMMPS2022,
  author  = {Thompson, Aidan P. and Aktulga, H. Metin and Berger, Richard and Bolintineanu, Dan S. and Brown, W. Michael and Crozier, Paul S. and in 't Veld, Pieter J. and Kohlmeyer, Axel and Moore, Stan G. and Nguyen, Trung Dac and Shan, Ray and Stevens, Mark J. and Tranchida, Julien and Trott, Christian and Plimpton, Steven J.},
  title   = {{LAMMPS} -- a flexible simulation tool for particle-based materials modeling at the atomic, meso, and continuum scales},
  journal = {Computer Physics Communications},
  volume  = {271},
  pages   = {108171},
  year    = {2022},
  doi     = {10.1016/j.cpc.2021.108171}
}

@article{PLUMED2,
  author  = {Tribello, Gareth A. and Bonomi, Massimiliano and Branduardi, Davide and Camilloni, Carlo and Bussi, Giovanni},
  title   = {{PLUMED} 2: New feathers for an old bird},
  journal = {Computer Physics Communications},
  volume  = {185},
  number  = {2},
  pages   = {604--613},
  year    = {2014},
  doi     = {10.1016/j.cpc.2013.09.018}
}

@article{DPA3zhang2025graph,
  author = {Zhang, Duo and
            Peng, Anyang and
            Cai, Chun and
            Li, Wentao and
            Zhou, Yuanchang and
            Zeng, Jinzhe and
            Guo, Mingyu and
            Zhang, Chengqian and
            Li, Bowen and
            Jiang, Hong and
            Zhu, Tong and
            Jia, Weile and
            Zhang, Linfeng and
            Wang, Han},
  title   = {A graph neural network for the era of large atomistic models},
  journal = {npj Computational Materials},
  year = {2026},
  volume = {12},
  pages = {276},
  doi = {10.1038/s41524-026-02146-2}
}

@article{DPGENZHANG2020107206,
title = {DP-GEN: A concurrent learning platform for the generation of reliable deep learning based potential energy models},
journal = {Computer Physics Communications},
volume = {253},
pages = {107206},
year = {2020},
issn = {0010-4655},
doi = {10.1016/j.cpc.2020.107206},
url = {https://www.sciencedirect.com/science/article/pii/S001046552030045X},
author = {Yuzhi Zhang and Haidi Wang and Weijie Chen and Jinzhe Zeng and Linfeng Zhang and Han Wang and Weinan E},

}

@misc{mace_omol_4m,
  author       = {{ACEsuit}},
  title        = {A {MACE} model trained on the {4M} training split of {OMol25}},
  year         = {2025},
  howpublished = {GitHub release asset, {ACEsuit}/mace-foundations},
  url          = {https://github.com/ACEsuit/mace-foundations/releases/tag/mace_omol_0}
}

@misc{dpgen2_github,
  author       = {{deepmodeling}},
  title        = {{DPGEN2}: 2nd generation of the Deep Potential GENerator},
  year         = {2024},
  howpublished = {GitHub repository},
  url          = {https://github.com/deepmodeling/dpgen2}
}

@article{AIMNet2,
  author  = {Anstine, Dylan M. and Zubatyuk, Roman and Isayev, Olexandr},
  title   = {AIMNet2: a neural network potential to meet your neutral, charged, organic, and elemental-organic needs},
  journal = {Chemical Science},
  volume  = {16},
  pages   = {10228--10244},
  year    = {2025},
  doi     = {10.1039/D4SC08572H}
}

@misc{dpa4,
      title={DPA4: Pushing the Accuracy-Cost Frontier of Interatomic Potentials with {EMFA} {SO(2)} Convolution}, 
      author={Tiancheng Li and Wentao Li and Anyang Peng and Jianming Xue and Linfeng Zhang and Duo Zhang and Han Wang},
      year={2026},
      note={arXiv:2606.02419},
      eprint={2606.02419},
      archivePrefix={arXiv},
      primaryClass={physics.chem-ph},
      url={https://arxiv.org/abs/2606.02419}, 
}
